\documentclass[referee]{raa}  
\usepackage[flushleft]{threeparttable}
\usepackage{blindtext}
\usepackage{amsmath}
\usepackage{mathtools}
\usepackage{multirow}
\usepackage{comment}
\usepackage{enumerate}
\usepackage{enumitem}
\usepackage{natbib} %\usepackage{natbib}
\usepackage{amssymb}

\let\oldAA\AA
\renewcommand{\AA}{\text{\normalfont\oldAA}}

\usepackage{amsmath,amssymb,bm}
\usepackage[table]{xcolor}
\usepackage{booktabs}
\usepackage{multirow}
\usepackage{threeparttable}

\newcommand{\tSD}{t_{\rm NS, SpD}}
\newcommand{\tQN}{t_{\rm QN}}

\newcommand{\tSNR}{t_{\rm SNR}}

\newcommand{\BHS}{B_{\rm HS}}
\newcommand{\Bc}{B_{\rm fossil, c}}
\newcommand{\Msun}{M_\odot}
\newcommand{\Mdec}{M_{\rm dec}}
\newcommand{\Pfast}{P_{\rm fast}}
\newcommand{\rhodec}{\rho_{\rm dec}}
\newcommand{\rhoc}{\rho_{\rm c}}
\newcommand{\rhoNuc}{\rho_{\rm nuc}}

\newcommand{\fdec}{f_{\rm dec}}
\newcommand{\ffoss}{f_{\rm foss}}

\newcommand{\flfbot}{f_{\rm lfbot}}
\newcommand{\fslsn}{f_{\rm slsn}}

\newcommand{\sigM}{\sigma_M}
\newcommand{\sigP}{\sigma_P}
\newcommand{\sigB}{\sigma_B}

\newcommand{\ffast}{f_{\rm fast}}

\providecommand{\plb}{Phys.\ Lett.\ B}
\providecommand{\rmp}{Rev.\ Mod.\ Phys.}
\providecommand{\prl}{Phys.\ Rev.\ Lett.}
\providecommand{\prd}{Phys.\ Rev.\ D}

\begin{document}

\title{Beyond Spin: QCD Magnetars}

%\volnopage{ {\bf 20XX} Vol.\ {\bf X} No. {\bf XX}, 000--000}
\setcounter{page}{1}

\author{Rachid Ouyed}
\institute{Department of Physics and Astronomy, University of Calgary, Calgary, AB, Canada; {\it rouyed@ucalgary.ca}}
 
\abstract{
We present a unified framework in which anomalous X-ray pulsars (AXPs), soft gamma-ray repeaters (SGRs), superluminous supernovae (SLSNe-I), luminous fast blue optical transients (LFBOTs), and fast radio bursts (FRBs) originate from quark deconfinement in the core of a massive neutron star (NS). Spontaneous ferromagnetism in the deconfined phase generates $\sim10^{18}$\,G core fields, producing a highly magnetized hybrid star (HS) -- the ``QCD magnetar'' -- whose surface field is set by NS mass, not birth spin. The Quark-Nova (QN) that forms the HS ejects $\sim10^{-2}\,M_\odot$ of neutron-rich outer layers, powering a kilonova and leaving the HS crustless for centuries. During this phase, magnetic instabilities at the hadron--quark interface release rising flux ropes that power X-ray-quiet FRBs; as the crust reforms, the source evolves into an X-ray-loud AXP\&SGR. Two parameters govern the model: a critical mass $\Mdec$ triggering deconfinement, and a critical period $P_{\rm fast}$ separating fast and slow rotators. Fast rotators inject spin-down energy into the QN ejecta, producing an LFBOT -- directly observable once the SN ejecta is optically thin, or via binary accretion with no preceding SN; otherwise the LFBOT is reprocessed by the SN ejecta into an SLSN-I. A Bayesian Monte Carlo population synthesis reproduces the observed rates of AXPs\&SGRs, SLSNe-I, and LFBOTs with $\Mdec\sim2.1\,M_\odot$ and $P_{\rm fast}\sim5.5$~ms, and predicts non-merger r-process signatures and kilonovae from isolated NSs, with or without an LFBOT. The QN ejecta also carries its own DM and RM, independent of environment, predicted to appear as excess dispersion and rotation measure in FRBs in addition to the Galactic, host, and intergalactic contributions. These provide direct observational tests of the hadron--quark phase transition and ferromagnetic ordering in dense quark matter.
\keywords{dense matter --  (stars:) supernovae: general -- (stars:) neutron -- interiors -- magnetic fields}}

\authorrunning{Rachid Ouyed}            %author_head in even pages
\titlerunning{Beyond Spin: QCD Magnetars}  % title_head in odd pages
\maketitle
% ════════════════════════════════════════════════════════════════════════

\section{Introduction}
\label{sec:intro}

Anomalous X-ray Pulsars (AXPs) and Soft Gamma Repeaters (SGRs) are
slowly rotating (with period $\sim2$--$12$~s), strongly magnetized neutron stars
(NSs) whose persistent X-ray luminosities ($\sim10^{33}$--$10^{36}$~erg~s$^{-1}$)
exceed what can be supplied by rotation alone \citep[e.g.,][]{Mereghetti1995,Kouveliotou1998,Hurley1999,Woods2006,
Mereghetti2008,Rea2011,Kaspi2017}.
Their defining observational hallmarks, including sporadic short bursts and occasional giant flares, are best explained by the release of magnetic energy stored in an ultra-strong surface magnetic field, $\sim10^{14}$--$10^{15}$~G, with the bursts triggered by magnetic stresses that crack (or plastically deform) the NS crust and/or drive magnetospheric reconnection \citep{Duncan1992,Thompson1993,Thompson1995,Lyutikov2003}.
These objects are collectively referred to as magnetars, and their
Galactic birth rate is estimated at $\sim5$--$15\%$ of the
core-collapse supernova (CCSN) rate \citep[e.g.,][]{GillHeyl2007,Beniamini2019}.

Superluminous supernovae (SLSNe) are among the most luminous stellar
explosions in the Universe, with peak luminosities $\sim10$--$100$
times greater than those of normal SNe
\citep{Ofek2007,Quimby2007a,Quimby2007b,Smith2008,Barbary2009,Young2010,Quimby2011,
Gal-Yam2012,Moriya2024}.
The hydrogen-poor subclass (SLSNe-I), relevant to this paper, 
are extremely rare, with a volumetric rate of roughly $1$ in every
$10^4$ CCSNe \citep{Quimby2013,Frohmaier2021}.
Luminous Fast Blue Optical Transients
\citep[LFBOTs;][]{Drout2014,Pursiainen2018,Metzger2022} constitute a
distinct class of engine-powered transients
characterised by rise times of a few days, blue colours, and peak
luminosities $> 10^{44}$ ~erg~s$^{-1}$.
The volumetric rate of LFBOTS is estimated to be $< 0.1\%$ of the CCSN
rate \citep{Coppejans2020,Ho2023}.

AXPs\&SGRs, SLSNe-I, and LFBOTs are observationally diverse but share
a common thread: each class appeals to the magnetar as the central engine.
The most energetic events require a rapidly rotating magnetar whose
spin-down luminosity powers the observed emission
\citep{Kasen2010,Woosley2010,Metzger2015}.
Fast Radio Bursts (FRBs; \citealt{Lorimer2007}) are also expected to be powered by magnetars
\citep[][to cite only a few]{Lyubarsky2014,Beloborodov2017,Margalit2018,Lu2020,Lyubarsky2021}.
Gamma-ray bursts (GRBs) provide another possible connection: the shallow
X-ray plateaus observed in a large fraction of \textit{Swift}
afterglows are well modelled by energy injection from a
millisecond magnetar spinning down on timescales of hours  \citep[e.g.,][]{Metzger2011}.

Despite decades of study, the physical origin of the extreme magnetic field 
remains an open question.  In the proposed models, the magnetar field is acquired during the
proto-NS (PNS) phase, which lasts only a few seconds, via amplification mechanisms. This includes: 
convective dynamo action driven by rotation
\citep{Thompson1993,Raynaud2020,Masada2022,White2022},
the magnetorotational instability (MRI; \citealt{BalbusHawley1991})
in a differentially rotating PNS
\citep{Obergaulinger2009,Siegel2013,Mosta2014,Kiuchi2015,Mosta2015,Rembiasz2017,ReboulSalze2022},
the Tayler-Spruit instability in the stably stratified stellar interior
\citep{Spruit2002,Barrere2023}, and the standing accretion shock
instability \citep{Endeve2012}. Each of these mechanisms requires special circumstances, particularly rapidly rotating 
PNS birth spin periods $P_0\lesssim6$~ms. Compression of a fossil progenitor magnetic field during core collapse
\citep{Ferrario2006} is another mechanism which however places strong
requirements on the progenitor field distribution and  cannot account
for the whole magnetar population (\citealt{Makarenko2021}). 

The fraction of rapidly rotating NSs  (with birth period $P_0 < 10$ ms) inferred from the measured 
radio pulsar birth distribution \citep{FaucherKaspi2006,Igoshev2022} does not exceed $\sim1\%$, far below the observed 
AXP\&SGR birth rate.
A second tension concerns the SNRs associated with known AXPs\&SGRs.
If the magnetic field in these objects is related to rapid spin, one would expect
their natal SNe to be over-energised relative to ordinary CCSNe.
Yet the SNRs associated with AXPs\&SGRs appear unremarkable in
their expansion velocities \citep{Vink2006}.
The coupling between rotational energy and SN ejecta therefore
appears negligible for AXPs\&SGRs, yet approaches almost optimum 
efficiency in the engine models required for the most energetic and superluminous transients. 

The thermalization efficiency of the pulsar wind nebula, which determines the coupling efficiency, depends on the wind magnetization and the SN ejecta optical depth, neither of which is expected to systematically differ between these progenitor classes. While processes such as gravitational-wave emission (e.g., \citealt{Dallosso2009,Jones2010}), neutrino cooling during the PNS  phase (e.g., \citealt{Thompson2004,Dvornikov2010,SurHaskell2021}), or jet breakout (e.g., \citealt{Bucciantini2009,Bromberg2011}) may remove rotational energy before it thermalizes in the ejecta, they are unlikely to  fully explain this inconsistency. 

Related work has also explored how strong magnetic fields affect the
structure and composition of HSs. Such fields can modify
the particle content, spin polarization, and magnetization of matter
across the hadronic, mixed, and quark phases \citep{Dexheimer2013},
while also shifting the deconfinement transition and affecting global
stellar properties such as the mass--radius relation and moment of
inertia \citep{Dexheimer2012,Mariani2019}. In these studies, the
magnetic field is generally prescribed through a density- or
chemical-potential-dependent profile, so that its strength is an input
rather than a quantity derived from the properties of the quark phase.
\citet{Franzon2016} instead treated the magnetic field and stellar
structure self-consistently through the coupled Maxwell--Einstein
equations for an assumed poloidal field geometry, demonstrating that
strong fields can substantially modify the internal structure of HSs.
These studies establish that strong magnetic fields can have important
consequences for HS structure and deconfinement, but they do not connect
the origin and magnitude of the field to the deconfinement transition
itself, nor do they use observed transient-population rates to constrain
the deconfinement threshold. The framework developed in this paper addresses these questions by
linking the ultra-strong field to the deconfined quark phase and by
using population-level transient rates to constrain the conditions
for deconfinement.

Here, we propose that rather than tying the field to the birth spin, we tie it to
\textit{mass}. NSs born above a critical gravitational mass $\Mdec$ have central
densities exceeding the quark deconfinement threshold value $\rhodec$ from
birth; the subscript ``dec" stands for deconfinement. This is the density when quark bubbles start to form in the NS core
but as we find here, these bubbles are contained until NS spin-down 
induces a pressure change in the core at time $t=\tSD$; the NS spin-down (SpD)  timescale.  
This triggers immediate and complete percolation of the quark phase converting the NS core in a matter of milliseconds.
 These massive NSs are therefore not ordinary NSs but pre-hybrid stars (pre-HSs):
objects whose cores are poised to convert to quark matter as soon as
the thermodynamic conditions allow it at around $t=\tSD$.

We refer to the NS-to-HS conversion as a Quark-Nova (QN). In the scenario considered here, the QN leads to a partial deconfinement transition in which only the central region of the NS converts into a $(u,d)$ quark core, producing a strongly magnetized HS \citep{Ouyed2025,Ouyed2026a,Ouyed2026b}. In principle, the QN may also trigger a complete conversion of the NS  into a quark star (\citealt{Ouyed2022a,Ouyed2022b}) composed of strange quark matter ({\it (u,d,s)} matter; \citealt{Itoh1970,Bodmer1971,Terazawa1979,Witten1984}). However, this possibility relies on the still unconfirmed stability of {\it (u,d,s)} matter and the existence of strange quark stars.  
In addition, despite assuming absolutely stable strange quark matter, 
three-dimensional hydrodynamic simulations
of the combustion of a NS into a quark star found that the conversion is partial 
 because the conversion front stops
when it reaches conditions under which the combustion is no longer exothermic \citep{Herzog2011}.
We therefore focus on the more conservative partial conversion scenario, which does not require these additional assumptions.

This partial conversion also allows {\it (u,d)} quark phase that can be ferromagnetic in nature and can generate an $B_{\rm QCD} \sim10^{18}$~G QCD magnetic
field spontaneously in the HS core
which translates to a $B_{\rm HS} \sim10^{15}$~G surface field (see \S \ref{sec:ferromag} and Appendix \ref{sec:Bqcd_origin}).
 Because of this, we refer to the QN compact remnant, interchangeably with the highly magnetized HS, as the ``QCD magnetar" where
 QCD stands for Quantum-Chromo-Dynamics.  
 
The magnetic free energy available at the hadron--quark interface is
estimated from the mismatch between the core field, $B_{\rm core}=B_\mathrm{QCD}\sim
10^{18}\,$G, and the $B_{\rm env}=B_\mathrm{had} \sim 10^{15}$ G at the base of the hadronic envelope field extrapolated inward from its surface
value ($\sim10^{12.5}\,$G for a typical NS).
Because $B_\mathrm{env}\ll B_\mathrm{core}$, the available energy
density is dominated entirely by the core field,
$B_\mathrm{core}^2/8\pi\approx4\times10^{34}\,\mathrm{erg\,cm^{-3}}$,
essentially independent of the precise envelope field value. Integrating
over an interface shell sitting at a radius $R_{\rm c}\sim 1$-3 km  and 
of thickness say $\Delta R_{\rm c}\sim 0.01R_{\rm c}$
yields a total available reservoir
$\sim 5\times10^{48}$--$10^{50}\,$erg (see \S\ref{sec:qcd} for the associated dynamics). 

This energy is released gradually over centuries through the continuous readjustment of the interface as the QCD magnetar spins down, thereby reducing the mismatch across the interface. In the context of FRBs, as discussed in \S~\ref{sec:frb}, this energy reservoir comfortably exceeds the total energetics inferred even for the most hyperactive known repeater, FRB,20240114A, whose estimated total isotropic energy exceeds approximately $86\%$ of the dipolar magnetic energy of a typical magnetar \citep{Zhang2025}, i.e., $\sim 10^{47}$ erg. The interface mismatch reservoir therefore surpasses the demonstrated energy output of the most extreme known repeater by several orders of magnitude, indicating that the available energy is not a limiting factor for sustaining even hyperactive repeating behavior in our model.

The QN expels approximately $M_{\rm QN}\sim10^{-2}M_\odot$ from the outer layers of the NS with a kinetic energy of $E_{\rm QN}^{\rm KE}\sim10^{50}$~erg (\citealt{Keranen2005}; see Appendix \ref{app:ejecta} for further details). Since the total mass of the NS crust (outer plus inner) is $\sim10^{-2}M_\odot$ (\citealt{ChamelHaensel2008}), this implies that the entire crust is ejected during the QN. In addition to providing a favorable site for r-process nucleosynthesis (Appendix \ref{app:kilonova}), the QN ejecta contributes its own dispersion measure (DM$_{\rm QN}$) and rotation measure (RM$_{\rm QN}$) (Appendix \ref{app:dm_rm}), both of which play an important role in the model presented here.

If the conversion event completely removes the crust and exposes hadronic matter at densities above the crust-envelope transition, a new crust does not automatically reform as the HS cools. The existence of a crust requires matter at subnuclear densities, where nuclei embedded in an electron gas are energetically favoured over uniform nuclear matter (\citealt{ChamelHaensel2008}). Therefore, cooling alone may not be enough to regenerate a crust and may require either the deposition of fresh material onto the stellar surface (e.g., through fallback or subsequent accretion) or the expansion and decompression of the exposed hadronic envelope outer layers to densities below the envelope-crust transition. Consequently, the crust formation timescale may be determined by a combination of  intrinsic property of the HS hadronic envelope and  astrophysical environment. In this proof-of-principle paper, we proceed with the crustless formation scenario and assume that the crust is acquired later. However, a plausible scenario is that of a distribution of the QN ejecta peaking at $M_{\rm QN}\sim 10^{-2}M_{\odot}$ while allowing for some crust retention in some QNe. This is discussed in \S\ \ref{sec:MQN-distribution}. I.e., in the current simplified version of our model, all QCD magnetars are born crustless, which has unique observational consequences.
The crust has no effect on the energy reservoir in our model, which arises from the lack of a global magnetic equilibrium shared between the quark core and the hadronic envelope (see Appendix~\ref{app:magnetic_avalanche}). 

 There are two astrophysical channels for the NS to HS conversion to give birth to a QCD magnetar in the QN model.
\textit{Channel~A} (our focus in this paper) is the CCSN channel:
a massive NS is born above $\Mdec$ in a core-collapse event and
subsequently undergoes a QN at $\tSD$.
\textit{Channel~B} is the accretion
channel (Section~\ref{sec:chanB}): a NS in a binary system accretes sufficient mass from a
companion to cross $\Mdec$, so the QN and the resulting QCD magnetar
are decoupled from any prior SN event.

 %%%%  SN TIMESCALES %%%%%%
In addition to the spin-down timescale $\tSD$ which is a property of the NS, 
Table \ref{table:timescales} lists the timescales relevant to Channel A 
namely: (i) the SN ejecta diffusion timescale ($t_{\rm SN, d}$); 
(ii) The SN ejecta optical transparency timescale ($t_{\rm SN, o}$); 
(iii) The SN ejecta radio transparency timescale ($t_{\rm SN, r}$); 
(iv) The SN remnant (SNR) dissipation timescale ($t_{\rm SNR}$); 
(v) The lifespan of the QCD magnetar as a crustless HS ($t_{\rm crust}$).
 In this scaling $t_{\rm SN}=0$ is the SN proper while the QN occurs at $t_{\rm QN}\sim \tSD$.

\begin{table}[t!]
\centering
\caption{Relevant timescales in our model}
\begin{tabular}{|c|c|c|}\hline
Description & time & Estimate \\\hline
 SN event & $t_{\rm SN}$ & 0.0\\\hline
 QN event  & $t_{\rm QN}$ & $
\sim \tSD$ \\\hline
NS Spin-down timescale $^1$ &$\tSD$ & days to Myrs \\\hline
SN diffusion timescale $^2$ & $t_{\rm SN, d}$ & months \\\hline
SN optical thinness $^2$ & $t_{\rm SN, o}$ & years \\\hline
SN radio  thinness $^3$ &  $t_{\rm SN, r}$ & decades \\\hline
SNR lifespan$^4$ & $t_{\rm SNR}$ & millenia \\\hline
\end{tabular}\\
$^1$\citealt{Arnett1982}; $^2$\citealt{Branch2017}; $^3$\citealt{Mezger1967}; $^4$\citealt{Chevalier1982}.
\label{table:timescales}
\end{table}
%%%% END SN TIMESCALES %%%%%%

Our framework has several consequences that directly address the
tensions mentioned above. \textit{First}, the magnetar fraction is no longer set by the tail
of the spin distribution but by the tail of the \textit{mass}
distribution above $\Mdec$.
\textit{Second}, the birth spin period acts only as an energy
reservoir and a selector determining which observable class the QN
produces: all NSs with $M>\Mdec$ eventually become AXPs\&SGRs,
whereas fast rotators ($P<\Pfast$, with $\Pfast$ defined as the
critical birth spin period separating QNe with sufficient rotational
energy to power SLSNe-I and LFBOTs from those without; see
\S\ref{sec:population}) produce SLSNe-I and LFBOTs.
\textit{Third}, the QCD magnetar forms after the SN proper, with a delay set by the NS spin-down timescale $\tSD$.
 In this regime, $PdV$ (pressure--volume) losses, i.e. the adiabatic
loss of injected energy due to the expansion of the SN ejecta, are
minimal compared to prompt energy injection, allowing a much higher
rotation-to-radiation conversion efficiency during the post-QN
spin-down phase.
In particular, when the QN occurs before the SN ejecta becomes transparent ($\sim$ months after the SN event), the energy injection occurs while the ejecta is still optically thick—crucial for explaining the most energetic SLSNe-I from the re-processing of LFBOT power.
\textit{Fourth}, when the QN is delayed until after the SN ejecta has already expanded significantly ($\tSD >> t_{\rm SN, o}$), the ejecta is at large radius and low density when the QN fires while still in the vicinity of star-forming regions. The QN signatures are then directly observable.
When $\tSD > t_{\rm SNR}$, as is the case for massive NSs born very slowly rotating and with weak magnetic field,
 the conversion occurs in complete isolation far from the NS birth site.

The paper is organised as follows.
Section~\ref{sec:qcd} describes the QCD origin of the magnetic field
in the core of the massive NS and the resulting surface field.
Section~\ref{sec:framework} presents the QN framework, its
outputs and the resulting observable events.
Section~\ref{sec:population} describes the population synthesis and
Bayesian analysis using the measured rates of AXPs\&SGRs, SLSNe-I,
LFBOTs and the rate of association of AXPs\&SGRs with SNRs as
constraints; here we derive the optimal values of $\Mdec$ and $\Pfast$.
Section~\ref{sec:frb} is dedicated to FRBs in our model and shows how the continuous mismatch at the core--envelope interface, sustained by the HS spin-down, powers crustless QCD magnetars as FRB sources over timescales of centuries.
In Section~\ref{sec:discussion}, we discuss QCD magnetars via a binary channel,
 the QN as an r-process site and the 
implications of our findings  for the physics of hadronic-to-quark matter conversion 
and the nature of the underlying ferromagnetic quark phase.
Section~\ref{sec:predictions} presents key testable and falsifiable
predictions of the model and its limitations. Here, we discuss the $M_{\rm QN}$ distribution scenario and explain how GRBs arise in this context.
We conclude in
Section~\ref{sec:conclusions}.

% ════════════════════════════════════════════════════════════════════════
\section{QCD Origin of the Magnetic Field}
\label{sec:qcd}
% ════════════════════════════════════════════════════════════════════════

We consider two magnetic field-generation mechanisms throughout this paper:
the QCD-induced field  the HS acquires after the conversion (the focus of this section) and 
what we refer to as the \textit{fossil} magnetic field inherited by the NS
from flux conservation and compression of the NS progenitor field. 
 The former subpopulation is the dominant one in our model yielding QCD magnetars while
  the  latter minority sub-population yields \textit{fossil magnetars} if the NS birth magnetic field is strong. 
  In double-humped SLSNe, in our model, the first hump is powered by spin-down from the fossil
  magnetar while the second hump is powered by the spinning-down QCD magnetar
  using the left-over rotational energy from the pre-QN phase (\citealt{Ouyed2025,Ouyed2026a,Ouyed2026b}).

\subsection{Ferromagnetic quark matter in the core}
\label{sec:ferromag}

The central departure from classical magnetar models is that the
magnetic field is generated by the quark phase itself, with no reference to rotation.
A simple energy argument establishes the natural field scale.
The characteristic QCD energy scale is
$\Lambda_{\rm QCD}\approx260$~MeV compared to the
$\sim 8.8$ MeV of the hadronic phase.
Equating the magnetic energy density to this scale (see Appendix \ref{sec:Bqcd_origin})
yields $B_{\rm QCD}\sim10^{18}$~G compared to the hadronic scale which yields
$B_{\rm had}\sim10^{15}$~G; i.e., $B_{\rm QCD}/B_{\rm had}\sim 10^3$.
This is not a fine-tuned result: it is the field strength that QCD
energetics demands, and that the quark phase can sustain, if it
orders magnetically. Whether it does is a question of the ground-state structure of dense quark matter.

A body of theoretical work has established that at
densities relevant to massive NS cores (a few times nuclear saturation
density $\rho_{\rm nuc}\simeq 2.8\times 10^{14}$ g cm$^{-3}$), the free energy of some
two-flavour $(u,d)$ quark phases can be minimised by a ferromagnetic state
in which a large-scale magnetic field emerges spontaneously
\citep{Tatsumi2000,Iwazaki2003,Iwazaki2005,Niegawa2005,Ebert2005,
Yoshike2015,Dvornikov2016,Ferrer2021}.
In one class of models the mechanism is analogous to itinerant
ferromagnetism in condensed matter: exchange interactions between
quarks at high density favour spin alignment, and the ground state
spontaneously breaks rotational symmetry.
In another class, the driving mechanism is colour ferromagnetism,
where the chromomagnetic field arises via the gluonic sector.
The response of the up and down quarks to this field yields the
observable field which can reach up to $10^{18}$ G (e.g., \citealt{Iwazaki2003,Iwazaki2005}).

Crucially, this QCD field is a genuine ground-state property of dense
quark matter. It does not require special initial conditions and depends mainly on
the local baryon number density $n_B$. For a core size $R_{\rm c}$ of a few kilometers, the energy requirement is
$B_{\rm QCD}^2/(8\pi)\times(\frac{4}{3}\pi R_{\rm c}^3)\sim10^{50}$--$10^{51}$~erg,
which represents a small percentage of the total energy released in the
NS-to-HS conversion $E_{\rm conv}= (M_{\rm c}/m_{\rm B})\times \Delta \epsilon_{\rm B}\sim 4\times 10^{53}$~erg$\times(M_{\rm c}/M_{\rm NS})$;
here $M_{\rm c}$ is the core mass with about $\Delta \epsilon_{\rm B}\sim 100$~MeV released per converted baryon
 of mass $m_{\rm B}$ (\citealt{Weber2005}).

\subsection{Emergence of the surface magnetic field}
\label{sec:transport}

The generation of an ultra-strong magnetic field in the quark core is a necessary but not sufficient condition. The HS must subsequently reconfigure its magnetic field on a global scale, establishing a new equilibrium between the core, the hadronic envelope and the surface field components. 

The  initial global re-configuration must occur on a timescale short enough to account for the observed transient phenomena.
Within the QN framework, two factors render this transport process both rapid and efficient.

The removal of the crust fundamentally changes the magnetic-field evolution.
In conventional NSs, global magnetic-field rearrangement is controlled
by slow crustal transport processes such as Hall drift and Ohmic diffusion
\citep{Pons2007,Vigano2013}. In a crustless QCD magnetar, this bottleneck is absent: the strong magnetic field
generated in the core is directly coupled to the residual hadronic envelope.

The sudden mismatch between the core magnetic field and the field supported at
the base of the hadronic envelope generates large magnetic stresses at the
core-envelope interface. The magnetic pressure
contrast $\Delta P_B \simeq
(B_{\rm core}^2-B_{\rm env}^2)/8\pi
\simeq B_{\rm core}^2/8\pi
\approx
4\times10^{34}\ {\rm erg\,cm^{-3}}$  exceed the 
matter pressure of the envelope
$P_{\rm env}\sim \rho_{\rm env}c_s^2\sim (2.5$-$7.5)\times10^{33}\ {\rm erg\,cm^{-3}}$ 
for $c_s=(0.1$-$0.3c$ ($c$ is the speed of light) and and envelope density 
of the order of the nuclear saturation value $\rho_{\rm env}\simeq \rho_{\rm nuc}\sim 2.8\times 10^{14}$ g cm$^{-3}$. I.e., 
a magnetized structure at the interface cannot remain
in simple hydrostatic balance without a compensating density
perturbation and an associated buoyant restoring force.
The instability at the interface persists well beyond the initial reconfiguration of the global magnetic field. As the HS continues to spin down, the interface is continuously driven toward instability, giving rise to the threshold-driven magnetic avalanche behavior described in Appendix~\ref{app:magnetic_avalanche}. This mechanism lies at the heart of the centuries-long FRB phase of the crustless QCD magnetar.

The magnetic energy stored in this stressed
configuration is expected to be released through Parker-like
(\citealt{Parker1966}) and/or Tayler-like instabilities
(\citealt{Tayler1973}; see also \citealt{Woltjer1958}), which reorganize the
interface field into coherent magnetic flux ropes; see Appendix \ref{app:magnetic_avalanche}. 
Hereafter we refer to these structures as ``magnetic bubbles" to be differentiated form the ``quark bubbles"
due to quantum nucleation in the core.

These structures transport
magnetic flux and helicity outward through magnetic buoyancy
\citep{Woltjer1958} re-arranging the global filed into a new dipole-like configuration.
The Alfv\'en speed
associated with the pre-existing envelope field, $B_{\rm env}\sim10^{15}$ G and density $\rho_{\rm env}\sim \rho_{\rm nuc}$, is
$v_{\rm env,A}=B_{\rm env}/\sqrt{4\pi\rho_{\rm env}}
\sim5.6\times10^{-4}c$,
so that an individual flux rope traverses the hadronic envelope on a timescale
\begin{equation}
t_{\rm buoy}<\frac{R_{\rm NS}}{v_{\rm buoy}}\sim0.05~{\rm s}\ ,
\label{eq:tbuoy}
\end{equation}
where the buoyancy velocity is $v_{\rm buoy}\sim v_{\rm env,A}$ to first order.

Thus, the global magnetic-field reconfiguration occurs on a sub-second
timescale, essentially contemporaneous with the QN event. Residual flux ropes
and buoyant magnetic structures can, however, continue to form at the
core-envelope interface, due to continuous HS spin-down,  until the system relaxes toward its new equilibrium,
sustaining the FRB-emitting phase of the crustless QCD magnetar
(see \S\ref{sec:frb}). 

 The outward transport of magnetic flux and helicity by the
buoyant structures provides the mechanism by which the initially
strongly non-uniform core--envelope magnetic configuration can
reorganize into a global field extending through the hadronic envelope
and coupling to the external magnetosphere. Importantly, the
$\sim10^{18}$~G field is not assumed to emerge unchanged at the
surface. Rather, we adopt the relaxed large-scale component to be
approximately dipolar, so that the field decreases outward according
to the dipolar radial scaling. Thus,
\begin{equation}
B_{\rm HS}\sim B_{\rm core}\left(\frac{R_c}{R_{\rm NS}}\right)^3 .
\end{equation}
For $B_{\rm core}\sim10^{18}$~G and our fiducial
$R_c/R_{\rm NS}\sim0.1$, this gives
$B_{\rm HS}\sim10^{15}$~G. The low density of the outer envelope does
not require the surface field to be generated locally at the surface:
the field is maintained by the global magnetic configuration anchored
in the strongly magnetized interior and transported through the
envelope. The $\sim10^{15}$~G value therefore represents the
large-scale field strength at the stellar surface after global
reconfiguration, rather than the local energy-density scale of the
surface layers.

% ════════════════════════════════════════════════════════════════════════
\section{The Quark-Nova Framework}
\label{sec:framework}
% ════════════════════════════════════════════════════════════════════════

%%%% QN TIMESCALES using prime %%%%%%
Before describing the features of the QN, we first summarize the QN-related
timescales relevant to our model. The timescales (denoted by a prime) are measured relative to the QN event.
In addition to the HS spin-down timescale, $t^\prime_{\rm HS,SpD}$, which is a
property of the HS, Table~\ref{table:HS-timescales} lists the timescales
associated with the QN event, namely:
(i) the QN ejecta diffusion timescale ($t^\prime_{\rm QN,d}$);
(ii) the QN ejecta optical transparency timescale ($t^\prime_{\rm QN,o}$);
(iii) the QN ejecta radio transparency timescale ($t^\prime_{\rm QN,r}$); and
(iv) the lifetime of the crustless QCD magnetar, which corresponds to the GRB
epoch in our model.

\begin{table}[t!]
\centering
\caption{Relevant QN timescales in our model.}
\begin{tabular}{|c|c|c|}
\hline
Description & Timescale$^{1}$  & Typical value \\
\hline
QN event  & $t^\prime_{\rm QN}$ & 0 \\
\hline
HS spin-down timescale & $t^\prime_{\rm HS,SpD}$ & days--years \\
\hline

Ejecta diffusion timescale & $t^\prime_{\rm QN,d}$ & days \\
\hline
Ejecta optical transparency  & $t^\prime_{\rm QN,o}$ & weeks \\
\hline
Ejecta radio transparency  & $t^\prime_{\rm QN,r}$ & years \\
\hline
Crust formation$^{2}$ & $t^\prime_{\rm crust}$ & centuries \\
\hline
\end{tabular}

\vspace{1mm}

{\footnotesize
$^{1}$  Primed times are measured relative to the QN event; 
$^{2}$ Derived in this  work.
}

\label{table:HS-timescales}
\end{table}
%%%% END QN TIMESCALES %%%%%%

\subsection{The triggering condition}
\label{sec:trigger}

A NS born with $M_0>\Mdec$ has a central density
$\rhoc>\rhodec$ from birth, where $\rhodec$ is the quark
deconfinement threshold density.
Thermal quark nucleation creates quark bubbles in the core
immediately, but the pressure jump at the hadronic--quark interface
satisfies $\Delta P\sim 0$: the bubbles are in pressure equilibrium with
their surroundings and cannot percolate.
As the star cools below $\sim 1$ MeV, quantum tunnelling becomes the relevant nucleation
channel, but the star remains in a metastable hadronic phase
(Appendix~\ref{sec:Bqcd_origin}).
This is the massive NS phase: physically a pre-HS,
observationally indistinguishable from an ordinary radio pulsar.

At $t=\tSD$, angular-momentum loss reaches a critical level,
 triggering a sudden pressure imbalance between the quark bubbles and the surrounding hadronic matter, which drives
immediate and complete percolation of the quark phase throughout the core: the
QN  fires on millisecond timescales, $\sim R_{\rm c}/c$, where $R_{\rm c}$ is the few kilometers NS core radius.
The trigger is therefore not a density threshold being crossed ---
the central density already exceeded $\rhodec$ at birth --- but a
\textit{pressure threshold} crossed at around the NS charactersitic spin-down timescale.
This distinction is physically important: the QN delay is set
entirely by $\tSD$ (i.e., by $P_0$ and $B_0$; $\tSD\propto P_0^2B_0^{-2}$ where $P_0$ and $B_0$ are the NS birth period and magnetic field), not by the nucleation
kinetics, and it implies that thermal nucleation and quantum
tunnelling are always quenched in massive NSs before spin-down is
complete. We return to this in Section~\ref{sec:implications}.

We consider a representative NS with 
birth spin period $P_0=0.2$~s, surface
dipole field $B_0=10^{12.5}$~G  (\citealt{FaucherKaspi2006,Igoshev2022}) and 
moment of inertia $I_{\rm NS}=2\times10^{45}$~g~cm$^2$. 
The spin-down formalism adopted here follows standard magnetic dipole
radiation for an aligned, force-free wind with braking index $n=3$ 
\citep{Michel1970,Manchester1977,Shapiro1983,Contopoulos1999,
Spitkovsky2006,Lorimer2004,Lyne1993,Lattimer2005}.

The NS rotational energy is
\begin{equation}
  E_{\rm NS,rot}
  \simeq 7.5\times10^{47}\times 
  \left(\frac{P_0}{0.2~{\rm s}}\right)^{-2}
  ~{\rm erg}\ ,
  \label{eq:ENSrot}
\end{equation}
with a  characteristic spin-down timescale \begin{equation}
  \tSD
  \simeq 6.3\times 10^4~{\rm yr} \times 
  \left(\frac{P_0}{0.2~{\rm s}}\right)^{2}
  \left(\frac{B_0}{10^{12.5}~{\rm G}}\right)^{-2}\ .
  \label{eq:tSD}
\end{equation}
Its spin-down luminosity is
\begin{align}
\label{eq:Lnspd}
L_{\mathrm{NS,SpD}}
&\sim
3.8 \times 10^{35}\,
\mathrm{erg\,s^{-1}}\times 
 \left( \frac{B_{\rm NS}}{10^{12.5}\ {\rm G}} \right)^{2}\,
\left( \frac{P_{\rm NS}}{0.2\ {\rm s}} \right)^{-4}\times \\\nonumber
&\times \left(1+\frac{t}{t_{\mathrm{NS,SpD}}}\right)^{-2}\ ,
\end{align}
and a period evolving as
\begin{equation}
  P_{\rm NS}(t) = P_0\times \left(1+\frac{t}{\tSD}\right)^{1/2}\ ,
  \label{eq:Pevol}
\end{equation}
so that the HS  inherits a spin period
$P_{\rm HS}\sim \sqrt{2}\,P_{\rm 0}$ when it is born at $t_{\rm QN}\sim \tSD$.  It has $E_{\rm HS,rot} \sim E_{\rm NS,rot}/2$
 in rotational energy leftover from the NS  phase;  $I_{\rm HS}\simeq I_{\rm NS}$ 
 since the ejected mass is $M_{\rm QN}<< M_{\rm NS}$.
  The HS spin-down timescale is 
  \begin{equation}
  t^\prime_{\rm HS, SpD}\sim 12.7\ {\rm yr}\times \left( \frac{P_{\rm HS}}{{\rm 0.28\ s}}\right)^2 \left(\frac{B_{\rm HS}}{{\rm 10^{15}\ G}}\right)^{-2}\ ,
   \label{eq:tHSSD}
  \end{equation}
  for a typical $P_{\rm HS}\sim 2^{1/2}\times 0.2\sim 0.28$ s.

The range of $\tSD$ across the birth distribution is enormous
--- from days (fast ms rotators with strong fossil birth fields $\sim 10^{15}$ G) to millions
of years (slow rotators with weak fields). It is this range
that naturally produces the diversity of observable outcomes in
our model (Section~\ref{sec:four_outcomes}).

As an illustrative example, a massive NS with $B_0 \sim 10^{12}$~G
and $P_0 \sim 1$~s converts into a QCD magnetar at
$\tSD \sim 10^7$ yr after its birth.
The resulting slowly rotating QCD magnetar is effectively isolated
and decoupled from its birth environment.
This late-conversion channel, as we show in this paper, may be relevant to Fast Radio Burst (FRB; \citealt{Lorimer2007}) models that invoke magnetars \citep[e.g.,][]{Kumar2017,Metzger2019,Lu2020,Margalit2020,Yang2021}. Assuming that magnetar-based FRB emission mechanisms also apply to QCD magnetars (see \S \ref{sec:frb}), the delayed formation at $\tSD$ relative to the SN event naturally places QCD magnetars in atypical host environments, including globular clusters and the outskirts of elliptical galaxies \citep{Tendulkar2017, Bhandari2022, Kirsten2022}.
Isolated QCD magnetars can also form via accretion in binaries, when a NS gains sufficient mass from a companion to exceed $M_{\rm dec}$ (Channel~B; Section~\ref{sec:chanB}), further broadening the diversity of FRB environments in our model.

\subsection{Outputs of the Quark Nova}
\label{sec:outputs}

The NS-to-HS branch of the QN model, adopted in this paper, produces the following outputs
\citep{Ouyed2025,Ouyed2026a,Ouyed2026b}.

\begin{enumerate}

\item \textit{Crustless QCD magnetar.}\\
The QN compact remnant --- a highly magnetised HS --- carries
a surface field $\BHS\sim10^{15}$~G, established promptly at $t=\tSD$ through
the core-to-surface transport described in
Section~\ref{sec:transport}.
This field is spin-independent; it is set by the baryon number
density in the quark core.

 Any fall-back material from the QN ejecta is tiny and occurs on  much longer timescales
 than the magnetic field reconfiguration sub-second  timescales (Eq. (\ref{eq:tbuoy})). Furthermore, the surface magnetic field $B_{\rm HS}\sim 10^{15}$ G 
 will overwhelm the fall-back ram pressure, propelling or channelling fallback material away from the HS.
 
Thus, for typical $M_{\rm QN}\sim10^{-2}M_{\odot}$ (comparable to the total mass of a typical inner and outer crust of a NS), the QCD magnetar is born as a genuinely crustless HS. During this crustless phase, it acts as an FRB-emitting engine before transitioning into the AXP\&SGR phase once a hadronic envelope re-establishes a crust on  timescales $t^\prime_{\rm crust}$ (see \S\ref{sec:frb}).

\item \textit{QN ejecta  as an r-process and kilonova site.}\\
The neutron-rich QN ejecta is the site of a robust r-process
nucleosynthesis capable of synthesis actinides and lanthanides \citep{Jaikumar2007,Kostka2014a,Kostka2014b}.
 With the expected  high opacity from the lanthanides  yield ($\kappa_{\rm QN}\sim1$--$100$~cm$^2$~g$^{-1}$; \citealt{Kasen2013}), 
 the QN ejecta can produce a red kilonova peaking in the near-infrared 
(Appendix~\ref{app:kilonova}). 

In Channel~A, the kilonova is directly observable only when
$\tSD \gg t_{\rm SN,o}$, so that the QN occurs after the SN ejecta has
become optically thin.
In Channel~B, the kilonova is always potentially observable since
the QN is decoupled from any SN event. 

Kilonovae associated with
QCD magnetars require no NS merger and can therefore occur in
field galaxies without a recent merger history. Conversely, directly
observable QN kilonovae can also occur near star-forming regions if
$\tSD>t_{\rm SN,o}$ but the NS has not migrated too far from its
birth site.

\item \textit{QN ejecta as an DM and RM screen.}\\
The expanding QN ejecta naturally contributes an intrinsic dispersion measure (DM$_{\rm QN}$), independent of any surrounding SN remnant or pulsar wind nebula, reaching values of $\sim10^2$--$10^5~{\rm pc\,cm^{-3}}$ during the first weeks after the QN (see Appendix~\ref{app:dm_rm}). Once the QN ejecta sweeps up an amount of ambient material comparable to its own mass, a forward shock develops, amplifying the magnetic field of the shocked ejecta and producing a rotation measure, RM$_{\rm QN}$, that remains as large as $\sim10^{5}$--$10^{7}~{\rm rad\,m^{-2}}$ for decades in dense environments. These quantities constitute the local DM and RM contributions of the QN during the FRB-active phase of the QCD magnetar lifetime (see \S\ref{sec:frb}).

\item \textit{An LFBOT.}\\
If the QCD magnetar inherits a millisecond spin period from its
parent NS ($P_{\rm HS}\sim 2^{1/2}P_{\rm NS}$), its spin-down luminosity thermalises in the optically
thick QN ejecta, producing a hot blue blackbody spectrum with
  a diffusion timescale $t_{\rm QN, diff} \sim (\kappa_{\rm QN}M_{\rm QN}/4\pi c v_{\rm QN})^{1/2}$ of days to a week, matching the defining properties of
LFBOTs \citep[see][for detailed light-curve modelling]{Ouyed2026a}. The derived peak luminosity,
$L_{\rm LFBOT}> 10^{44}$~erg~s$^{-1}$, is also consistent with observations \citep{Drout2014,Ho2023}.  
 
 A rapidly rotating QCD magnetar with $M_0>\Mdec$ and $P_0<\Pfast$ is thus associated with an LFBOT in addition to a kilonova. The latter is the direct outcome of the QN ejecta and is therefore produced regardless of the QCD magnetar birth period. In this sense, a \emph{quiet} QN is one in which the resulting HS (i.e. the QCD magnetar) is born with $P>\Pfast$, so that spin-down power does not significantly energize the ejecta and no LFBOT is produced.

\item \textit{A gravitational-wave transient.}\\
A non-spherical conversion front during the QN may produce a
millisecond-duration gravitational-wave (GW) burst
\citep{Staff2012}.
This prompt burst would be distinct from the quasi-periodic GW emission
produced by interface oscillation modes (e.g., $g$-modes; \citealt{Jaikumar2013}) of the HS after conversion; the former is the trigger, the latter
the ringdown. The precise frequency content, strain amplitude, and waveform
morphology of both components depend on the geometry of the
conversion front and the post-conversion oscillation modes of
the HS; a dedicated numerical relativity calculation is
deferred to future work.  This is the least explored aspect of the
NS-to-HS branch of the QN modelling and will only be mentioned its possible
concurrence with other outputs discussed here.

\end{enumerate}

All outputs listed above are produced simultaneously at the same
location, at time $\tQN\sim\tSD$ after the birth of the massive NS (in general whenever $M_{\rm dec}$ is reached) in
both channels. Their simultaneous appearance, decoupled from compact binary merger
environments, constitutes a unique multi-messenger signature of
the QN (see \S \ref{sec:predictions}) and its compact remnant, the QCD magnetar.

%------------------------------------------------------------
\subsection{Observable outcomes}
\label{sec:four_outcomes}
%------------------------------------------------------------

We now discuss the observable events
emanating from the interaction of the QN with its environment and how it affects the QN outputs.

The physical discriminant in this framework is not the birth period
alone but the spin-down timescale,
$\tSD\propto P_0^2 B_0^{-2}$ (Equation~\ref{eq:tSD}).
In Channel~A, what matters observationally is the
\textit{relationship} between $\tSD$ and two external timescales
set by the SN ejecta: the photon diffusion time
$t_{\rm SN, d}\sim$ month,
which sets the window during which spin-down power can be
thermalised and radiated, and the SN ejecta optical transparency time
$t_{\rm SN, o}\sim$ 1 yr, beyond which the ejecta is
optically thin while still  freely expanding; 
see Table \ref{table:timescales}.

The main observable outcomes are determined by the pair
$(P_0, B_0)$ for all NSs with $M_0>\Mdec$. We define $P_{\rm fast}$ as the
critical period (to be derived) separating the slowly and rapidly rotating QCD magnetars.
 All each of events listed below is potentially 
 accompanied by  a kilonova (or at least by r-processed elements) and by signatures of the QN ejecta interacting with its surrounding medium when shedding
its kinetic energy $E_{\rm QN}^{\rm KE}\sim 10^{50}$ erg.

\begin{enumerate}

\item \textit{Crustless QCD magnetars as FRB sources:}\\
A QCD magnetar is born crustless (see \S~\ref{sec:frb}).
As it relaxes toward its new magnetized equilibrium, magnetic flux ropes (we refer to as ``bubbles") are generated at the hadron--quark interface by the large magnetic stresses arising from the mismatch between the core and (the base of) the envelope magnetic fields, and buoyantly rise through the hadronic envelope. Because the magnetic energy stored at this interface is transported through the envelope with negligible dissipation, it is released upon reaching the stellar surface, powering FRB emission. During this early crustless phase, lasting centuries, QCD magnetars therefore manifest as active FRB sources (see \S\ref{sec:frb}).

\item \textbf{Crusted QCD magnetars as AXPs\&SGRs:}\\
Any QCD magnetar that subsequently re-forms a crust during its evolution, irrespective of the
NS birth period or magnetic field, becomes an AXP\&SGR in our model. Although the magnetic
activity at the hadron--quark interface is identical in both crustless and crusted objects, the
presence of even a relatively thin crust causes much of the magnetic energy carried by the
buoyant flux tubes to be dissipated before reaching the stellar surface.  In this case, a
smaller fraction of the released magnetic energy is available to produce a rapid magnetospheric
disturbance and coherent radio emission, while energy deposited in the crust is expected to
emerge predominantly through thermal and non-thermal X-ray activity. The model therefore
predicts that crust formation substantially reduces the efficiency of FRB production rather
than strictly forbidding it. Crusted QCD magnetars therefore naturally exhibit predominantly
the defining phenomenology of AXP\&SGRs (e.g., \citealt{Kaspi2017} and references therein).

\item \noindent\textit{LFBOTs:}\\
QCD magnetars born as fast rotators with $P_0<\Pfast$ and undergoing the QN only after the SN ejecta has become optically thin (i.e., $\tSD > t_{\rm SN,o}$) deposit their spin-down energy into the expanding QN ejecta, producing a directly observable LFBOT (\citealt{Ouyed2026a}). This includes the late-converting isolated channel ($\tSD > t_{\rm SNR}$; Section~\ref{sec:trigger}), in which the LFBOT and the accompanying kilonova occur without an associated CCSN and far from any compact-binary merger environment.

\item \noindent\textit{SLSNe-I:}\\
Here we focus exclusively on SLSNe-I. They arise from stripped-envelope massive stars, show little evidence for strong early-time circumstellar-medium (CSM) interaction, and have well-characterized photometric evolution and volumetric rates over a broad redshift range \citep{Gal-Yam2009,Quimby2011}. Their preference for low-metallicity environments \citep[e.g.,][and references therein]{Nicholl2021} also favors the formation of rapidly rotating, massive NS remnants, as required by our model. Although the SLSN-I volumetric rate remains uncertain, current estimates based on existing samples (\citealt{Quimby2013,Prajs2017,Frohmaier2021}) provide the observational constraints adopted in the population and Bayesian analysis presented in \S\ref{sec:population}.

SLSNe-I arise from fast rotators with $P_0<\Pfast$ and
$\tSD<t_{\rm SN,o}$. The QN therefore occurs while the SN ejecta is
still optically thick, causing the spin-down-powered QN ejecta (i.e., the LFBOT) to be
reprocessed into a SLSN-I  (\citealt{Ouyed2025,Ouyed2026b}). The condition
$\tSD<t_{\rm SN,o}$ translates to

\begin{equation}
B_0 > B_{\rm fossil,slsn}
\sim10^{13}\ {\rm G}\times 
\left(\frac{P_0}{5\,{\rm ms}}\right),
\label{eq:B0slsn}
\end{equation}
for $t_{\rm SN,o}\sim1$ yr; adopting  a spin period of 5~ms as representative of rapidly rotating QCD magnetars.

Using Eq.~(\ref{eq:Lnspd}), this
corresponds to a minimum pre-QN spin-down luminosity of

\begin{equation}
L_{\rm NS,SpD}
>6\times10^{42}\ {\rm erg\,s^{-1}}\times 
\left(\frac{P_0}{5\,{\rm ms}}\right)^{-3}\ .
\end{equation}

Every fast-rotating massive NS ($M_0>\Mdec$, $P_0<\Pfast$)
experiences two distinct energy-injection episodes: a pre-QN
fossil-magnetar spin-down phase inside the SN ejecta, followed by a
post-QN QCD-engine phase. Consequently, all SLSNe-I in Outcome~3 are
powered by the same two-stage engine. The distinction between the single-humped 
and double-humped SLSNe-I discussed below is therefore observational rather than physical:
it depends solely on whether the pre-QN fossil-powered component is
bright enough to be detected above the rising QCD-engine emission.

\item \noindent\textit{Single-humped SLSNe-I
($B_{\rm fossil,slsn}<B_0<B_{\rm fossil,c}$):}\\
The pre-QN spin-down phase, powered by the inherited fossil magnetic
field, is physically present but contributes only a subdominant
luminosity. Consequently, the observed light curve appears
single-peaked, being dominated by the post-QN QCD-engine emission
reprocessed through the SN ejecta. A subtle change in the slope of
the rising light curve, marking the transition from the dim
fossil-powered phase to the QCD-engine phase, may nevertheless be
detectable statistically as a shoulder in large samples of SLSNe-I.

\item \noindent\textit{Double-humped SLSNe-I
($B_0>B_{\rm fossil,c}$):}\\
NSs born with the shortest periods in the $P_0<\Pfast$ range and with 
fossil magnetic fields exceeding the critical value,
$B_{\rm fossil,c}\sim10^{14}$\,G, produce a prominent and directly
observable pre-QN hump. In these systems, the rotational energy
budget is sufficient to power both the pre-QN hump and the subsequent
QCD-powered SLSN-I. The rate of double-humped events is therefore
determined by the high-$B_0$ tail of the fossil-field distribution,
which is directly constrained by the Bayesian population analysis
presented in \S\ref{sec:population}
(\citealt{Ouyed2026b}).

The threshold $B_0>B_{\rm fossil,c}$ is therefore an observational
detectability criterion rather than a physical bifurcation. The
transition between single- and double-humped morphologies is
continuous, being controlled by the ratio
$B_0/B_{\rm fossil,c}$.

\item  \noindent\textit{QN ejecta interaction:}\\
As it propagates away from the HS, the QN ejecta sweeps up ambient material, producing a forward-reverse 
shock system observable as radio and X-ray emission  \citep{Ouyed2026a}.
In Channel A, where the QN occurs within the expanding SN ejecta, the QN ejecta ($v_{\rm QN}\sim0.1c$)
overtakes the slower SN ejecta, producing late-time post-peak bumps in SLSN-I light curves \citep{Ouyed2025}.

\end{enumerate}

% ════════════════════════════════════════════════════════════════════════
\section{Population Synthesis and Bayesian Analysis}
\label{sec:population}
% ════════════════════════════════════════════════════════════════════════

\subsection{Birth distributions}
\label{sec:distributions}

Throughout this section $M_0$, $P_0$, and $B_0$ denote the NS birth
mass, spin period, and surface dipole field. 
All NSs are drawn from log-normal, statistically independent,  birth distributions,
regardless of whether they subsequently undergo QN conversion:
\begin{align}
  \log_{10}(M_0/\Msun)    &\sim \mathcal{N}(\log_{10}(M_{\rm 0, pk}/\Msun),\;\sigM^2)\,, \notag\\
  \log_{10}(P_0/{\rm ms}) &\sim \mathcal{N}(\log_{10}(P_{\rm 0, pk}/{\rm ms}),\;\sigP^2)\,,  \notag\\
  \log_{10}(B_0/{\rm G})  &\sim \mathcal{N}(\log_{10}(B_{\rm 0, pk}/{\rm G}),\;\sigB^2) \ .
  \label{eq:distributions}
\end{align}
The peak values we used are averages of the observed radio pulsar
population \citep{FaucherKaspi2006,Igoshev2022}; i.e. $M_{\rm 0, pk}=1.35M_{\odot}, P_{\rm 0, pk}=200\ {\rm ms}, B_{\rm 0, pk}=10^{12.5}$ G.
This reduces the number of free parameters during the population and bayesian analysis
we present here; $\sigM = \sigma_{\log_{10}(M_0)},\sigP = \sigma_{\log_{10}(P_0)},\sigB = \sigma_{\log_{10}(B_0)}$. 
For this proof-of-principle paper, we adopt log-normal distributions for all three quantities, as they are physically 
motivated and ensure that each variable is positive-definite. Other distributional choices will be explored elsewhere.

We define three fractions from the universal birth distributions:
\begin{align}
  \fdec  &= \int_{\Mdec}^{\infty} p(M_0)\,dM_0 \ ,
  \label{eq:fdec}\\
  \ffast &= \int_{P_{\rm 0, min}}^{\Pfast}    p(P_0)\,dP_0 \ ,
  \label{eq:ffast}\\
  \ffoss &= \int_{\Bc}^{\infty}  p(B_0)\,dB_0 \ .
  \label{eq:fhigh}
\end{align}
Here:\\

 $- \fdec$ is the fraction of NSs massive enough to undergo the QN.
 Recall that by construction all NSs with
$M_0>\Mdec$ become AXPs\&SGRs regardless of their birth spin period
or fossil field strength.

 Although rapid rotation reduces the central density through
centrifugal support, the effect is modest ($\lesssim10\%$ even for
millisecond periods; \citealt{Hartle1967,Staff2006}). In our model,
stars with $M_0>M_{\rm dec}$ are assumed to be born with
$\rho_c>\rho_{\rm dec}$. Consequently, the rotational correction does
not alter whether the deconfinement threshold is exceeded. 
Models in which deconfinement is triggered by spin-down alone
require the birth central density to lie very close to
$\rho_{\rm dec}$, whereas we consider stars that are already above
the threshold at birth.\\
 
$-\ffast$ is the fraction of NSs born with spin periods below $\Pfast$.
The lower limit $P_{0,\rm min}$ is imposed by centrifugal breakup and,
more importantly, by r-mode instabilities that efficiently remove
angular momentum through gravitational-wave emission. These effects are
generally expected to limit the birth spin period of rapidly rotating
NSs to $P_{0}\gtrsim 1$ ms
\citep[e.g.,][]{Andersson1998,Haskell2015} so we set $P_{0,\rm min}=1$ ms.

QCD magnetars born with $P_0>\Pfast$ produce quiet QNe with no bright transients (i.e. no LFBOTs or SLSNe-I).
This channel dominates the event rate in our model, contributing $\sim\fdec$ of all CCSNe (see Section~\ref{sec:method}).\\

 $-\ffoss$ is the fraction born with a fossil field above the 
threshold $\Bc$, which governs the double-humped SLSN rate and
involves no timescale. The fossil-field threshold $\Bc$, which enters the
double-humped SLSN fraction via $\ffoss$
(Equations~\ref{eq:fhigh} and~\ref{eq:rslsn2}), is held fixed during the Bayesian analysis (\S \ref{sec:population});
varying it over $[10^{13.5},10^{14.5}]$~G does not affect the
posteriors on $\Mdec$ or $\Pfast$ at the $10\%$ level (see \S \ref{sec:limitations}).

The physical discriminant between LFBOTs and SLSNe-I is whether the
QN fires before or after the SN ejecta becomes optically thin, i.e.\
whether $\tSD < t_{\rm SN,o}$ or $\tSD > t_{\rm SN,o}$.
The SLSN rate is thus 
\begin{equation}
f_{\rm slsn}
=
\int_{P_{\rm 0, min}}^{\Pfast} p(P_0)
\left[
\int_{B_{\rm fossil, slsn}}^{\infty} p(B_0)\, dB_0
\right] dP_0 \ ,
\end{equation}
with $B_{\rm fossil, slsn}$ given in Eq. (\ref{eq:B0slsn}) and 
$\ffast = \fslsn + \flfbot$. 
The LFBOT progenitors are the complementary dominant  subset 
since $\flfbot = \ffast - \fslsn \simeq  \ffast$; most NSs with $M>\Mdec$
will have $\tSD>t_{\rm SN,o}$ because of the mean magnetic field $B_{\rm 0, pk}\sim 10^{12.5}$.  
I.e.,  SLSNe-I are rarer than LFBOTs according to our model.

Focusing on Channel~A, and neglecting the small fraction of core-collapse events that do not produce NSs, such that $\mathcal{R}_{\rm NS}\sim \mathcal{R}_{\rm CCSN}$, the rates of the relevant observable events can be approximated as (derived more rigorously by direct Monte Carlo (MC) counting; see Appendix \ref{sec:mcmc}):
\begin{align}
  \mathcal{R}_{\rm AXP}          &\sim \mathcal{R}_{\rm CCSN}\times\fdec \ ,
  \label{eq:raxpfull}\\
   \mathcal{R}_{\rm LFBOT}        &\sim \mathcal{R}_{\rm CCSN}
                                    \times\fdec\times\ffast \ ,
  \label{eq:rlfbot_approx}\\
  \mathcal{R}_{\rm SLSN-I}  &\sim \mathcal{R}_{\rm CCSN}
                                    \times\fdec\times \fslsn \ ,
  \label{eq:rslsn1}\\
  \mathcal{R}_{\rm SLSN-I}^{\rm DH}  &= \mathcal{R}_{\rm AXP}
                                    \times\ffast\times\ffoss \ .
  \label{eq:rslsn2}
\end{align}
The subscript AXP
includes both AXPs and SGRs while the superscript $DH$ stands for ``Double-Humped".

For completeness, NSs born with $M_0<\Mdec$ but with a magnetar-strength
fossil field are fossil AXPs\&SGRs (identified with subscript ``fAXP"). Compared to QCD magnetars, their rate is
\begin{equation}
  \frac{\mathcal{R}_{\rm fAXP}}{\mathcal{R}_{\rm AXP}}
  = \frac{(1-\fdec)\times\ffoss}{\fdec} \approx \frac{\ffoss}{\fdec} << 1\ .
  \label{eq:rfaxp}
\end{equation}

\subsection{Observational constraints, likelihood and rate predictions}
\label{sec:method}

The five constraints used in the likelihood $\mathcal{L}(\theta)$ are listed in
Table~\ref{tab:constraints} with the parameter vector being 
$\theta=(\Mdec,\sigM,\Pfast,\sigP,\sigB)$. 
The first four constraints are occurrence rates relative to the CCSN rate;
the fifth is the fraction of AXPs\&SGRs, $f_{\rm SNR}$, associated with
SNRs, which constrains the tail of the
spin-down timescale distribution independently of the rate
constraints.

The analysis is a forward model. For a given parameter vector,
we draw $N_{\rm MC} =10^{7}$ NSs from the three birth distributions,
classify each into one of the observable outcomes of
Section~\ref{sec:four_outcomes} by direct counting, and compute
five predicted rates. All fractions are evaluated by direct MC counting over
the $N$ draws rather than by numerical integration,
which avoids any approximation and naturally propagates the
full sampling uncertainty; at this resolution, the rarer
channels (SLSN-I, DH-SLSN-I, LFBOT) carry $\mathcal{O}(10\%)$ relative
shot noise per evaluation (Appendix~\ref{app:likelihood}).

The Markov Chain MC (MCMC) sampler then explores the posterior
$p(\theta\mid{\rm data})\propto\mathcal{L}(\theta)\,p(\theta)$
and returns the joint distribution of all five parameters.
The two physically interesting outputs are $\Mdec$ and $\Pfast$;
the remaining three ($\sigM$, $\sigP$, $\sigB$) are nuisance
parameters that characterise the birth distributions and are
marginalised over. Details including the explicit forms used for penalties 
 are given in Appendix~\ref{app:likelihood}.

\begin{table*}[t!]
\centering
\caption{The likelihood constraints and posteriors.
\label{tab:constraints}}
\begin{tabular}{lll}
\toprule
Constraint & Observed value (references) & Model\\
\midrule
$\mathcal{R}_{\rm AXP}/\mathcal{R}_{\rm CCSN}$
  & $0.05$--$0.15$ (\citealt{GillHeyl2007,Beniamini2019})
  &  0.11 \\
$\mathcal{R}_{\rm SLSN-I}/\mathcal{R}_{\rm CCSN}$
  &  (2-4)$\times 10^{-4}$ (\citealt{Frohmaier2021})
  &  $3.2\times 10^{-4}$ \\
$\mathcal{R}_{\rm SLSN-I}^{\rm DH}/\mathcal{R}_{\rm SLSN-I}$
  & $< 0.1$ (\citealt{Angus2019, Chen2023})
  & $0.03$ \\
$\mathcal{R}_{\rm LFBOT}/\mathcal{R}_{\rm CCSN}$
  & $< 10^{-3}$ (\citealt{Coppejans2020,Ho2023})
  &  $1.1\times 10^{-3}$\\
$f_{\rm SNR}$
  & $8/30\simeq  0.267$ (\citealt{Olausen2014})
  & 0.30\\
\bottomrule
\end{tabular}
\end{table*}

\subsection{Results}
\label{sec:results}

\begin{figure*}[t!]
\centering
\includegraphics[width=\textwidth]{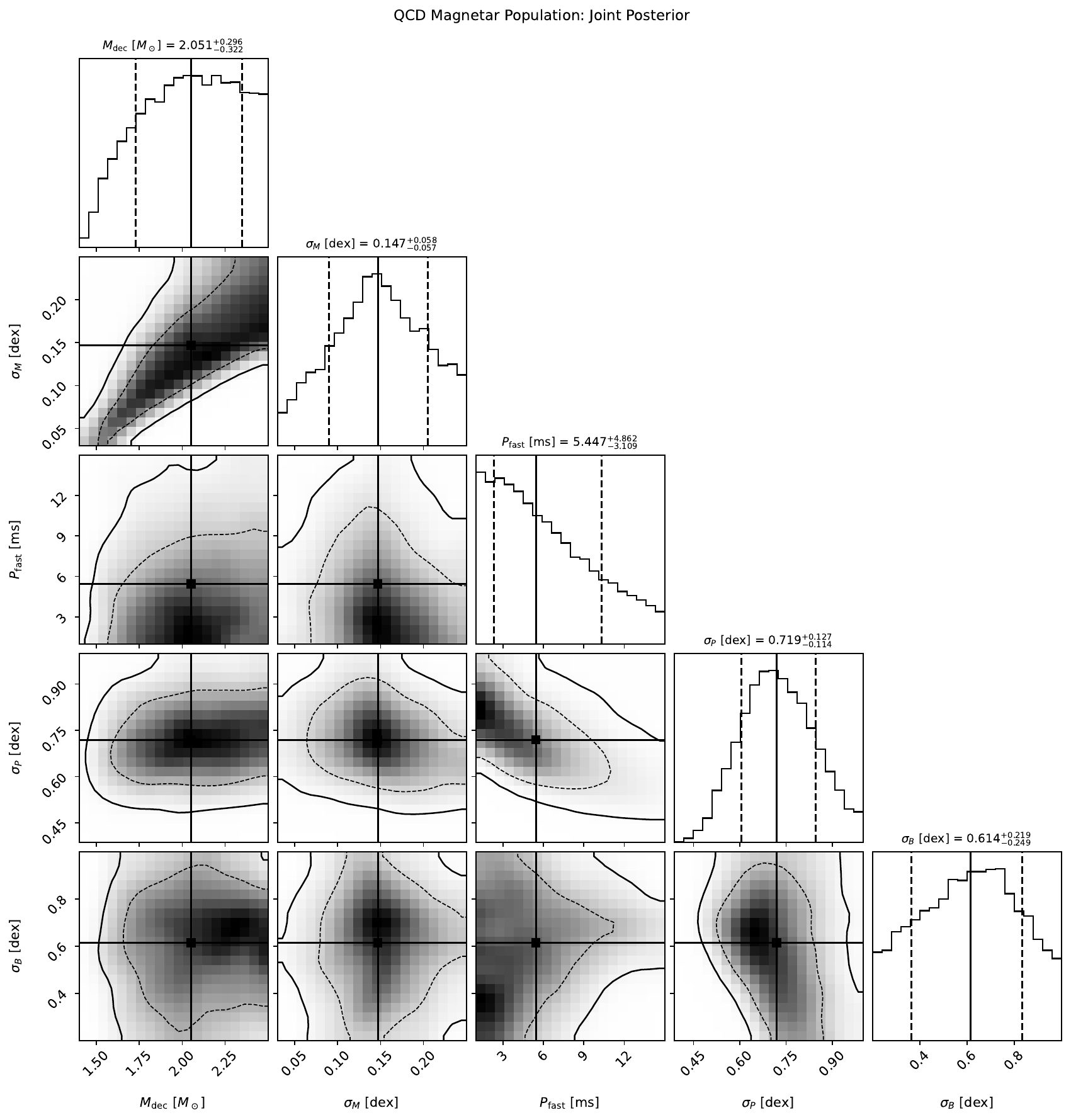}
\caption{Joint posterior distribution of the five free parameters
$(\Mdec,\sigM,\Pfast,\sigP,\sigB)$.
Diagonal panels show marginalised one-dimensional posteriors;
off-diagonal panels show two-dimensional joint posteriors with
$1\sigma$ (dark regions) and $2\sigma$ (light regions) credible
contours. Dashed vertical lines mark the 16th, 50th, and 84th percentiles.
The broad upper uncertainty in $\Pfast$ reflects
a genuine degeneracy between $\Pfast$ and $\sigP$
(see \S\ref{sec:population}).
\label{fig:corner}}
\end{figure*}

Table \ref{tab:constraints} list the resulting likelihood constraint and posteriors
while Table \ref{tab:posterior} lists the corresponding marginalized posterior values. 
The corner plot (Figure~\ref{fig:corner}) shows the joint
posterior on all five parameters. 
The marginals for $\Mdec$ and $\Pfast$ are well-constrained
with nearly symmetric distributions; the nuisance parameters
$\sigM$, $\sigP$, and $\sigB$ show broader posteriors
reflecting the weaker constraining power of the rate data
on the birth distribution widths.
No strong degeneracies are present between $\Mdec$ and
$\Pfast$, confirming that the two physically interesting
parameters are independently constrained by the data.

\begin{table}[t!]
\centering
\caption{Marginalised posterior constraints (median and 68\%
credible intervals). Provisional, low-resolution run
($N_{\rm MC}=10^7$); see text.
\label{tab:posterior}}
\begin{tabular}{lll}
\toprule
Parameter & Posterior & Units \\
\midrule
$\Mdec$  & $2.05^{+0.30}_{-0.32}$ & $\Msun$ \\
$\sigM$  & $0.15^{+0.06}_{-0.06}$ & dex \\
$\Pfast$ & $5.45^{+4.86}_{-3.11}$ & ms \\
$\sigP$  & $0.72^{+0.13}_{-0.11}$ & dex \\
$\sigB$  & $0.61^{+0.22}_{-0.25}$ & dex \\
\bottomrule
\end{tabular}\\
$\Mdec$ and $\Pfast$ are the parameters of
primary physical interest.
Flat priors are used throughout (see Table~\ref{tab:priors};  Appendix \ref{app:likelihood}).
\end{table}

The deconfinement mass threshold is constrained to
$\Mdec\sim 2.1\,\Msun$, above the well-measured
two-solar-mass pulsars \citep{Demorest2010,Antoniadis2013,
Cromartie2020} and consistent with current upper limits on the
maximum hadronic NS mass \citep[e.g.,][and references therein]{Annala2018,Annala2020}.

The fraction $\fdec$ relevant to the present model refers to
$P(M_{\rm birth}>\Mdec)$, i.e., the fraction of the NS
\emph{birth-mass distribution} above $\Mdec$, rather than to the
fraction of currently observed NSs with precisely measured current
masses above this threshold. For the adopted birth-mass distribution,
$\log_{10}(M_0/\Msun)\sim\mathcal{N}(\log_{10}(1.35),,0.15^2)$, the
threshold $\Mdec\approx2.05,\Msun$ lies at approximately $1.2\sigma$
above the peak, corresponding to $\fdec\approx11$--$12\%$ under this
distribution. The three NSs with the most precisely measured masses
above $2,\Msun$ \citep{Demorest2010,Antoniadis2013,Cromartie2020} are
recycled millisecond pulsars in NS--white-dwarf binaries, whose present
masses need not coincide with their birth masses because their
evolutionary histories include a phase of mass transfer and recycling.
More generally, the precision-mass sample is small and not a
representative draw from the NS birth-mass distribution. Consequently,
the observed rarity of precisely measured $\gtrsim2,\Msun$ NSs cannot
by itself be taken as evidence against the birth-mass tail fraction
inferred in our model.

The fast-spin threshold is constrained to $\Pfast\sim 5.5$~ms.
This constraint is driven primarily by the LFBOT rate  because it depends sensitively on the combination of $\Pfast$
and the $\tSD$ vs.\ $t_{\rm SN, o}$ split.
The inferred value corresponds to the millisecond tail of the
pulsar birth spin distribution, 
confirming that the LFBOT channel requires genuine millisecond
rotation at the time of QN conversion.

\begin{figure}[t!]
\centering
\includegraphics[scale=0.9]{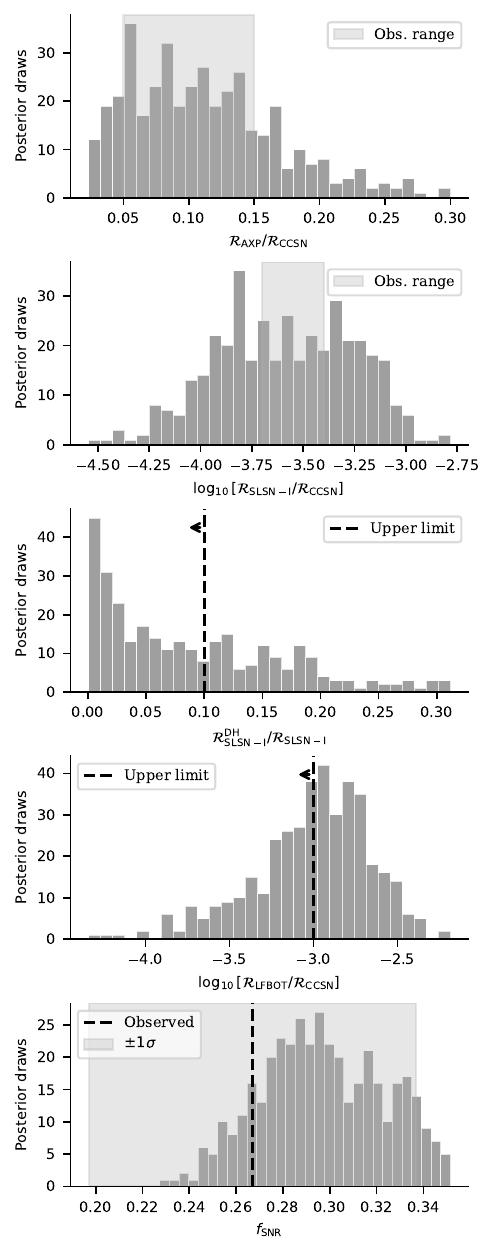}
\caption{Posterior predictive distributions of the five
observational constraints, evaluated from 400 draws from the
joint posterior. The bands show the observed ranges or $\pm1\sigma$ intervals
from Table~\ref{tab:constraints}. The double-humped SLSNe distribution is not shown since it is an upper limit.
All five constraints are simultaneously satisfied for the
median posterior parameters.
\label{fig:rates}}
\end{figure}

The posterior predictive distributions are shown in Figure~\ref{fig:rates}. Four of
the five constraints (AXP\&SGR, SLSN-I, DH SLSN-I, $f_{\rm SNR}$) are
satisfied across the bulk of the posterior width; in particular, the
DH SLSN-I fraction ($\sim0.03$) sits well below the adopted upper
limit ($<0.1$; see Table~\ref{tab:constraints}). The LFBOT
constraint shows a mild tension: the model
prediction ($\sim 1.1\times 10^{-3}$) sits at the observational
upper limit ($<10^{-3}$; see Table \ref{tab:constraints}).

The posterior yields birth distribution widths  $\sigP \sim 0.7$~dex, $\sigB \sim 0.6$~dex, and $\sigM \sim 0.15$~dex.
 The model therefore suggests that the widths of the distributions are consistent with those required by the observed high-energy transient populations \citep{FaucherKaspi2006,Igoshev2022,Ozel2016}, without requiring any modification of the underlying NS birth population.
A non-trivial internal consistency check of the framework
is that the three recovered birth distribution widths --- the only free parameters
not directly tied to the transient phenomenology ---
are claimed to be consistent with independent pulsar population
studies, despite having been constrained entirely by the
AXP\&SGR, SLSN, LFBOT rates, and the fraction of AXPs\&SGRs associated with SNRs. 

\subsection{SNR age consistency}
\label{sec:mcgill}

Of the 30 confirmed magnetars in the McGill catalogue
\citep{Olausen2014}, 8 are associated with SNRs whose ages
have been estimated independently from their dynamics or
thermal emission.
In the QN framework, a magnetar is associated with a visible
SNR only if $\tSD<\tSNR\sim 10^4$ yrs, i.e.\ the QN fired before the
remnant faded (calibrated to reproduce the observed association
fraction; see Appendix~\ref{app:likelihood}). Furthermore, the post-conversion characteristic age $\tau_c=P_{\rm HS}/(2\dot{P}_{\rm HS})$ measures only the
time since QN conversion under $\BHS\sim10^{15}$~G; it does
not include the Phase~1 lifetime (the pre-QN evolution of the NS
between the birth SN and quark deconfinement) and is therefore systematically
smaller than $t_{\rm SNR}$, which measures the time since the birth SN.

At the population-mean field $\sim 10^{12.5}$~G,
the implied maximum birth period $P_0^{\rm max}$ for the
best-constrained SNR-associated magnetars
(\citealt{Olausen2014,Vink2006})
ranges from $\sim13$ to $\sim51$~ms, all within $1$--$2\sigma$
of the birth distribution peak.
The fraction of the Monte Carlo population satisfying
$\tSD<10^{4}$~yr is $f_{\rm SNR}\sim 0.322$,
consistent with the observed $8/30=0.267$ at
better than $1\sigma$
(Appendix~\ref{app:likelihood}).

%%%%%%--------------------------
%% FIGURE1
%% END FIGURE 1
%%%%%%--------------------------

%%%%%%%%%%%%%%%%%%%%
%%%% Connection to FRBS
% -----------------------------------------------------------------------
\section{QCD Magnetars as Potential X-ray Quiet FRB Sources}
\label{sec:frb}
% -----------------------------------------------------------------------

The crustless, highly magnetized HS provides a distinct physical engine
for repeating FRBs. In our model, the quark core sustains a
volume-filling ferromagnetic field of order $10^{18}\,$G, whereas the
surrounding hadronic envelope retains a much weaker inherited field of
order $\sim10^{15}\,$G. This strong magnetic discontinuity across the
hadron--quark interface represents a long-lived reservoir of magnetic
free energy. Maintaining force balance across such a sharp transition
requires substantial magnetization or screening currents at the
interface, making this region naturally susceptible to magnetic
instabilities, reconnection, and localized failures of the interface.

The interface is continuously driven away from equilibrium by the
secular evolution of the HS. As the star spins down, the reduction in
centrifugal support modifies the pressure balance and shifts the
equilibrium position of the hadron-quark interface. If the magnetic
field is frozen into the ferromagnetic quark core, the interface cannot
remain static and must continually readjust toward its evolving
equilibrium configuration. Consequently, even after the initial
spin-down epoch, when the rotational energy loss approaches its
asymptotic ${t^\prime}^{-2}$ evolution, the interface remains in a state of
quasi-continuous disequilibrium. Rather than responding smoothly to
this secular driving, the interface is expected to undergo
threshold-triggered, avalanche-like relaxation events: slow stress
accumulation followed by rapid magnetic reconnection and localized
energy release once a critical stability threshold is exceeded.

 We discuss this mechanism in Appendix \ref{app:magnetic_avalanche} where we argue that the magnetic energy stored at the interface 
 is released by constant adjustment of the interface  through
Parker-like \citep{Parker1966} or Tayler-like instabilities
\citep{Tayler1973,Woltjer1958} into coherent magnetic flux ropes (the ``magnetic bubbles"), which
propagate through the hadronic envelope on approximately the Alfv\'en
crossing timescale, reaching the stellar surface on sub-second timescales while remaining
sufficiently coherent. Spin-down controls the long-term evolution of the
system while stress build up and release the duration of an active episode. 
The cadence of magnetic bubbles formation at the interface (and subsequently the 
triggered FRBs at the HS surface) is determined by
the fast local reconnection and relaxation physics at the interface.
We give a rough estimate of the energetics and relevant timescales in Appendix \ref{app:magnetic_avalanche}
including the  threshold-loading time between avalanche
events while cautioning that this remains a qualitative description.

While the HS lacks a solid lattice, transient crustal-fracture heating is absent, and
the bursts produced at the surface are X-ray quiet by construction --
a distinctive signature differentiating the mechanism we propose here from
conventional crust-fracture magnetar scenarios. This X-ray-quiet phase
is transient rather than permanent: as the HS evolves, the
crustless surface is eventually expected to acquire a solid crust at a
timescale $t^\prime_{\rm crust}$ (see \S \ref{sec:dtcrust}), after which the star transitions into a
conventional, X-ray-loud AXP\&SGR. 

\subsection{Crustless-phase duration, $t^\prime_{\rm crust}$}
\label{sec:dtcrust}

We assume that magnetic bubbles (or rising flux ropes) emerging through the HS surface perturb the magnetosphere, preferentially in the vicinity of the magnetic polar caps where the field lines remain open. The rapid reconfiguration of the magnetosphere is expected to induce strong electric fields, transient particle acceleration, and pair creation, producing relativistic electron--positron outflows along the open field lines. Such conditions are favorable for coherent radio emission at GHz frequencies through one or more of the magnetospheric mechanisms proposed for pulsars and FRBs, including coherent curvature radiation from charge bunches streaming along curved magnetic field lines \citep{GoldreichJulian1969,RudermanSutherland1975,Melikidze2000,Gil2004,Kumar2017,Yang2018,Lu2020}. In the present work we do not specify the microphysical emission process; rather, we assume that a fraction of the magnetic energy released by the emerging bubbles is converted into coherent radio emission. The open-field-line region above the polar caps provides a natural beaming geometry, while the burst energetics, activity timescale, and repetition rate are determined primarily by the magnetic activity at the interface rather than by the details of the emission mechanism.

As the HS enters the spin-down phase, $P_{\rm HS}\propto {t^\prime}^{1/2}$ for $t^\prime>t^\prime_{\rm HS,SpD}$, the interface continually readjusts to the evolving pressure balance (\S\ref{app:magnetic_avalanche_threshold}), generating successive magnetic bubbles. These provide a sustained reservoir of electromagnetic energy capable of powering repeated FRBs throughout the active crustless phase, $t^\prime_{\rm crust}$.

We can estimate $t^\prime_{\rm crust}$ by imposing the
$P_{\rm AXP}\sim2$--12 seconds period of AXPs and SGRs so that
\begin{equation}
P_{\rm AXP}\sim  2^{1/2} P_0\times \left(1+ \frac{t^\prime_{\rm crust}}{t^\prime_{\rm HS, SpD}}\right)^{1/2}\ .
\end{equation}
In the late-time spin-down regime, $t^\prime\gg t^\prime_{\rm HS,SpD}$,
the above expression reduces to
\begin{equation}
t^\prime_{\rm crust}\sim 32\ {\rm yr}\times \left( \frac{P_{\rm AXP}}{s} \right) \left( \frac{10^{15}\ {\rm G}}{B_{\rm HS}} \right)\ .
\end{equation}
This suggests the crust forms at $t^\prime_{\rm crust}\sim$ (64-384) yr; I.e. QCD magnetars act
as FRB sources for centuries.

We emphasize that $t^\prime_{\rm crust}$ is not intended as a
microscopic crust-nucleation timescale derived from cooling alone.
Rather, it represents the characteristic evolutionary time over which
the crustless hadronic envelope can evolve toward conditions permitting
crust re-formation. Physically, crust formation requires matter to reach
subnuclear densities, where nuclei embedded in an electron gas are
energetically favored. This may occur through decompression of the
exposed outer hadronic layers and/or through deposition of fresh matter
by fallback or subsequent accretion. The actual onset of crust
formation may therefore depend on the structure and evolution of the
hadronic envelope and on the astrophysical environment. The
$64$--$384$~yr timescale should consequently be regarded as a
characteristic evolutionary timescale adopted in this
proof-of-principle model, rather than as a first-principles prediction
of the microscopic crust-formation process.

\subsection{FRB Rate}

The FRB source formation rate is
\begin{equation}
\mathcal{R}^{\rm source}_{\rm FRB} \sim f_{\rm dec}\times\mathcal{R}_{\rm CCSN}\ ,
\label{eq:source_rate_corrected}
\end{equation}
with $\mathcal{R}_{\rm CCSN}\sim 10^5\ {\rm Gpc^{-3}\,yr^{-1}}$
\citep{Strolger2015}.

The observed all-sky FRB rate is roughly
$\mathcal{R}_{\rm FRB}^{\rm obs} \sim 7\times10^4\ {\rm Gpc^{-3}\,yr^{-1}}$
above a pivot energy of $10^{39}\,$erg and scattering $<10\,$ms at
600 MHz \citep{Shin2023}. I.e., the expected burst count per QCD
magnetar is
\begin{equation}
\langle N_{\rm bursts}\rangle \sim \frac{7}{f_{\rm pc}}\ ,
\label{eq:nbursts_corrected}
\end{equation}
where $f_{\rm pc}$ accounts for beaming effects arising from the preferential emission along the polar caps (pc), as described below.

For a dipolar HS, the polar cap half-opening angle is approximately
\begin{equation}
\theta_{\rm pc} \simeq
\left(\frac{R_{\rm HS}}{R_{\rm LC}}\right)^{1/2}
=
\left(\frac{2\pi R_{\rm HS}}{c P_{\rm HS}}\right)^{1/2},
\end{equation}
where $R_{\rm HS}=R_{\rm NS}$ is the HS radius and $R_{\rm LC}=c/\Omega_{\rm HS}$
the HS light cylinder, with $\Omega_{\rm HS}$ the angular spin
frequency. The corresponding solid-angle fraction is
\begin{equation}
f_{\rm pc} = \frac{\pi \theta_{\rm pc}^2}{4\pi} \propto \frac{1}{P_{\rm HS}}\ ,
\end{equation}
with $P_{\rm HS} \sim \sqrt{2}\,P_{0}$, the post-conversion period.
 Rapidly rotating QCD magnetars  (i.e., with wide polar cap) are far more likely to have a
substantial fraction of their intrinsic activity beamed toward any
given observer; slowly rotating ones (narrow polar cap) are far
more likely to have their activity beamed away from us entirely,
appearing as non-detections or, if a rare alignment occurs, as apparent
one-off bursts.

\subsubsection{Population-average normalization}

The universal true event rate is fixed using the \emph{population
average} of $f_{\rm pc}$ over the $P_{\rm HS}=2^{1/2}P_0$ distribution (with $P_0$ the lognormal,
median 0.2~s, $\sigma_P=0.71$~dex; \S~\ref{sec:population}). I.e., $\langle f_{\rm pc}\rangle_{\rm pop} \approx 8.5\times10^{-4} $ so that 
\begin{equation}
\langle N_{\rm bursts}\rangle_{\rm pop} = \frac{7}{\langle f_{\rm pc}\rangle_{\rm pop}} \approx 8.2\times10^{3}\ .
\end{equation}

For an individual source with its own $f_{\rm pc,i}$, the observed
lifetime burst count scales \emph{proportionally}:
\begin{equation}
N_{{\rm obs},i} = \langle N_{\rm bursts}\rangle_{\rm pop}\times
\frac{f_{\rm pc,i}}{\langle f_{\rm pc}\rangle_{\rm pop}}\ .
\end{equation}

\begin{table*}[t!]
\centering
\begin{tabular}{lccccc}
\hline
$P_{0}$ (NS) & $P_{\rm HS}=2^{1/2}P_{0}$ & $t^\prime_{\rm HS, SpD}$ & $f_{\rm pc,i}$ & $N_{{\rm obs},i}$ (ratio to pop.\ mean) & $t_{\rm wait}\sim 0.01t^\prime_{\rm HS, SpD}$\\
\hline
Typical (0.2 s)         & 283\,ms  & $\sim$ 13 yr  & $2.22\times10^{-4}$ & $2.16\times10^3$ (0.26$\times$) &  $\sim$ 47 d\\
$P_{\rm fast}$ (5.5 ms)  & 7.78\,ms & $\sim$ 3.6 d  & $8.08\times10^{-3}$ & $7.85\times10^4$ (9.5$\times$) &  $\sim$ 52 mn\\
$P_{0,\rm min}$ (1\,ms) & 1.41\,ms & $\sim$ 2.8 hr   & $4.44\times10^{-2}$ & $4.32\times10^5$ (52$\times$) &  $\sim$ 100 s\\\hline
\end{tabular}
\caption{Total lifetime burst count $N_{{\rm obs},i}$ for
representative sources, from polar-cap visibility
$f_{\rm pc}\propto P_{\rm HS}^{-1}$ alone, relative to the population mean.
This spans $\sim2.3$~dex from a typical source to the extreme
fast-rotator tail; see Table~\ref{tab:frb_rate} for the corresponding
peak observed rate, which scales more steeply
($\propto P_{\rm HS}^{-3}$) once the energy-budget effect is included.}
\label{tab:frb_geometric}
\end{table*}

This geometric argument accounts for a $\sim2.3$~dex spread in activity
from typical to the extreme fast-rotator tail as shown in Table \ref{tab:frb_geometric}.

\subsubsection{Peak Rate}

For a  $B_{\rm HS}\sim10^{15}\,$G that varies little among QCD magnetars, the
period $P_{\rm HS}$ is the primary control on $t^\prime_{\rm HS,SpD}$. 
The conditions in the cores of the massive NSs are expected to be similar meaning that 
 the resulting magnetic free energy available to
power the FRB-active phase, $E_{\rm mag}$, is approximately the same
across sources (set by $B_{\rm QCD}$, which is essentially
independent of spin). In this picture,  a QCD magnetar with shorter
$t^\prime_{\rm HS,SpD}$ -- i.e.\ smaller $P_{\rm HS}$ -- exhausts the same
reservoir on a correspondingly shorter timescale, and therefore shows a
higher intrinsic activity rate (see Appendix \ref{app:magnetic_avalanche}),
\begin{equation}
\dot N_{\rm intrinsic} \propto \frac{E_{\rm mag}}{t^\prime_{\rm HS,SpD}}
\propto P_{\rm HS}^{-2}\ .
\end{equation}
This is a distinct effect from the polar-cap visibility fraction
$f_{\rm pc}\propto P_{\rm HS}^{-1}$,
and the two combine differently depending on which observable is
considered. 

For the \emph{instantaneous (peak) observed rate}, beaming and the energy-budget effect combine
multiplicatively,
\begin{equation}
\dot N_{{\rm peak},i} = \dot N_{\rm intrinsic,i}\times f_{\rm pc,i}
\propto P_{\rm HS}^{-3}\ .
\label{eq:rate_scaling}
\end{equation}

Combined with the polar-cap visibility fraction $f_{\rm pc}\propto
P_{\rm HS}^{-1}$, the peak observed rate
$\dot N_{\rm peak}=\dot N_{\rm intrinsic}\times f_{\rm pc}\propto P_{\rm HS}^{-3}$ yields a wider spread in
activity as shown in Table \ref{tab:frb_rate}.

\begin{table}[t!]
\centering
\begin{tabular}{lccc}
\hline
$P_{0}$ (NS)  & $P_{\rm HS}$ & rate ratio ($\propto P_{\rm HS}^{-3}$) & dex \\
\hline
Typical         & 283\,ms  & 1$\times$              & 0    \\
$P_{\rm fast}$  & 7.78\,ms & $4.8\times10^{4}$      & 4.68 \\
$P_{0,\rm min}$ (1\,ms) & 1.41\,ms & $8.1\times10^{6}$ & 6.91 \\
\hline
\end{tabular}
\caption{Peak observed rate ratio relative to a typical source, from
the combined beaming ($P_{\rm HS}^{-1}$) and energy-budget
($P_{\rm HS}^{-2}$) effects, Eq.~(\ref{eq:rate_scaling}).}
\label{tab:frb_rate}
\end{table}

FRB\,20240114A, the most hyperactive known repeater, released bursts
at a peak rate of $729\,{\rm hr^{-1}}$ over 214 days
\citep{Zhang2025}, an excess of $\sim4.7$~dex relative to a typical
source in our model -- closely matched by the $P_{\rm fast}$ entry in
Table~\ref{tab:frb_rate}, consistent with this source being drawn from
the fast-rotator tail of the $P_0$ distribution.  FRB~121102's  peak rate $122\,{\rm hr^{-1}}$
\citep{Li2021} is about 6$\times$ lower than FRB\,20240114A and is more
consistent with an intermediate $P_{\rm HS}$, illustrating that peak
rate and total lifetime count need not track one another under this
model. We caution
that the $E_{\rm mag}$-universality assumption underlying
Eq.~(\ref{eq:rate_scaling}) is a working hypothesis, not an
independently established result, and that both quantities in
Table~\ref{tab:frb_rate} remain relative ratios rather than absolutely
normalized predictions.

\subsubsection{Loading time $t_{\rm wait}$}
\label{sec:twait}

The threshold-loading time $t_{\rm wait}$ between bursts episodes is derived in Appendix \ref{app:magnetic_avalanche}
with  
\begin{equation}
t^\prime_{\rm
wait}\sim0.01\,t^\prime_{\rm HS,SpD},\propto\
P_{\rm HS}^{2}\ ;
\end{equation}
see Table \ref{tab:frb_geometric}. We note here that
burst cadence is related
to the onset and growth of the instability at the interface
(Appendix \ref{app:magnetic_avalanche_alfven}), but whether it can reproduce
the observed cadence in FRBs remains to be confirmed.

Taking $t^\prime _{\rm wait}$ literally as the interval between bursting episodes   would imply an average cadence far below the peak rate
of $729\ {\rm hr^{-1}}$ reported for FRB\,20240114A
\citep{Zhang2025} (one burst every $\sim5$ s) 
even for the $P_{\rm HS}\sim P_{\rm fast}$ source
($t_{\rm wait}\sim52$~min, Table~\ref{tab:frb_geometric}): matching
the observed rate would require each threshold-crossing avalanche to
release a clustered storm of order $10^{2}$--$10^{3}$ individual
sub-bursts, rather than a single event. This is not obviously
unreasonable --- clustered, multi-component burst morphology within
short active windows is a well-documented feature of hyperactive
repeaters --- and would be a natural consequence of a single
large-scale interface instability fragmenting into many smaller
reconnection/flux-rope-detachment events as expected in 
self-organized criticality avalanche model (see Appendix \ref{app:magnetic_avalanche}).

% -----------------------------------------------------------------------
\subsection{Periodicity}
\label{sec:frb_periodicity}
% -----------------------------------------------------------------------

A natural question is whether the threshold-loading time $t^\prime_{\rm
wait}$ derived above can account for the $\sim16.35$-day
periodicity observed in the activity of FRB\,180916.J0158+65, whose
bursts are confined to a $\sim5$-day active window recurring every
$16.35\pm0.15$~days \citep{CHIME2020}, and possibly analogous longer
periodicities such as the $\sim157$--161-day modulation reported for
FRB\,121102 \citep{Rajwade2020,Cruces2021}.

Solving $t_{\rm wait}\sim0.01\,\tau_{\rm HS,SpD}=16$~days for $P_{\rm HS}$ (at
fixed $B_{\rm HS}=10^{15}$~G) gives $P_{\rm HS}\approx164$~ms --
comfortably inside our modeled period range, between the Typical
$P_{\rm HS}$ (283~ms) and $P_{\rm HS, fast}$ (7.78~ms) benchmarks of Table~5. In this
narrow sense, a $\sim16$-day recurrence timescale is not exotic or
fine-tuned within our framework: it sits naturally within the range of
$t_{\rm wait}$ spanned by the underlying $P_0$ distribution.

This does not, however, mean our mechanism explains the observed
periodicity. $t_{\rm wait}$ is a threshold-loading time in a
stochastically driven avalanche-like process
(\S5.2.1--5.2.3); such processes generically produce a
\emph{broad distribution} of waiting times of order $t_{\rm wait}$,
not a sharp, clock-like recurrence. The observed periodicity, by
contrast, is measured to $\sim1\%$ precision and remains phase-coherent
over dozens of cycles spanning years \citep{CHIME2020}. Reproducing
this level of regularity from a stochastic threshold process would
require additional structure we have not derived -- e.g.\ a
narrowly peaked (rather than broad) waiting-time distribution -- and we
do not claim to have such a mechanism. We therefore do not identify
$t_{\rm wait}$ with the 16-day period; at best it shows that
week-to-month recurrence is a natural \emph{scale} in our model, not
that our model produces strict periodicity at that scale.

A further question is why periodicity has been confirmed in only a
handful of repeaters. This is partly an observational selection
effect: robust period detection requires dozens of well-timed bursts
and a monitoring baseline spanning several cycles, a bar met by only
the most prolific, longest-monitored sources. It is also consistent
with our framework: if the recurrence timescale governing burst
clustering varies as steeply across the population as
$t^\prime_{\rm HS,SpD}\propto P_{\rm HS}^2$ (Table~5), only sources whose
characteristic period happens to fall within the few-year baseline of
current surveys -- roughly days to months -- would show a detectable
signal; much shorter periods blur into apparently continuous activity,
and much longer ones would require decades of monitoring to reveal
even one cycle.

Finally, we note that $P_{\rm HS}$ itself (millisecond-scale) is not
expected to appear directly in the data, and this is not unique to
our model: even for FRB\,200428, whose source SGR\,1935+2154 has an
independently measured $3.245$-s rotation period from X-ray timing,
that period is not seen in its burst activity \citep{Katz2021}. In our
picture the avalanche trigger is a rare, stochastic threshold-crossing
event rather than a beam sweeping past the observer every rotation, so
individual burst arrival times are not expected to phase-lock to
$P_{\rm HS}$; with $P_{\rm HS}$ itself evolving rapidly
($P_{\rm HS}\propto {t^\prime}^{1/2}$) during the youngest, most active phase,
any residual periodicity at this timescale would in any case be
difficult to fold and detect with realistic burst samples.

%%%%%%--------------------------
%% FIGURE 2
%% END FIGURE 2
%%%%%%--------------------------

% -----------------------------------------------------------------------
\subsection{QN DM and RM}
\label{sec:frb_dmrm}
% -----------------------------------------------------------------------

The QN ejecta always produces a local DM$_{\rm QN}$, intrinsic to the
expanding QN ejecta itself and independent of any surrounding
environment. Furthermore, the shock between the QN ejecta and the
surrounding ambient medium of density $n_{\rm amb}$ induces a magnetic
field yielding a rotation measure that is also intrinsic to the QN
ejecta. 

The QN ejecta becomes transparent to GHz radio emission at
time $t^\prime_{\rm QN,r}\sim54$~days for fiducial parameter values, while
the transition from free expansion (with negligible RM$_{\rm QN}$) to
the Sedov-Taylor phase \citep{Sedov1959,Taylor1950} at $t^\prime_{\rm QN,tr}$ depends on $n_{\rm amb}$; see
Appendix~\ref{app:dm_rm} for details. We have

\begin{align}
\label{eq:QN-DM}
{\rm DM}_{\rm QN} \sim
\left\{
    \begin{aligned}
         & \frac{2.6\ {\rm pc\ cm}^{-3}}{{t^\prime}_{\rm yr}^2}  \quad & t^\prime \le t^\prime_{\rm QN, tr} \\
         & 1.2\times 10^{-2}\ {\rm pc\,cm^{-3}}\times \\
&\times \left(\frac{n_{\rm amb}}{1\ {\rm cm}^{-3}}\right)^{2/3}
 \left(\frac{t^\prime}{t^\prime_{\rm QN, tr}}\right)^{-4/5} \quad & t^\prime >  t^\prime_{\rm QN, tr}                 
    \end{aligned}
\right. \ ,
\end{align}
where ${t^\prime}_{\rm yr}$ is time in years and $t^\prime_{\rm QN, tr} \sim
14.5\ {\rm yr}\times (n_{\rm amb}/1\ {\rm cm}^{-3})^{-1/3}$; RM$_{\rm
QN}(t^\prime)$ is given by Eq. (\ref{eq:RM_QN_TR}):
\begin{equation}
{\rm RM}_{\rm QN}(t^\prime) \sim 75\ {\rm rad\,m^{-2}}\times
\left(\frac{n_{\rm amb}}{1\ {\rm cm}^{-3}}\right)^{7/6}
\left(\frac{t^\prime}{t^\prime_{\rm QN, tr}}\right)^{-7/5}\ .
\label{eq:RM_QN_TR}
\end{equation}

Both DM$_{\rm QN}$ and RM$_{\rm QN}$ depend on $n_{\rm amb}$, which is
set by the environment the QN ejecta expands into. Setting
$t^\prime_{\rm QN,r}\sim t^\prime_{\rm QN,tr}$ identifies a critical
density, $n_{\rm amb}\sim10^6\,{\rm cm^{-3}}$, that separates two
regimes. Below this value, the QN ejecta clears to GHz transparency
\emph{before} the shock forms: DM$_{\rm QN}$ is seen first, in the
free-expansion branch of Eq.~(\ref{eq:QN-DM}), with negligible
RM$_{\rm QN}$ until the shock catches up later. Above this value, the
shock forms \emph{while} the ejecta is still opaque, so that by the
time the source first becomes observable it is already in the
post-shock regime, and DM$_{\rm QN}$ and RM$_{\rm QN}$ appear together
from the outset.

Reaching $n_{\rm amb}\sim10^6\,{\rm cm^{-3}}$ requires a genuinely
dense environment: a canonical SN ejecta ($M_{\rm
SN}=10\,M_\odot$, $E_{\rm SN}=10^{51}$~erg) reaches this density at an
age of $\simeq14$~yr. This is only available if the QN occurs within
the surrounding SN ejecta itself, within roughly this same $\sim14$~yr
window (Channel~A); Channel~B has no associated SN ejecta and cannot
reach this density. Table ~\ref{tab:namb} shows the sensitivity to $n_{\rm amb}$.

\subsubsection{Effect of the surrounding SN ejecta (Channel A)}
\label{sec:dm_rm_sn_screen}

The high-$n_{\rm amb}$ configuration only matters observationally
 if the SN ejecta itself is also transparent to the resulting
radio emission -- a separate condition from the QN ejecta's own
transparency. Applying the same free-free
criterion (Eq.~\ref{eq:tqnr}), but now to a uniform-density sphere of SN ejecta, $V_{\rm
SN}=\tfrac{4\pi}{3}(v_{\rm SN}t)^3$, with $n_e\approx M_{\rm
SN}/(\mu_e m_H V_{\rm SN})$ for standard (solar) composition,
$\mu_e\approx1.18$, gives
\begin{align}
\label{eq:tSNr}
t_{\rm SN,r} &\sim 28.8\ {\rm yr} \times  \left(\frac{M_{\rm SN}}{10\,M_\odot}\right)^{0.4}
\left(\frac{v_{\rm SN}}{10^4\,{\rm km\,s^{-1}}}\right)^{-1}\times \\\nonumber
&\times \left(\frac{T_e}{10^4\,{\rm K}}\right)^{-0.27} \left(\frac{\nu}{1\,{\rm GHz}}\right)^{-0.42}\ ,
\end{align}
where $T_{\rm e}$ is the electron temperature. This is comparable to
 the $\sim14$~yr timescale on which the SN
ejecta's own density falls below $10^6\,{\rm cm^{-3}}$: by the time
the SN ejecta has cleared enough to let any radio signal through, the
ambient density it can supply to the QN shock has already dropped well
below the fiducial value. Channel~A therefore does not generically
produce an observable high-RM, high-DM young QN-FRB through this
specific mechanism. From Eq.~(\ref{eq:QN-DM}), $n_{\rm amb}<10^6\,{\rm
cm^{-3}}$ correspondingly caps DM$_{\rm QN}\lesssim100\,{\rm
pc\,cm^{-3}}$ once the source becomes visible.

The two limiting values of $n_{\rm amb}$ therefore bracket the
resulting RM. At the low end, Channel~B ($n_{\rm amb}\sim1\,{\rm
cm^{-3}}$) is the more natural candidate for an early-time, large-DM,
low-RM young QN-FRB. At the high end, $n_{\rm amb}\sim10^6\,{\rm
cm^{-3}}$, Eq.~(\ref{eq:RM_QN_TR}) gives a strong RM sustained at
$10^4$--$10^8\,{\rm rad\,m^{-2}}$ out to decades (Table~\ref{tab:dm_rm})
-- comparable to the extreme RM measured for known persistent
repeaters (FRB~121102, ${\rm RM}\sim10^5\,{\rm rad\,m^{-2}}$,
\citealt{Michilli2018}; FRB~190520B, \citealt{AnnaThomas2023}) --
though, as shown above, Channel~A sources are unlikely to show this
correspondingly large RM until well after their own high-$n_{\rm amb}$
epoch has already passed.

Because the crustless, FRB-active phase of the QCD magnetar persists
for $t^\prime_{\rm crust}\sim$~centuries (\S\ref{sec:dtcrust}), a
single repeating source can in principle be monitored over a large
fraction of the DM/RM evolutionary window above (Table~\ref{tab:dm_rm}):
its DM and RM are not fixed quantities but should secularly decline as
the source ages, steepest in the first decade and largely flat
thereafter. Channel-B sources -- observable from the start, per the
above -- are therefore the most promising targets for tracking this
early-time DM/RM evolution directly.

\begin{table}[h]
\centering
\begin{tabular}{ccc}
\hline
$t^\prime$ & DM$_{\rm QN}$ [pc cm$^{-3}$] & RM$_{\rm QN}$ [rad m$^{-2}$] \\
\hline
1 day                    & $3.47\times10^{5}$ (FE) & pre-shock \\
30 days                  & $3.85\times10^{2}$ (FE) & pre-shock \\
55 days ($t^\prime_{\rm QN, tr}$)   & $1.17\times10^{2}$      & $7.31\times10^{8}$ \\
1 yr                     & $25.7$                  & $5.17\times10^{7}$ \\
10 yr                    & $4.07$                  & $2.05\times10^{6}$ \\
100 yr                   & $0.645$                 & $8.18\times10^{4}$ \\
\hline
\end{tabular}
\caption{QN ejecta DM and RM vs. time since the QN for $n_{\rm amb}=10^6\,{\rm cm^{-3}}$.  Here, FE refers to the free-expansion
regime, prior to shock formation.}
\label{tab:dm_rm}
\end{table}

%%%%%%%%%%%%%%%%%%%%
%%%% End Connection to FRBS

% ════════════════════════════════════════════════════════════════════════
\section{Discussion}
\label{sec:discussion}
% ════════════════════════════════════════════════════════════════════════

We now discuss the complementary accretion channel, the
 chemical enrichment signatures of the QN as well as the implications of our findings
to QCD  before summarising its limitations.

%% -----------------------------------------------------------------------
\subsection{Channel~B: Accretion-induced QN}
\label{sec:chanB}
% -----------------------------------------------------------------------
In Channel~B, a NS born with $M_0<\Mdec$ accretes from a binary
companion until $\rhoc$ crosses $\rhodec$. Because accretion torque
dominates over dipole spin-down while mass transfer is ongoing, $\tSD$
for the resulting QCD magnetar is counted from the end of accretion,
not from NS birth.
A NS born near the peak of the mass distribution
($M_0\sim 1.5\,\Msun$) must accrete $\sim 0.5\,\Msun$ to
reach $\Mdec$.
Whether stable mass transfer from a hydrogen-rich or helium
companion can deliver this amount without triggering common-envelope
evolution or NS collapse is uncertain
\citep[e.g.,][and references therein]{Tauris2017}. Channel~B may therefore
preferentially involve NSs born already massive,
for which the required accretion is modest.
There are some evidence that some NSs in binaries are born massive
for which the required accreted mass would be reasonable \citep{Bhattacharya1991,Tauris2012,Antoniadis2013}.
Those could constrain the rate of QCD magnetars in Channel~B.

If the massive NS is spun up during accretion, the resulting QCD magnetar at $\tSD$ (from the end of accretion) would give birth
to an isolated LFBOT. In this case, the Channel~B LFBOT maximum rate is 
\begin{equation}
  \mathcal{R}_{\rm LFBOT}^{\rm B}
  \sim \mathcal{R}_{\rm CCSN}
  \times f_{\rm bound}
  \times f_{M_0,{\rm reach}}
  \times f_{\rm MT}\,,
  \label{eq:rlfbot_B}
\end{equation}
where $f_{\rm bound}\sim 0.05$ is the fraction of core-collapse NSs
retaining a bound companion immediately post-SN, consistent
across independent population-synthesis codes
\citep{Chrimes2022}; $f_{M_0,{\rm reach}}\approx0.21$ is the fraction
of NSs born within $0.5\,\Msun$ of $\Mdec$ (using the birth $M_0$
distribution of Eq.~7), i.e.\ close enough that plausible accretion
can close the gap; and $f_{\rm MT}$
is the fraction of those binaries achieving sufficient mass transfer
to drive the NS above $\Mdec$.

As a consistency check, we compare to the volumetric rate of
radio-confirmed LFBOTs, which selects specifically for an embedded
compact-object central engine and is therefore the most directly
comparable observational quantity to our model 
\citep{Sharma2025},
$\mathcal{R}_{\rm LFBOT}/\mathcal{R}_{\rm CCSN}\lesssim2\times10^{-4}$.
Requiring $\mathcal{R}_{\rm LFBOT}^{\rm B}$ not exceed this observed
rate gives
\begin{equation}
  f_{\rm MT} < \sim 10^{-2}\,; 
\end{equation}
i.e., Channel~B is viable only if well under 1\% of mass-reachable,
bound, post-SN NS binaries achieve sufficient mass transfer to reach
$\Mdec$. This is an order-of-magnitude plausibility bound, not a tight
observational constraint; a dedicated binary population synthesis is
required to predict $f_{\rm MT}$ directly.

The key distinction from Channel~A is environmental.
Without a confining SN ejecta shell, all QN outputs are
directly visible: the LFBOT is unobscured, the r-process
kilonova is uncontaminated, and the gravitational-wave burst
is unattenuated.
Channel~B events can occur wherever important binary mass transfer is
possible.

% -----------------------------------------------------------------------
\subsection{Chemical enrichment signatures}
\label{sec:chem}
% -----------------------------------------------------------------------

 We reiterate that a single NS (as it converts to a HS) can contribute to the r-process yield in the Universe. More generally, as discussed below, QN ejecta as an r-process site can complement the yields from NS mergers and help better account for the observed chemical abundances in the Universe.

\paragraph{The Galactic r-process budget.}
The QN r-process yield per event (up to $\sim0.01\,\Msun$) exceeds the CCSN
contribution ($\sim10^{-6}\,\Msun$ \citealt[e.g.,][for a recent review]{Cowan2021}) by roughly four orders of magnitude.
The Galactic r-process mass injection rate from QN events is
\begin{equation}
  \dot{M}_{\rm r}^{\rm QN}
  \sim Y_{\rm r}^{\rm QN}\,\mathcal{R}_{\rm AXP}\,,
  \label{eq:Mdot_r}
\end{equation}
where $Y_{\rm r}^{\rm QN}$
is the r-process yield per event.
With $\mathcal{R}_{\rm CCSN}\sim1$--$3$~century$^{-1}$ and
$\mathcal{R}_{\rm AXP}/\mathcal{R}_{\rm CCSN}\sim0.1$, the maximum injection rate is
\begin{equation}
  \dot{M}_{\rm r}^{\rm QN}
  \sim 10^{-5}
  \,\Msun\,{\rm yr}^{-1}\,.
  \label{eq:Mdot_r_num}
\end{equation}
The above is sufficient to account for the total r-process production rate required
by Galactic chemical evolution models,
$\sim10^{-7}$--$10^{-6}\,\Msun\,{\rm yr}^{-1}$
\citep{Sneden2008,Cowan2021}.

Because massive NSs form preferentially from massive progenitors,
QN events are expected to be more frequent when massive
(Population~III) stars dominate the stellar mass function.
QN r-process enrichment therefore begins in the early Universe and may account for
early enrichment before the NS-NS mergers contribution 
\citep{Ouyed2009}.

\paragraph{R-process signatures in AXP\&SGR-associated SNRs.}
r-process nucleosynthesis in the QN ejecta produces elements such as Eu, Ba, La, Ce, PU and other 
potentially long-lived actinides; these may provide the most accessible tracers from QNe (\citealt{Jaikumar2007,Kostka2014a,Kostka2014b}).
Several confirmed SNR-associated AXPs\&SGRs may already retain
detectable enrichment signatures.

\paragraph{R-process pollution of binary companions.}
In Channel~B, r-process material from the QN ejecta may
pollute the atmosphere of the binary companion with anomalous (i.e., non-SN) 
r-process abundances and elements.
A binary system showing r-process enhancement in the companion
alongside an identified LFBOT counterpart would
support Channel~B directly and is a clean prediction of the
model.

% ════════════════════════════════════════════════════════════════════════
\subsection{Implications for the Hadron–Quark Transition and Ferromagnetic Quark Matter}
\label{sec:implications}
% ════════════════════════════════════════════════════════════════════════
\subsubsection{The hadron-to-quark transition}
\label{sec:hadron_quark}
A massive NS born with $M_0>\Mdec$ has $\rhoc>\rhodec$ at birth,
yet our population and Bayesian analysis supports the notion that full conversion of the core 
 does not occur until $t\sim \tSD$. It suggest that  nucleation is contained 
 and that the pressure difference  at the hadronic-quark interface of quark bubbles throughout
Phase~1 is $\Delta P\approx0$. As the HS spins down, centrifugal support decreases and the
central pressure rises. At $\tSD$ the pressure differential at the interface
crosses a threshold set by the competition between the surface tension $\sigma$ and
the available free energy, causing the quark bubbles to percolate
and the entire core to convert on the dynamical timescale of milliseconds (Appendix \ref{sec:Bqcd_origin}).
The conversion is therefore not driven by a density threshold being crossed --- the central density already exceeded $\rhodec$ at birth --- but by a \textit{pressure threshold} reached during spin-down. This inference places a non-trivial constraint on the hadron--quark transition itself: deconfinement at $\rhoc>\rhodec$ cannot immediately trigger global conversion. Instead, the transition must permit a long-lived metastable state in which quark matter remains confined to isolated regions for timescales comparable to the stellar spin-down time (i.e. as long as $t<\tSD$). Models that predict prompt conversion once $\rhoc>\rhodec$ is exceeded are therefore incompatible with the delayed-transition picture favored by our Bayesian analysis.

\subsubsection{Constraining the Ferromagnetic Quark Phase}
\label{sec:ferromagnet}

The Bayesian posterior $\Mdec\sim 2\,\Msun$ constrains
the deconfinement density through the NS mass--central
density relation. For EOS models consistent with the
observed $2\,\Msun$ mass limit \citep{Demorest2010,
Antoniadis2013}, the central density 
spans $\rhodec\sim (3\text{--}5)\,\rhoNuc$
\citep[e.g.][and references therein]{Lattimer2001}, where the range reflects
EOS uncertainty rather than measurement precision.

This is directly relevant to the ferromagnetic quark
phase. QCD ferromagnetism theory predicts a field
strength $B_{\rm QCD}$ that scales with the quark matter
density at deconfinement (see Appendix \ref{sec:Bqcd_origin}).
The inferred density range $\rhodec\sim(3\text{--}5)\,\rhoNuc$
therefore selects a specific density window within which
$B_{\rm QCD}$ must be evaluated, providing an
astrophysical constraint on the ferromagnetic quark
matter phase that is independent of laboratory
experiments. I.e., our posterior on
$\Mdec$ can in principle rule out models that predict
either no ferromagnetic instability or too weak a field
at densities $\sim(3\text{--}5)\,\rhoNuc$ to account for
the observed AXP\&SGR, SLSN-I, and LFBOT rates
simultaneously.

% ════════════════════════════════════════════════════════════════════════
\section{Predictions, Falsifiable Tests  and Limitations}
\label{sec:predictions}
% ════════════════════════════════════════════════════════════════════════

We emphasize again that every QCD magnetar may be born spatially and temporally associated with a kilonova. If the progenitor is rapidly rotating ($P_0<\Pfast$), the QCD magnetar also injects spin-down power into the QN ejecta, producing an accompanying LFBOT in addition to the kilonova. Such events are directly observable when the QN occurs in isolation, either in Channel~A with $\tSD > t_{\rm SNR}$ or in Channel~B after the NS reaches $\Mdec$ through binary accretion. 

The derived birth distributions (see \S \ref{sec:population}) imply $\fdec\sim 0.12, \ffast\sim 1.3\times 10^{-2}, \ffoss \sim 4.7\times 10^{-3}$. 
With these numbers in mind, we start by listing a set of predictions. In no particular order:

\begin{enumerate}

\item \textit{QCD magnetars occupy the high-mass end of the NS mass distribution:}\\
AXPs\&SGRs are identified as HSs formed through the conversion of NSs born above the critical mass threshold ($M_0 > M_{\rm dec}$). 
NSs  born below this threshold cannot undergo a QN transition unless they subsequently increase their mass through accretion. This prediction differs from classical magnetar formation scenarios, in which the generation of ultra-strong magnetic fields is not expected to depend on NS mass. Although observationally challenging, mass measurements of AXPs\&SGRs should reveal that the dominant magnetar population is systematically biased toward high masses. 

\item \textit{A sub-population of AXPs\&SGRs whose natal
SNRs carry SLSN-level kinetic energies:}\\
Fast rotators with $M_0>\Mdec$ and $P_0<\Pfast$, whose QN occurs
while the SN ejecta remain optically thick ($\tSD<t_{\rm SN,o}$),
are classified as SLSNe-I.
After their rotational energy is depleted,
the remnant HS becomes a slowly rotating QCD magnetar and will eventually contribute 
to the AXP\&SGR population.
Its natal SNR, however, should retain the imprint of the spin-down energy injected
during the SLSN phase, making it more energetic than a typical
AXP\&SGR SNR.
The expected fraction of such objects within the overall
AXP\&SGR population is
$\mathcal{R}_{\rm SLSN}/\mathcal{R}_{\rm AXP} < \ffast\sim 1.3\times 10^{-2}$.

\item \textit{Fossil AXPs\&SGRs:}\\
A small minority of the observed AXP\&SGR population may consist of fossil magnetars:
NSs born with $M_{\rm NS}<\Mdec$ and $B_0>\Bc$, whose magnetic fields
are acquired through flux conservation alone, without undergoing QN conversion.
Their expected rate is
$\mathcal{R}_{\rm fAXP}/\mathcal{R}_{\rm AXP}
=(1-\fdec)\times\ffoss/\fdec < \sim 3.9\times 10^{-2}$.

Their defining observational properties, which distinguish them from QCD
magnetars, are:
(i)~the absence of any associated kilonova or r-process enhancement at any epoch;
(ii)~a surface magnetic field that does \textit{not} correlate with mass;
and (iii)~a characteristic age equal to the true SNR age without the systematic overestimate
introduced by the two-phase spin-down evolution of QCD magnetars.

\item \textit{The first hump of double-humped SLSNe-I
traces the fossil field distribution:}\\
In Channel~3b ($M_{\rm NS}>\Mdec$, $P_{\rm NS}\sim P_{\rm 0, min}$,
$B_{\rm NS}>\Bc$), the first optical peak is powered by spin-down
of the rapidly rotating fossil magnetar before QN conversion.
The distribution of first-hump peak luminosities across a
double-humped SLSN-I sample should therefore be indicative of the high-$B$
tail of the birth fossil field distribution, convolved with the
fast-rotator sub-population.
The expected rate is $\mathcal{R}_{\rm SLSN-I}^{\rm DH}/\mathcal{R}_{\rm SLSN-I}
\simeq \ffoss \sim 4.7\times 10^{-3}$.

\item \textit{QCD magnetars and their kilonovae in atypical host environments:}\\
Massive NSs with weak birth fields ($B_0\sim10^{12}$~G) and
slow birth spins ($P_0\sim 0.2$~s) have $\tSD$ of the order of millions of years.
Assuming a natal kick velocity of $\sim 300$ km s$^{-1}$ they can travel up
to a few kilo-parsecs from their birth place by the time they convert.
They may reside in globular clusters,
the outskirts of elliptical galaxies, or other quiescent
environments.

A QCD magnetar together with its associated kilonova appearing in
such an environment, with no associated CCSN and no NS merger
history, would be a strong indicator of this late-converting
channel.
The classical magnetar model requires a young stellar population
unless novel mechanism for magnetar formation in quiescent environments
can be confirmed; e.g., the accretion-induced collapse of merged white dwarfs model (e.g. \citealt{King2001,Levan2006}).

\item \textit{Systematic characteristic age anomaly in
SNR-associated magnetars:}\\
In the two-phase spin-down picture, the characteristic age
$\tau_c = P_{\rm HS}/(2\dot{P}_{\rm HS})$, from the HS phase, is computed entirely from the
post-conversion phase under $\BHS\sim10^{15}$~G.
It therefore measures only the time since QN conversion, not
the total stellar age, which also includes the Phase~1 spin-down
under the birth field $B_0$.

Every SNR-associated magnetar in the QN framework should therefore
show $\tau_c > t_{\rm SNR}$: the characteristic age should
\textit{systematically exceed} the independent SNR kinematic age.

\item \textit{The QN--SN ejecta collision signature:}\\
In Channel~A, the faster QN ejecta may catch up with and collide with the slower SN ejecta.
In SLSNe, this interaction produces  late-time bumps in the post-peak
light curve, with a total radiated energy not exceeding
$E_{\rm QN}^{\rm KE}\sim 10^{50}$ erg \citep{Ouyed2025,Ouyed2026b}.
In isolated QNe, the QN ejecta plow into the surrounding medium, generating radio and X-ray
emission that would be associated with an LFBOT if the NS was born with $P_0 < \Pfast$ \citep{Ouyed2026a}.

\item \textit{SLSNe-I may harbour hidden FRB repeaters:}\\
Every SLSN-I leaves behind a crustless QCD magnetar that enters an FRB-repeating phase lasting for centuries. Although the compact remnant becomes an active FRB source immediately following the QN, the surrounding SN ejecta is initially opaque to GHz radio emission because of free--free absorption. The ejecta becomes transparent after $t_{\rm SN,r}\sim$ a few decades (Eq.~\ref{eq:tSNr}), allowing the FRB activity to become observable at
$t \gtrsim \max(t_{\rm SN,r},\,\tSD)$. From that point onward, the source is expected to exhibit the characteristic temporal evolution of the local dispersion measure, DM$_{\rm QN}$, and rotation measure, RM$_{\rm QN}$ (see \S\ref{sec:frb_dmrm}).

Several well-studied SLSNe discovered in the mid-2000s may therefore begin to reveal FRB activity in the coming years, including SN\,2005ap \citep{Quimby2007a,Quimby2007b} ($t\approx21$~yr), SN\,2006gy \citep{Smith2007,Ofek2007} ($t\approx20$~yr), and SN\,2007bi \citep{Gal-Yam2009} ($t\approx19$~yr). Targeted GHz monitoring of these remnants over the next several years would provide a direct test of this prediction.

\item \textit{Partially crystallized QCD magnetars: simultaneous FRB and X-ray activity:}\\
The transition from an FRB repeater to an AXP\&SGR at $t_{\rm crust}$ need not be abrupt. As a crust re-forms through crystallization of the outer hadronic layers or is rebuilt through accretion, an intermediate phase may exist in which only a fraction of the stellar surface is crusted while the remainder remains fluid. In the crusted regions, ascending magnetic flux ropes dissipate their energy beneath the surface, producing X-ray bursts, whereas in the fluid regions they can emerge coherently and power FRBs. As the crust progressively covers the surface, the area available for coherent flux-rope emergence decreases, leading to a corresponding reduction in the FRB activity until the source eventually exhibits only AXP\&SGR-like behavior.

This scenario may explain SGR1935+2154, which produced a radio burst several orders of magnitude weaker than typical extragalactic FRBs simultaneously with X-ray bursts \citep{Chime2020,Bochenek2020}. Such behaviour is naturally expected from a partially crystallized QCD magnetar, where both emission channels operate simultaneously.

\end{enumerate}

% -----------------------------------------------------------------------
\subsection{Falsifications}
\label{sec:falsifications}
% -----------------------------------------------------------------------

\begin{enumerate}

\item \textit{Millisecond birth period confirmed in
an AXP\&SGR:}\\
If a confirmed AXP or SGR is found with a millisecond birth
period --- inferred from a young SNR whose kinematic age matches
a spin-down age $\tau_c$ consistent with $P_0\lesssim\Pfast$ ---
the decoupling of field origin from spin is broken for that
object. This falsification requires such objects to constitute a percentage exceeding 
$\ffast$ of the full AXP\&SGR population, since a single
object could in principle be a fossil magnetar.

\item \textit{No r-process excess in any LFBOT:}\\
If deep late-time NIR photometry of a statistical sample of
LFBOTs shows no excess above the optical power-law decline in
any event, the lanthanide-rich QN ejecta hypothesis is ruled out;
but not necessarily the QN as an r-process site.
A non-detection in a single event is insufficient; the prediction
is statistical.

\item \textit{A confirmed hadronic-EOS NS above $\Mdec$:}\\
The posterior $\Mdec \sim 2.1\Msun$ requires that no NS with a confirmed hadronic EOS exists above $\Mdec$. A robust mass measurement $M > \Mdec$ for an object whose radius and tidal deformability are consistent with purely hadronic matter would place tension on the model that cannot be relieved by adjusting $\sigM$ alone, since $M_{\rm max}^{\rm hadronic} > \Mdec$ would force $\Mdec$ to shift beyond its $2\sigma$ posterior range. 

In principle, this threshold is directly testable using the handful of pulsars with robustly measured masses near or above $2\,M_\odot$, most notably PSR J0740+6620 ($2.08 \pm 0.07\,M_\odot$; \citealt{Fonseca2021}). However, current data cannot cleanly execute this test due to a fundamental observational selection bias: precise mass measurements via pulsar timing (e.g., Shapiro delay) require binary systems, which systematically selects for fast, recycled spin states. 

In the case of PSR J0740+6620 ($P \sim 2.9$\,ms), accretion-driven spin-up has restored significant centrifugal support. Although its white-dwarf companion implies a system age of $\gtrsim 5$\,Gyr (e.g., \citealt{Beryonya2019}), this recycling history decouples the chronological age from the rotational state. Just as recycling invalidates the standard spin-down age proxy ($\propto P/2\dot P$), it prevents this system from probing the slow-spin, centrifugally unsupported regime relevant to our deconfinement criterion. This limitation reflects current data constraints rather than an intrinsic model loophole. It remains achievable through future independent mass-measurement channels targeting isolated, non-recycled systems, such as next-generation X-ray pulse-profile modeling or gravitational-wave analysis.

\item \textit{A confirmed $\tau_c < t_{\rm SNR}$ in
a magnetar with $M>M_{\rm dec}$:}\\
Every SNR-associated QCD magnetar should show
$\tau_c > t_{\rm SNR}$.
A single well-measured case with $\tau_c < t_{\rm SNR}$ 
from an object whose mass is independently estimated to exceed
$\Mdec$  would falsify the two-phase spin-down picture and
the QCD magnetar origin simultaneously.

\end{enumerate}

% -----------------------------------------------------------------------
\subsection{Limitations}
\label{sec:limitations}
% -----------------------------------------------------------------------

The framework presented here has several limitations that
should be borne in mind when interpreting the results.

\begin{enumerate}

\item \textit{The identification $\tQN=\tSD$:}\\
As discussed in Appendix~\ref{sec:Bqcd_origin}, the conversion
occurs when the central pressure differential $\Delta P$, between the hadronic and quark phases, reaches
a critical value induced at $t_{\rm QN}\sim \tSD$
but not identically equal to it.
The approximation is accurate at the population level because
of the extreme exponential sensitivity of the nucleation rate
to $\Delta P$, but introduces an
order-unity uncertainty in $\tQN$ for any individual star.
This does not affect the rate predictions, which depend only
on the statistical distribution of $\tSD$.  We plan to explore this further
where we free $\tQN$ from $\tSD$.

\item \textit{The constant $B_{\rm HS}=10^{15}$ G:}\\
Because QCD magnetars are expected to attain essentially identical thermodynamic conditions in their cores 
(i.e., the same quark-matter phase), they should acquire a nearly universal QCD-generated magnetic field.
However, variations in transport processes and in NS structure (converting core size, envelope
density profile, geometry, and the exact crust mass ejected)
introduce scatter in the resulting surface field $B_{\rm HS}$. For example, adopting a dipole field configuration, $B_{\rm HS}\sim B_{\rm core}(R_{\rm c}/R_{\rm NS})^3$, the expected variation in $R_{\rm c}/R_{\rm NS}$ would naturally produce a distribution of surface magnetic fields (e.g., log-normal) peaking at $\sim10^{15}$~G for an average value of $R_{\rm c}/R_{\rm NS}\sim0.1$.

\item \textit{The Channel~B rate is unconstrained:}\\
Equation~(\ref{eq:rlfbot_B}) contains  factors that are not yet
computed from first principles for the specific mass range
required by the model.
The Channel~B contribution to the observed LFBOT rate is
therefore uncertain, and the population analysis of
Section~\ref{sec:population} attributes all LFBOTs to
Channel~A.
If Channel~B contributes significantly (see \S \ref{sec:chanB}), the inferred $\Pfast$
would shift downward to compensate.

\item \textit{The birth distributions with fixed peaks:}\\
The peak values of the mass, spin, and field distributions are
fixed to observed radio pulsar values, which reflect the
\textit{current} population rather than the birth population.
Secular evolution (spin-down, field decay, and observational
selection) may shift the peaks relative to their birth values.
We have verified that moderate changes ($\pm0.1$~dex) in the
peak values do not affect the posteriors on $\Mdec$ or $\Pfast$
at the $1\sigma$ level, but this assumption deserves more
systematic scrutiny in future work; a more rigorous treatment
would evolve the inferred distributions forward in time and
compare them to the current observed population.

\item \textit{The DH-SLSNe-I rate:}\\
We adopt $\mathcal{R}^{\rm DH}_{\rm SLSN-I}/\mathcal{R}_{\rm SLSN-I}< 0.10$,
based on the fraction of double-humped events relative to the
full (not early-cadence-selected) SLSN-I samples of
\citet{Angus2019}  and \citet{Chen2023}, rather than
the higher fractions (20-33\%) reported among the subset of
objects with adequate early photometric coverage. This 
value is likely to be revised as
higher-cadence, wider-field surveys improve
early-time completeness.

\item \textit{The fossil-field threshold $\Bc$:}\\
The fossil-field threshold, $\Bc$, enters the double-humped SLSN-I fraction through $\ffoss$ (Equations~\ref{eq:fhigh} and~\ref{eq:rslsn2}). It defines the fraction of NSs born with fossil magnetic fields exceeding $\Bc$ and therefore directly affects the predicted double-humped SLSN rate.  Repeating the population synthesis and Bayesian inference while varying $\Bc$ over the range $10^{13.5}$--$10^{14.5}$~G changes the posterior estimates of $\Mdec$ and $\Pfast$ by less than $10\%$. The inferred parameters are therefore relatively insensitive to the precise choice of $\Bc$ within this range. A more rigorous treatment would include $\Bc$ as a free parameter, although doing so would require an additional observational constraint (e.g. beaming corrected GRB rate) to break the resulting degeneracies. We defer such an analysis to future work.

\item \textit{Tayler and Parker instabilities:}\\
A caveat of the present model concerns the mechanism by which magnetic flux is transported from the quark core into the hadronic envelope. For the field strengths considered here ($B_{\rm core}=B_{\rm QCD}\sim10^{18}$ G and $B_{\rm env}\sim10^{15}$ G), a large amount of magnetic free energy is stored at the core--envelope interface. However, the dominant instability responsible for its release remains uncertain. In stratified media, Parker (magnetic buoyancy) instability can develop when magnetic pressure opposes gravity, causing magnetic field lines with a significant horizontal component to rise buoyantly (\citealt{Parker1966,Acheson1978}). In our case, the strong toroidal component of the core field may instead favour the Tayler instability (\citealt{Tayler1973}), while Parker-like buoyancy and magnetic interchange may contribute to the subsequent outward transport of magnetic flux. 

Since the magnetic pressure associated with $B_{\rm QCD}\sim10^{18}$ G is comparable to or exceeds the matter pressure (i.e. a plasma beta $\beta\lesssim1$), the classical Parker picture is unlikely to apply in isolation, and the system may instead undergo a global magnetic reconfiguration on approximately the Alfvén timescale
in the envelope, $t_{\rm env, A}\sim R_{\rm NS}/v_{\rm env, A}$. Determining which instability dominates, and the fraction of magnetic energy ultimately released, requires multidimensional MHD simulations and is beyond the scope of the present work.

\item \textit{The QCD field generation and theoretical uncertainties:}\\
The ferromagnetic ground state of two-flavour quark matter is
established by several independent calculations
(Appendix~\ref{sec:Bqcd_origin}), but identifying the proper quark matter phase and the precise value of
$B_{\rm QCD}$ remains a challenge.
Astrophysical constraints of the kind derived here may provide
the most direct observational handle on these unknowns.

\item \textit{The FRB activity:}\\
The long-term evolution of the residual magnetic substructure at the hadron--quark interface, which powers the FRB activity in our model, remains to be quantified through multidimensional magnetohydrodynamic simulations. Although the global magnetic field relaxes to its new equilibrium on a timescale of seconds, the deconfinement transition is expected to leave behind localized non-potential magnetic structures that retain a reservoir of free magnetic energy. The gradual relaxation of these structures through discrete buoyant flux-rope eruptions provides a physically plausible mechanism for sustaining FRB activity (as the HS spins-down) over much longer timescales. Quantifying how this magnetic energy is stored, transported, and released, as well as the resulting burst statistics and longevity of the FRB-active phase, requires dedicated numerical simulations and lies beyond the scope of the present work.

\end{enumerate}

\subsection{The $M_{\rm QN}$ distribution scenario}
\label{sec:MQN-distribution}

Throughout this work we have adopted a single fiducial value,
$M_{\rm QN}\sim10^{-2}M_\odot$, sufficient to eject the entire NS
crust (\citealt{ChamelHaensel2008}) and produce a crustless QCD
magnetar at birth. In reality, $M_{\rm QN}$ is more plausibly drawn
from a distribution peaking near this value, with individual QN events
ejecting somewhat more or less mass depending on the details of the
conversion. Since the crust mass sets a fixed threshold, such a
distribution naturally implies that not every QN event fully strips
the crust: events with $M_{\rm QN}\gtrsim10^{-2}M_\odot$ produce a
crustless HS as treated in the main text, while events with
$M_{\rm QN}<10^{-2}M_\odot$ would leave some residual crust in
place, plausibly yielding a QCD magnetar that is AXP\&SGR-like from
birth rather than passing through an initial crustless, FRB-active
phase.

Under this extension, the deconfinement rate $f_{\rm dec}\times
\mathcal{R}_{\rm CCSN}$ used throughout this work is shared between the
two channels rather than assigned entirely to the crustless case.
The AXP\&SGR formation rate would then receive contributions from
both channels: objects born with a crust directly, and objects born
crustless that later acquire a crust at $t^\prime_{\rm crust}$
(\S\ref{sec:frb_dmrm}). The present
work corresponds to the limit  in which essentially every deconfining star
is born crustless and the entire AXP\&SGR population is supplied by the
delayed-crust channel. This simplification is a genuine assumption of
the proof-of-principle model presented here, not a derived result: the
width of the $M_{\rm QN}$ distribution is currently
unconstrained by any independent measurement of QN ejecta-mass
scatter, and we do not attempt to fit it in this work. A narrow
distribution  would already imply a non-negligible born-crusted population, providing an
alternative or complementary origin for AXPs\&SGRs alongside the
delayed-crust-formation channel developed here, and a natural target
for future refinement of the model once $M_{\rm QN}$ can be
independently constrained.

\subsubsection{GRBs as the tail of the $M_{\rm QN}$ distribution}
\label{sec:grb_plateau}

Magnetar-powered X-ray plateaus are a standard afterglow signature of
spin-down energy injection \citep{Metzger2011}. In the CCSN channel,
where spin-down occurs while the bulk SN ejecta is still present, the
thermal energy deposited in the hadronic envelope following the
hadron-to-quark conversion in the core is identical across events
($E_{\rm T}\sim10^{51}$~erg; Eq.~\ref{eq:ET}, Appendix~\ref{app:ejecta}):
only $M_{\rm QN}$ differs, setting its velocity via
$v_{\rm QN}\sim\sqrt{2E_{\rm T}/M_{\rm QN}}$.

For sub-relativistic ejecta near the peak of the $M_{\rm QN}$
distribution ($M_{\rm QN}\sim10^{-2}M_\odot$), the spin-down power of
the rapidly rotating QCD magnetar is efficiently thermalized within the
optically thick QN ejecta, powering an LFBOT. In contrast, for
relativistic ejecta ($M_{\rm QN}<10^{-4}M_\odot$, i.e. an ejecta's Lorentz factor 
$\Gamma_{\rm QN}>100$) -- two decades into the low-mass tail of the
same distribution -- the weaker coupling between spin-down power and
ejecta, together with the rapid transition to optical thinness,
prevents efficient thermalization; no LFBOT is produced. Instead, the
relativistic QN ejecta collides with the still-expanding SN ejecta to
produce a prompt GRB \citep{Ouyed2020}, while the unthermalized
spin-down power escapes through the transparent ejecta and is observed
as an X-ray plateau \citep{Staff2007}.

\smallskip\noindent\textit{The relativistic tail.} A QN event occurs
at a rate $\sim f_{\rm dec}\times\mathcal{R}_{\rm CCSN}$, of which a
fraction $f_{\rm fast}$ are born spinning fast enough to power a
luminous ($10^{44}$--$10^{45}\,{\rm erg\,s^{-1}}$) X-ray plateau. We
denote the fraction of these falling in the relativistic
($M_{\rm QN}<10^{-4}M_\odot$) tail $f_{\rm rel}$; since $f_{\rm rel}$
is set by the (currently unconstrained) width of the
$M_{\rm QN}$ distribution, it remains a free parameter. The GRB rate
follows as $\mathcal{R}_{\rm GRB}\sim f_{\rm dec}f_{\rm fast}f_{\rm
rel}\,\mathcal{R}_{\rm CCSN}$, estimable only once $f_{\rm rel}$ is
independently constrained.

\smallskip\noindent\textit{Orphan X-ray plateaus (GRB-less).} The GRB
requires a collision target: SN ejecta that has not yet dissipated.
This fails in the isolated/binary channel, where the QN forms long
after the original core-collapse SN ejecta has dissipated, so no
collision site exists and no prompt GRB is produced, regardless of
$v_{\rm QN}$. If the ejecta in this channel is relativistic, spin-down
power still escapes directly through it by the same physics as above,
but unaccompanied by a GRB -- an orphan X-ray plateau: an X-ray
transient lacking both a gamma-ray trigger and an optical/thermal
counterpart.

% ════════════════════════════════════════════════════════════════════════
\section{Conclusions}
\label{sec:conclusions}
% ════════════════════════════════════════════════════════════════════════

We have presented the QN framework, in which a NS born above the critical mass $M_{\rm dec}\sim2.1\,M_\odot$
eventually undergoes quark deconfinement in its core. Spontaneous
quark-matter ferromagnetism in the $(u,d)$ quark core generates a
$\sim10^{18}\,$G magnetic field, whose strength depends on the core
baryon density rather than the NS's birth spin -- translating to a HS
surface field of $\sim10^{15}\,$G, reminiscent of magnetar values.

Conversion of the NS to the highly magnetized HS (the QCD
magnetar) is triggered at $\tSD$ by the pressure change induced at the
hadron-quark surface of nucleated quark bubbles, converting the NS core
on millisecond timescales: the QN event.  The QN ejects $M_{\rm QN}\sim10^{-2}M_\odot$ of the NS outer layers at
$v_{\rm QN}\sim0.1c$ -- comparable to the total NS crust mass -- so
that the resulting HS is  born crustless and acts as an intrinsic FRB repeater: magnetic
flux ropes generated at the quark-hadron interface rise buoyantly
through the hadronic envelope and dissipate at the bare surface as
FRBs. This phase persists for $t^\prime_{\rm
crust}\sim$ centuries, after which the star acquires a crust and
evolves observationally into an AXP\&SGR; crusted QCD magnetars,
whether born that way or having transitioned after the crust re-forms,
dissipate the same interface energy via Hall drift, powering AXP\&SGR
X-ray and soft gamma-ray emission.

The wide range of FRB host environments follows naturally from the
distribution of the parent NS spin-down timescale $\tSD$ (ranging from days to millions of years),
permitting QCD magnetar formation in both young star-forming regions
and old stellar populations, including globular clusters and
elliptical galaxies. QCD magnetars forming via the binary channel,
where the NS gains mass from a companion to reach $\Mdec$, provide a
further route to unusual host environments in our model.

If the progenitor NS is born rapidly rotating ($P_0<\Pfast\simeq5.5$
ms), the HS inherits the fast spin ($P_{\rm HS}\sim2^{1/2}P_{\rm 0}$),
and its spin-down energy is injected into the QN ejecta, producing an
LFBOT. This is directly observable only if the spin-down timescale
exceeds the SN ejecta transparency time, $\tSD>>t_{\rm SN,o}$;
otherwise, for $\tSD<t_{\rm SN,o}$, the same engine is observed as an
SLSN-I since the emerging QN LFBOT emission is reprocessed by the still-optically-thick
SN ejecta. 

The QN's neutron-rich ejecta offers a potential additional site for
r-process nucleosynthesis, complementing the role of NS mergers and
potentially contributing to the kilonova population and to heavy-element
enrichment. Because this production channel originates from isolated converting massive 
NSs rather than binaries, it is linked directly to ongoing star
formation, and could have been particularly important during the
Population~III era, prior to significant contributions from NS
mergers.

Key predictions include: (i) a correlation of AXP\&SGR surface fields
with mass rather than spin history; (ii) an r-process kilonova signature
 in all well-observed LFBOTs arising from isolated QCD magnetars,
independent of NS mergers; (iii) association of FRB repetition with
polar beaming (i.e., the QCD magnetar's spin period), with the most
active repeaters belonging to fast rotators (wider open polar caps);
(iv) As FRB sources,  crustless QCD magnetars should show no associated persistent or bursting X-ray
counterpart, in contrast to conventional crustal-fracture magnetar
models, which generically predict X-ray/radio coincidence; (v) For epochs when the medium surrounding the QN becomes transparent to GHz radio emission in Channel A, or for QNe occurring in Channel B, the QN ejecta contribution should manifest as excess DM and a large RM contribution in some FRBs.

The spin-down trigger constrains quark-nucleation physics itself. In
massive NSs with $\rho_c>\rho_{\rm dec}$, any initial quark seeds are
confined and contained until spin-down drives the NS core beyond a
critical pressure threshold, triggering global conversion; our model
supports pressure-driven percolation as the dominant deconfinement
mechanism near the phase boundary. The derived $M_{\rm dec}\sim 2.1M_\odot$
constrains the core density at conversion, potentially offering
astrophysical access to the equation of state at a few times nuclear
saturation density. Observationally, the surface field $\BHS$, inferred
from $P_{\rm HS}$ and $\dot{P}_{\rm HS}$ under a dipolar assumption,
provides an indirect probe of the core field and thus of $B_{\rm QCD}$; in concert
with $\rho_c$, this may help discriminate between competing models of
magnetically ordered quark phases.

In summary, every sufficiently massive NS is a QCD magnetar in waiting:
its magnetic field is encoded in the ground state of dense quark
matter, revealed once spin-down triggers the QN transition. What
distinguishes an AXP\&SGR, an SLSN-I, an LFBOT, a GRB, and an FRB
 is not five different physical mechanisms but two numbers
acting on one engine: the QN ejecta mass relative to the NS crust mass,
which decides whether the star is born crustless or crusted, and the
spin period relative to $P_{\rm fast}\sim 5.5$ ms, which decides how much
rotational energy the resulting QCD magnetar must dispose of -- into
whatever environment, thick or thin ejecta, present or absent, it
happens to find itself in. The observational diversity is therefore a
diversity of disguise, not of origin: each channel is the same QCD
magnetar engine, viewed during a different stage of its life cycle, in
a different amount of surrounding material.

%\begin{acknowledgments}
% 
%\end{acknowledgments}

%

% ── Appendix ─────────────────────────────────────────────────────────

\appendix

\section{Quark matter nucleation and origin of the QCD field scale}
\label{sec:Bqcd_origin}

%------------------------------------------------------------
\subsection{Quark nucleation physics}
%------------------------------------------------------------

The nucleation of quark matter  proceeds through two channels:
thermal fluctuations and quantum tunnelling \citep{Horvath1992,Olesen1993,Heiselberg1993,Iida1997,
Harko2004,Bombaci2004}.
The thermal nucleation rate per unit volume (in natural units $\hbar=c=1$
throughout this appendix) is
\begin{equation}
  \Gamma_T \sim T^4\,
  \exp\!\left[
    -\frac{16\pi\sigma^3}{3\,(\Delta P)^2\,T}
  \right],
  \label{eq:nuc_rate_T}
\end{equation}
and the quantum tunnelling rate is \citep{Olesen1993}
\begin{equation}
  \Gamma_Q \sim \mu_q^4\,
  \exp\!\left[
    -\frac{16\pi\sigma^3}{3\,(\Delta P)^2\,\mu_q}
  \right],
  \label{eq:nuc_rate}
\end{equation}
where $\mu_q$ is the quark chemical potential, $\sigma$ is the
surface tension of the hadronic-quark interface, and $\Delta P$
is the pressure difference between deconfined quark matter and
hadronic matter at the interface.
For cold, deleptonised NS cores with $T\lesssim1$~MeV and
$\mu_q\sim300$~MeV, both the prefactor ($T^4\ll\mu_q^4$) and
the exponent (larger by a factor $\mu_q/T\sim300$) suppress
$\Gamma_T$ relative to $\Gamma_Q$ by many orders of magnitude.
Nucleation in the pre-QN phase  is therefore dominated by quantum
tunnelling.

Converting the hadronic core to quark matter requires the
formation of at least one quark-matter droplet of critical size.
The nucleation timescale --- the time required for one critical
bubble to form in the core volume
$V_c=\frac{4}{3}\pi R_c^3$ with $R_c$ the radius --- is
\begin{equation}
  \tau_{\rm nuc} \sim \frac{1}{\Gamma_Q\,V_c}.
  \label{eq:tau_nuc}
\end{equation}

The nucleation rate is extraordinarily sensitive to $\sigma$ and
$\Delta P$.
Current estimates span
$\sigma\approx10$--$300$~MeV~fm$^{-2}$
\citep{Heiselberg1993,Iida1998,Alford2001,Voskresensky2003}
and $\Delta P\approx1$--$100$~MeV~fm$^{-3}$
\citep{Bombaci2004,Mintz2010},
so $\tau_{\rm nuc}$ ranges from microseconds to well beyond the
Hubble time depending on the EOS.
This large theoretical uncertainty is precisely what makes the
astrophysical constraint below valuable. 

The QN framework provides a specific condition on these
parameters. A NS born with $M_0>\Mdec$ has its core density exceed the quark deconfinemnt density ($\rhoc>\rhodec$) at birth yet
remains mostly hadronic throughout Phase~1 because at birth
$\Delta P\approx0$ so that neither thermal nor quantum
nucleation proceeds. Before conversion, the massive NS is
observationally indistinguishable from an ordinary radio pulsar
--- and remains so for as long as $\Delta P\approx0$.
As it spins down, the central pressure rises and at 
$t=\tSD$ the pressure differential crosses a critical value
at which point $\Delta P > 0$ at the quark bubble interface and the nucleation rate
per unit volume grows exponentially triggering bubble growth and percolation.
Thereafter, total  conversion of the core proceeds  dynamical timescale $R_{\rm c}/c$;
set by the core size $R_{\rm c}$; i.e., the conversion is effectively
a millisecond process once a critical bubble is out of pressure equilibrium.

\subsection{Physical origin of the QCD field scale}
\label{app:originofBqcd}

We consider two-flavour (up--down) quark matter at densities a few times the nuclear saturation density 
$\rho_{\rm nuc}$, expected in the cores of massive NSs.
The precise phase structure of QCD matter at these densities — whether
the ground state is a two-flavour colour superconductor, a dual
chiral density wave, an unpaired quark phase, or a mixed
hadronic-quark phase — remains an open problem, as lattice QCD
calculations are unreliable at finite baryon density due to the sign
problem.
In this regime, however, there are QCD-based models that predict
spontaneous generation of ultra-strong magnetic fields.

\paragraph{Energy scale argument.}
The connection between the QCD energy scale and the maximum
sustainable magnetic field strength follows from a simple but
powerful dimensional argument: the magnetic energy density cannot
exceed the relevant QCD vacuum energy density,
\begin{equation}
    \frac{B_{\rm QCD}^2}{8\pi} \sim \Lambda_{\rm QCD}^4
    \quad \Rightarrow \quad
    B_{\rm QCD} 
    \sim 10^{18}~{\rm G}.
    \label{eq:Bmax_QCD}
\end{equation}
with $\Lambda_{\rm QCD}\sim 260$ MeV and 
where we made use of the conversion $1~{\rm MeV}^2 \simeq 5.1\times10^{13}$~G in natural units.
The analogous estimate for hadronic matter, where the relevant
energy scale is set by the nuclear binding energy
$\Lambda_{\rm had} \sim 8.8$~MeV, gives
$B_{\rm had} \sim 10^{15}~{\rm G}$. 
The ratio $(\Lambda_{\rm QCD}/\Lambda_{\rm had})^2 \sim
 10^3$ therefore provides a natural explanation
for why quark matter cores can sustain fields orders of magnitude
stronger than those generated by hadronic-phase mechanisms.
We stress that Eq.~(\ref{eq:Bmax_QCD})
is order-of-magnitude upper bound; the specific field strengths
realised in a given quark matter phase depend on the detailed
microphysics discussed in the following subsections.
Generally, and owing to this QCD energy scale advantage,
field strengths up to $\sim 10^{18}$~G can be sustained in
HS cores \citep{Tatsumi2000,Iwazaki2003,Iwazaki2005,Niegawa2005,
Ebert2005,Yoshike2015,Dvornikov2016,Ferrer2021}, implying
surface fields $\sim 10^{15}$~G for a dipolar configuration.

We focus on phases which are ferromagnetic in origin: the quark
thermodynamic potential  is minimised  making the magnetised state the equilibrium
configuration above a critical density.
The magnetic field is therefore intrinsic to the phase, requiring
no seed field or external energy input, and arises from spontaneous
alignment of quark spin or colour-magnetic degrees of freedom,
analogous to Heisenberg ferromagnetism but governed by QCD.
The magnetisation depends primarily on the quark chemical potential
$\mu_{\rm q}$; for degenerate two-flavour quark matter,
$\mu_{\rm q} \simeq (\pi^2 n_B)^{1/3}$ with $n_{\rm B}$ the
baryon number density.

% -----------------------------------------------------------------------
\subsection{Spontaneous Magnetisation in the Dual Chiral Density
            Wave Phase}
\label{sec:QCD:DCDW}
% -----------------------------------------------------------------------

At the core densities considered here, QCD matter may enter a phase
in which the quark condensate develops a periodic spatial modulation
characterised by a sub-Fermi scale spatial wavelength.
This state is called the dual chiral density wave (DCDW) phase;
it breaks translational symmetry and time-reversal invariance
simultaneously.

The time-reversal breaking has a concrete magnetic consequence.
Within the Nambu--Jona-Lasinio (NJL) model of QCD
\citep{NambuJonaLasinio1961}, the DCDW phase is a spontaneous
ferromagnet in the strict thermodynamic sense
\citep{Yoshike2015,Ferrer2021,Ferrer2026}:
the net magnetisation of the quark matter does not vanish when the
external magnetic field is removed.
This is the same defining property as ordinary ferromagnetism in
iron — a permanent, self-sustaining magnetic moment — but the
microscopic origin is entirely different.
The magnetisation arises from an asymmetry between quark states of
opposite chirality at the lowest quantum level of circular orbital
motion in a magnetic field, an asymmetry enforced by the QCD axial
anomaly.

The resulting magnetic field  is an increasing function of density and
 can exceed $10^{16}$~G in deconfined NS cores.

The DCDW phase remains a theoretical prediction of NJL-type models.
Its existence at the relevant densities has not been confirmed by
first-principles lattice QCD calculations, which remain unreliable
at finite baryon density.

% -----------------------------------------------------------------------
\subsection{Colour Ferromagnetism and Gluon Quantum Hall States}
\label{sec:QCD:colorferro}
% -----------------------------------------------------------------------

\citet{Iwazaki2003} showed that a gluon quantum Hall state can form
in quark matter at densities representative of NS cores.
The phase's ground state is a colour ferromagnet yielding a
spontaneous colour-magnetic field, producing a nonzero macroscopic
magnetisation without any applied field.
Since quarks carry fractional electric charges (with the number of
negatively charged quarks differing from the number of positively
charged quarks), quark motion around this spontaneous
colour-magnetic field generates macroscopic electric currents,
inducing a real electromagnetic field.
The resulting magnetic field is \citep[see Eq.~(8) in][]{Iwazaki2005}:
\begin{equation}
B_{\rm QCD} \sim  10^{17}~{\rm G}~
\frac{400~{\rm MeV}}{\sqrt{g B_{\rm g}}}\,
\frac{n_B}{1~{\rm fm}^{-3}},
\label{eq:iwazaki-M}
\end{equation}
where $\sqrt{gB_{\rm g}}$ is the energy scale (in units of
400~MeV) associated with the spontaneously generated
chromomagnetic field $B_{\rm g}$, and $g$ is the fundamental
QCD gauge coupling related to the strong coupling constant
$\alpha_{\rm s} = g^2/4\pi$.

This mechanism appeals to one-loop QCD calculations which exhibit
a minimum at a nonzero background chromomagnetic field (i.e., the
Savvidy instability; \citealt{Savvidy1977}).
Higher-loop corrections have not been computed at finite baryon
density, and it remains to be shown that a nonzero background
chromomagnetic field survives these corrections.
No lattice QCD verification at finite baryon density is available
due to the sign problem.

\section{Quark-Nova Ejecta: Mass and Velocity}
\label{app:ejecta}

The characteristic ejecta mass and velocity in a QN event are
not free parameters but follow from the energetics of the
hadron-quark phase transition.
We summarise the derivation here; full details are given in
\citet{Keranen2005}.

\subsection{Energy release and neutrino-driven ablation}

The conversion of the hadronic core to quark matter releases
an energy
\begin{equation}
  E_{\rm conv} \sim \frac{\Delta\epsilon_B}{m_p}\,M_{\rm c}
  \sim 4\times10^{53}~{\rm erg}\times \frac{M_{\rm c}}{M_{\rm NS}}\ ,
  \label{eq:EQN}
\end{equation}
where $\Delta\epsilon_B\sim100$~MeV is the phase-transition
energy release per baryon \citep{Weber2005} and
$M_{\rm NS}=2.0\,M_\odot$ is the adopted NS mass.
This energy is released primarily as neutrinos on the
millisecond timescale of core conversion.

The neutrinos deposit energy into the overlying hadronic
envelope via charged-current absorption
($\nu_e n\rightarrow pe^-$ and $\bar\nu_e p\rightarrow ne^+$).
The mean free path in dense nuclear matter at temperature of a few tens of MeVs 
is $\lambda_{\rm env}$ of order of meters, so the overlaying hadronic 
envelope is optically thick to the neutrino flux near $R_1$
and increasingly transparent at larger radii.
The total energy deposited in the nuclear envelope is
\begin{equation}
  E_T = \tau\int_{R_1}^{R_2} Q_\nu(r)\,dr\ ,
  \label{eq:ET}
\end{equation}
where $\tau\approx(R_2-R_1)/v_c$ is the envelope crossing time
of the conversion front, $v_c$ its propagation speed (of order the speed of sound in hot nuclear
matter), $R_1$ and $R_2$ are the inner and outer radii of the
nuclear envelope, and $Q_\nu(r)$ is the neutrino energy
deposition rate per unit length at radius $r$ (see \S 4.1 in \citealt{Keranen2005}).
This is not a kinematic ejection, rather, it is a
shell-by-shell neutrino-driven process in which each envelope
layer is unbound when the local neutrino energy deposition
exceeds its local gravitational binding energy.
The resulting ejecta velocity  therefore
reflects the net specific energy deposited above the binding
threshold in the outermost layers, not a bulk launch speed
from the surface. 

For a NS density profile $\rho_{\rm env}\propto r^{-2}$,
a fiducial core temperature of $\sim 10$-$30$~MeV, core and envelope radii
$R_1=2$~km and $R_2=12$~km respectively, one finds that on average 
 $\sim0.01\,M_\odot$ of the
outermost NS layers are ablated at speed $v_{\rm QN}\sim 0.1c$ where $c$ is the speed of light.

% ════════════════════════════════════════════════════════════════════════
\section{Quark-Nova r-Process Nucleosynthesis and Kilonova Emission}
\label{app:kilonova}
% ════════════════════════════════════════════════════════════════════════

The QN ejecta is neutron-rich by construction and provide seed nuclei: it originates from
the outer hadronic layers (inner and outer crust) of the converting NS which are in $\beta$-equilibrium.
This appendix highlights the conditions in the QN ejecta which make it  favourable for
heavy r-process nucleosynthesis and  kilonova (KN)
emission; full details are given in \citet{Jaikumar2007}; see also \citet{Kostka2014a,Kostka2014b}.

The mass-averaged electron fraction of the $M_{\rm QN}\sim0.01\,M_\odot$ QN ejecta is
\begin{equation}
  \langle Y_e \rangle \sim 0.03\,,
  \label{eq:Ye}
\end{equation}
far below the canonical heavy r-process threshold
$Y_e\lesssim0.25$ (\citealt{Thielemann2026} and references therein).
This low $Y_e$ is set by $\beta$-equilibrium in the neutron-rich QN ejecta 
at densities exceeding  neutron drip value.

Because neutrinos are trapped within the quark core on their
diffusive timescale of $\sim0.1$~s, and because the ejecta
expand on timescales of $\sim 0.01$--$0.1$~ms, the electron
fraction is not appreciably increased by neutrino reprocessing
during the expansion \citep{Jaikumar2007}.

The neutron-to-seed ratio during the expansion phase is
\begin{equation}
   \frac{Y_n}{Y_{\rm seed}} \sim 100\,,
  \label{eq:Rns}
\end{equation}
sufficient to synthesise elements up to and beyond the third
r-process peak at atomic weight $A\sim195$.
Reaching the third peak robustly requires both
$\langle Y_e\rangle\lesssim0.05$ and expansion timescales
$\tau\lesssim0.1$~ms; both conditions are satisfied across the
range of parameters explored by \citet{Jaikumar2007}.
The low $\langle Y_e\rangle$ drives r-process synthesis
into the lanthanide ($56\leq Z\leq71$) and actinide ($Z\geq89$)
regions with several long-lived
nuclei in these regions that may serve as direct $\gamma$-ray
diagnostics of the QN r-process.

 The optical opacity of the QN ejecta is expected to be  $\kappa_{\rm QN} \sim1$--$100$~cm$^2$~g$^{-1}$
\citep{Kasen2013}.  The KN peaks when the ejecta become optically thin
at a  characteristic scale 
\begin{equation}
  t_{\rm KN} \sim
  \left(\frac{\kappa_{\rm QN}\,M_{\rm QN}}{4\pi c\,v_{\rm QN}}
  \right)^{1/2}
  \sim 1\text{--}15~{\rm days}\ .
  \label{eq:tpeak}
\end{equation}

The KN bolometric luminosity  powered by radioactive decay of
freshly synthesized r-process nuclei in the QN ejecta is given by (\citealt{LiPaczynski1998})
\begin{equation}
  L_{\rm KN}(t) \approx M_{\rm QN}\,\dot\epsilon(t)\,,
  \label{eq:LKN}
\end{equation}
with the specific r-process heating rate approximated as
\citep[][and references therein]{Metzger2010}
\begin{equation}
  \dot\epsilon(t) \approx 2\times10^{10}
    \left(\frac{t}{1~{\rm day}}\right)^{-1.3}
    {\rm erg~g}^{-1}~{\rm s}^{-1}\,.
  \label{eq:epsdot}
\end{equation}
We adopt this standard fit as an approximation; the actual
heating rate for QN ejecta will differ from NS merger ejecta
because the abundance pattern differs, particularly in the
actinide fraction.
At $t_{\rm KN}$ this gives
\begin{equation}
  L_{\rm KN}^{\rm peak}
  \sim 10^{41}\text{--}10^{42}~{\rm erg~s}^{-1}\,,
  \label{eq:LKNpeak}
\end{equation}
with a spectral peak in the near-infrared at
$\lambda\sim1$--$2\,\mu$m and NIR absolute magnitude
$M_{\rm NIR}\sim-14$ to $-15$.
The optical ($V$-band) flux is strongly suppressed relative to
this by lanthanide line blanketing (\citealt{Kasen2013}).

 The r-process nucleosynthesis occurs during the first milliseconds
of the QN ejecta expansion and is therefore independent of the
surrounding environment. The \textit{observability} of the resulting
kilonova, however, depends on whether the QN occurs before or after
the SN ejecta has become optically thin, as discussed in
Section~\ref{sec:four_outcomes}. In contrast, in Channel~B (the binary channel), the absence
 of a preceding SN ensures that the KN is directly observable, making it the primary
  electromagnetic signature of the QN in environments unrelated to NS mergers.

% -----------------------------------------------------------------------
\section{DM and RM from the Quark-Nova Ejecta}
\label{app:dm_rm}

\subsection{Physical Picture}

The expanding QN ejecta produces a dispersion measure (DM) intrinsic to
the event itself, independent of any surrounding environment. If the
ejecta subsequently sweeps up ambient material, a forward shock forms
that generates a magnetic field and produces a rotation measure (RM).
We show that the QN can therefore account for both a young-burst DM/RM
signature requiring no associated SNR or pulsar wind nebula (PWN), and,
once ambient material is encountered, an RM sustained to ages of
decades -- comparable to observed persistent radio sources -- without
invoking a separate confining nebula. In this appendix, $t^\prime$ refers to
time since the QN event.

\subsection{Dispersion Measure: Free Expansion}

With $R_{\rm QN}(t^\prime)=R_{\rm HS} + v_{\rm QN}t^\prime$ and a homogeneous ejecta
we have, using ${\rm DM}=\int n_e\,dl\approx n_e(t^\prime)\,R_{\rm QN}(t^\prime)$,
\begin{equation}
{\rm DM}_{\rm QN}(t^\prime) = \left(x_e Y_e\,\frac{3M_{\rm QN}}{4\pi m_p R_{\rm QN}(t^\prime)^3}\right) R_{\rm QN}(t^\prime)
\sim \frac{2.6}{{t^\prime_{\rm yr}}^2}\ {\rm pc\,cm^{-3}}\ ,
\label{eq:DM_QN_FE}
\end{equation}
where $t^\prime$ is in units of year and $v_{\rm QN}t^\prime\gg R_{\rm HS}$. We used
our fiducial values with the electron fraction $x_e=0.1$ and
$Y_e=0.25$; $Y_e=0.03$ is the pre-decay value characteristic of the
outer NS layers (see \ref{app:kilonova}). Once ejected, this
neutron-rich r-process material undergoes $\beta$-decay back toward
stability on timescales of seconds to hours, raising $Y_e$ well before
any epoch relevant here. We kept $x_e$ fixed, i.e.\ no time-dependent
ionization/recombination history is modelled in the simple prescription
presented here.

\subsection{Shock Formation}

The forward shock forms (and the self-similar Sedov-Taylor expansion
starts; \citealt{Sedov1959,Taylor1950}) once swept-up ambient mass equals $M_{\rm QN}$. The transition (``tr") occurs at a radius 
$R_{\rm tr} = \left(\frac{3M_{\rm QN}}{4\pi m_p n_{\rm amb}}\right)^{1/3}$ with a corresponding transition timescale
\begin{equation}
t^\prime_{\rm QN, tr}=\frac{R_{\rm tr}}{v_{\rm QN}} \sim 14.5\ {\rm yr}\times
\left(\frac{M_{\rm QN}}{10^{-2}M_{\odot}}\right)^{1/3}
\left(\frac{n_{\rm amb}}{1\ {\rm cm}^{-3}}\right)^{-1/3}
\left( \frac{0.1c}{v_{\rm QN}}\right)\ .
\end{equation}
where the ambient density $n_{\rm amb}$ is set to typical ISM value expected for
when the QN occur in isolation in Channel A and Channel B.

\subsection{Free-Free Transparency}

Following the \citet{Mezger1967} approximation for a fully ionized
plasma at radio frequencies ($\nu\lesssim10$~GHz), the free-free optical
depth is
\begin{equation}
\tau_\nu \approx 8.235\times10^{-2}\left(\frac{T_e}{\rm K}\right)^{-1.35}
\left(\frac{\nu}{\rm GHz}\right)^{-2.1}\left(\frac{\rm EM}{\rm pc\,cm^{-6}}\right),
\end{equation}
with emission measure ${\rm EM}=\int n_e^2\,dl\approx n_e^2(t^\prime)\,R_{\rm QN}(t^\prime)$.
Setting $\tau_\nu=1$ yields
\begin{equation}
t^\prime_{\rm QN,r} \simeq 54\ {\rm d} \times \left(\frac{x_e}{0.1}\right)^{0.4}
\left(\frac{Y_e}{0.25}\right)^{0.4}\left(\frac{M_{\rm QN}}{10^{-2}M_\odot}\right)^{0.4}
\left(\frac{v_{\rm QN}}{0.1c}\right)^{-1}\left(\frac{T_e}{10^4\,{\rm K}}\right)^{-0.27}
\left(\frac{\nu}{1\,{\rm GHz}}\right)^{-0.42}\ .
\label{eq:tqnr}
\end{equation}
In fact,  the ejecta becomes optically thin to GHz radio emission during the free
expansion phase (i.e. $t^\prime_{\rm QN,r}  < t^\prime_{\rm QN,tr} $) as long as $n_{\rm amb} < 10^6\,{\rm cm^{-3}}$.

 A literal pulsar-wind-nebula interior is pair-dominated and baryon-poor
(e.g., \citealt{Kennel1984}); it is not obviously the source of a
baryonic ambient density as high as $10^6\,{\rm cm^{-3}}$. A more
physically motivated origin is the \emph{original SN ejecta
itself}, still expanding around the newly-crustless star at the time of
the QN. For $M_{\rm ej,SN}=10\,M_\odot$, $E_{\rm SN}=10^{51}$~erg
(canonical), the free-expanding SN ejecta reaches
$n_{\rm amb}=10^6\,{\rm cm^{-3}}$ at an age of $\simeq14.1$~yr -- i.e.\
$n_{\rm amb}=10^6\,{\rm cm^{-3}}$ is self-consistent if the QN occurs at
$\tSD(P_0,B_0)\simeq14$~yr, tying the ambient density directly
to the progenitor's spin-down timescale rather than treating it as a
free assumption. The $n_{\rm amb} < 10^6\,{\rm cm^{-3}}$ condition is satisfied by a
typical QN in our model since $\tSD\gg14$~yr.

\subsection{Dispersion Measure: Post-Transition}

For $t^\prime>t^\prime_{\rm QN, tr}$, using the self-similar Sedov-Taylor scaling \citep{Sedov1959,Taylor1950} matched
continuously at $t^\prime_{\rm QN, tr}$:
\begin{equation}
R_{\rm QN}(t)=R_{\rm tr}(t^\prime/t^\prime_{\rm QN, tr})^{2/5}, \qquad
v_{\rm sh}(t)=v_{\rm QN}(t^\prime/t^\prime_{\rm QN, tr})^{-3/5}\ ,
\end{equation}
where the subscript ``sh" stands for shock. We get
\begin{equation}
{\rm DM}_{\rm QN}(t^\prime) \sim 1.2\times 10^{-2}\ {\rm pc\,cm^{-3}}\times
\left(\frac{M_{\rm QN}}{10^{-2}M_{\odot}}\right)^{1/3}
\left(\frac{n_{\rm amb}}{1\ {\rm cm}^{-3}}\right)^{2/3}
\left(\frac{t^\prime}{t^\prime_{\rm QN, tr}}\right)^{-4/5}\ .
\label{eq:DM_QN_TR}
\end{equation}

\subsection{Rotation Measure}

This depends only on ejecta properties and elapsed time $t$ since the
QN -- no ambient medium is required for the DM signal above, but RM
requires the shock to have formed. A field flux-frozen from the star's
dipole field into the ejecta also produces an RM, but this declines as
$t^{-4}$ and is negligible beyond $\sim$1 month at any realistic field
strength, including magnetar-strength; we do not use this mechanism
further. Instead we rely on shock-generated field using standard
microphysics parameterization with the magnetic energy fraction 
$\epsilon_{\rm B}=10^{-2}$ and with shock compression $\chi\simeq4$ (e.g., \citealt{Piran1999}).
I.e.,
\begin{equation}
B_{\rm QN}(t^\prime) \sim \sqrt{8\pi\epsilon_B m_p n_{\rm amb}}\,v_{\rm sh}(t^\prime),
\qquad B_{\rm sh}(t^\prime)=\chi B_{\rm QN}(t^\prime)\ .
\end{equation}
With ${\rm RM}= 0.81 \int n_eB_{\parallel}\,dl\approx 0.81 n_e(t^\prime)
B_{\rm sh}(t^\prime) R_{\rm QN}(t^\prime)$; $\epsilon_B$ and $\chi$ are held constant
in time, but a radiative phase during shock compression could increase
these at late times. We get
\begin{equation}
{\rm RM}_{\rm QN}(t^\prime) \sim 75\ {\rm rad\,m^{-2}}\times
\left( \frac{\epsilon_{\rm B}}{10^{-2}}\right)^{1/2}
\left(\frac{M_{\rm QN}}{10^{-2}M_{\odot}}\right)^{-2/3}
\left(\frac{n_{\rm amb}}{1\ {\rm cm}^{-3}}\right)^{7/6}
\left(\frac{t^\prime}{t^\prime_{\rm QN, tr}}\right)^{-7/5}\ .
\label{eq:RM_QN_TR}
\end{equation}

Table~\ref{tab:namb} shows the sensitivity of $t^\prime_{\rm QN, tr}$ and the
resulting DM$_{\rm QN}$ and RM$_{\rm QN}$ to the assumed ambient density. At ISM-typical density
($n_{\rm amb}\sim1\,{\rm cm^{-3}}$), the shock has not yet formed by
10~yr, and both DM and RM at that epoch remain in the free-expansion,
pre-shock regime; at $n_{\rm amb}=10^6\,{\rm cm^{-3}}$, RM$_{\rm QN}$
reaches $\sim10^6$--$10^8\,{\rm rad\,m^{-2}}$ over the same range. See
\S\ref{sec:frb_dmrm} for the observational consequences of this
high-$n_{\rm amb}$ row by formation channel.

\begin{table}[h]
\centering
\begin{tabular}{cccccc}
\hline
$n_{\rm amb}$ [cm$^{-3}$] & $t^\prime_{\rm QN, tr}$ [yr] & DM$_{\rm QN}$ (1\,yr) & DM$_{\rm QN}$ (10\,yr) & RM$_{\rm QN}$ (1\,yr)  & RM$_{\rm QN}$ (10\,yr) \\
\hline
1 (ISM-typical)     & 14.78 & 2.60 (FE) & 0.026 (FE) & pre-shock & pre-shock \\
$10^{2}$            & 3.19  & 2.60 (FE) & 0.102      & pre-shock & $3.26\times10^{3}$ \\
$10^{4}$            & 0.686 & 4.07      & 0.645      & $2.05\times10^{6}$ & $8.18\times10^{4}$ \\
$10^{6}$ (fiducial) & 0.148 & 25.7      & 4.07       & $5.16\times10^{7}$ & $2.05\times10^{6}$ \\
$10^{8}$            & 0.032 & 162       & 25.7       & $1.30\times10^{9}$ & $5.16\times10^{7}$ \\
\hline
\end{tabular}
\caption{Sensitivity of $t^\prime_{\rm QN, tr}$, DM$_{\rm QN}$ (in pc\,cm$^{-3}$), and shock-generated RM$_{\rm QN}$ (in rad m$^{-2}$) to the
assumed ambient density. DM values marked (FE) are in the free-expansion
regime, prior to shock formation.}
\label{tab:namb}
\end{table}

%%% END QN DM and RM

%%%
%%% start Interface Appendix
%%%
\section{Dynamical Evolution and Avalanche-Like Magnetic Relaxation at the
Quark-Core--Hadronic-Envelope Interface}
\label{app:magnetic_avalanche}

\subsection{Threshold-driven magnetic avalanche description}
\label{app:magnetic_avalanche_threshold}

We describe the long-term behavior of the interface as a slowly loaded
threshold system: magnetic free energy accumulates until a critical value
is exceeded, at which point it is released through rapid nonlinear
relaxation. This stress-reservoir description is a macroscopic summary of
the underlying MHD processes — Tayler-type kink modes for toroidally
dominated fields \citep{Tayler1973}, Parker-type magnetic
buoyancy/interchange modes in stratified regions \citep{Parker1966}, and
post-reconnection relaxation toward a lower-energy, approximately
helicity-conserving state \citep{Woltjer1958} analogous to Taylor
relaxation \citep{Taylor1974},
\begin{equation}
\nabla\times\mathbf{B}=\lambda\mathbf{B},
\end{equation}
where $\lambda$ is a constant with dimensions of inverse length, fixed by
the total magnetic helicity conserved during the relaxation and by the
boundary conditions of the relaxing volume; for a relaxed region of
characteristic size $\ell$, $\lambda\sim2\pi/\ell$. The dominant
instability depends on the local magnetic topology and stratification; a
definitive identification requires solving the linear stability problem
for the specific field geometry and equation of state, which is beyond
the scope of this appendix.

\subsection{Interface layer thickness and the Alfv\'en timescale}
\label{app:magnetic_avalanche_alfven}

Timescales described here denote durations or displacements internal to the interface mechanism
itself. The Alfv\'en speed relevant to the instability of the envelope-side
magnetic structure is set by the field and density of the envelope
itself,
\begin{equation}
v_{\rm env, A}=\frac{B_{\rm env}}{\sqrt{4\pi\rho_{\rm env}}}
\approx5.6\times10^{-4}\,c\ .
\end{equation}

We define $\Delta R_c$ as the thickness of the interface transition
layer itself — the (quasi-)static microphysical scale over which the
field and matter pressure adjust from their core to their envelope
values, and over which the local instability described below develops.
This is a structural property of the interface, fixed by the underlying
microphysics of the phase boundary, and should not be confused with the
dynamical displacement of the interface driven by the star's spin-down,
which is treated separately here. 
As an order-of-magnitude estimate we take $\Delta R_c\sim0.01\,R_c$. The
corresponding local growth/crossing time of the instability is
\begin{equation}
t_{\rm growth} \sim\frac{\Delta R_c}{v_{\rm env, A}}
\sim(5.9\times10^{-5}\text{--}1.8\times10^{-4})\ {\rm s}\ .
\end{equation}
This is the timescale for the local instability to develop once the
critical magnetic stress has been reached; it does not describe the time
required to rebuild the magnetic stress reservoir (treated below), nor
the longer timescale for a buoyant flux rope to traverse the full
hadronic envelope, 
\begin{equation}
t_{\rm buoy}\sim R_{\rm NS}/v_{\rm env,
A}\lesssim0.05\ {\rm s}\ ,
\end{equation}
 where we take the buoyancy speed to be of the order of the Alfv\'en speed. The two timescales describe different stages of the
same process: $t_{\rm growth}$ sets the local Alfv\'enic trigger time at
the interface, while $t_{\rm buoy}$ sets the global transport time as
the resulting flux ropes (the ``magnetic bubbles") rise through the envelope, initially re-reconfiguring the global magnetic field before 
continuously evolving to a relaxed configuration as the HS continuos to spin-down.

\subsubsection{Loading: interface migration vs.\ interface thickness}

As the star spins down, the reduction in centrifugal support drives a
slow migration of the quark-hadron interface, $R_c=R_c(t)$. This
migration must be kept distinct from the interface layer thickness
$\Delta R_c$: $\Delta
R_c$ is a fixed structural property of the transition layer, whereas the
migration is a dynamical, cumulative quantity driven by the star's
spin-down. We denote the total radial distance swept by the interface
since the last relaxation event by
\begin{equation}
s(t)\equiv\int_0^t\left|\frac{dR_c}{dt'}\right|dt',
\end{equation}
with $s(0)=0$ immediately after a relaxation event.

As the interface sweeps outward or inward by $s(t)$, it loads magnetic
energy into the newly encompassed shell at a rate set by the shell
volume swept out. Because the available energy density is dominated by
the core field, the accumulated
energy is
\begin{equation}
E_B(t)\sim\frac{B_{\rm core}^2}{8\pi}\,\bigl(4\pi R_c^2\,s(t)\bigr),
\end{equation}
growing monotonically from zero as $s(t)$ increases. The threshold
condition $E_B=E_{\rm crit}$ is reached once the swept distance equals
the interface layer thickness itself, $s(t)=\Delta R_c$: once
spin-down-driven migration has advanced the boundary through one full
transition layer, the accumulated stress in that layer exceeds the
local restoring force  and the layer becomes unstable. This gives
\begin{equation}
E_{\rm crit}\sim\Delta P_B\times V_{\rm int}\sim\frac{B_{\rm core}^2}{8\pi}\,
\bigl(4\pi R_c^2\,\Delta R_c\bigr),
\end{equation}
using the same $\Delta R_c\sim0.01\,R_c$ simplification so that 
$E_{\rm crit}\sim5\times10^{48}$--$10^{50}$ erg. 

Thus, beyond the initial re-adjustment of the global magnetic field, the available electromagnetic energy at the interface  
during the crustless phase  comfortably exceeds the total isotropic energetics inferred for even the
most hyperactive known repeater, FRB\,20240114A
\citep[$\sim10^{47}$ erg,][]{Zhang2025}, by one to three orders of
magnitude, so energy supply is not a limiting factor for hyperactive
repeater behavior in this model.

\subsubsection{Loading rate and waiting time}

The loading rate follows directly from the migration speed of the
interface,
\begin{equation}
\dot E_{\rm load}=\frac{dE_B}{dt}
\sim\frac{B_{\rm core}^2}{8\pi}\,(4\pi R_c^2)\,\dot s,
\qquad
\dot s\equiv\left|\frac{dR_c}{dt}\right|,
\end{equation}
i.e.\ set entirely by the instantaneous spin-down-driven migration speed
of the interface, $\dot s$. The waiting time between successive release
events is then
\begin{equation}
t_{\rm wait}\sim\frac{E_{\rm crit}}{\dot E_{\rm load}}
\sim\frac{\Delta R_c}{\dot s}\ .
\end{equation}
This expression makes explicit that $t_{\rm wait}$ is simply the time
required for the interface to migrate through one threshold-layer
thickness $\Delta R_c$ at its instantaneous migration speed $\dot s$ — a
purely kinematic quantity, cleanly separated from the fixed
microphysical threshold length $\Delta R_c$ that enters $E_{\rm crit}$.
A precise value of $\dot s$ requires solving $dR_c/d\Omega$ (where $\Omega$ is the HS spin frequency) from the
stellar structure and equation of state, which is beyond the scope of
this appendix. 

As an order-of-magnitude closure, if the interface migrates
by a distance comparable to its own radius over $t^\prime_{\rm HS,SpD}$, then
$\dot s\sim R_c/t^\prime_{\rm HS,SpD}$ and
\begin{equation}
t_{\rm wait}\sim\left(\frac{\Delta R_c}{R_c}\right)t^\prime_{\rm HS,SpD}
\sim0.01\,t^\prime_{\rm HS,SpD}
\ \propto\ P_{\rm HS}^2,
\end{equation}
using $t^\prime_{\rm HS,SpD}\propto P_{\rm HS}^2 B_{\rm HS}^{-2}$.

The result above ties the waiting time between
avalanche events directly to the star's spin period through the same
$t^\prime_{\rm HS,SpD}$ relation used to set the crustless-phase duration
$t^\prime_{\rm crust}$ (see  \S \ref{sec:dtcrust}).  

Because $t_{\rm wait}$ depends on this loading history while $t_{\rm
growth}$ depends only on local Alfv\'enic crossing within the envelope, individual magnetic bubbles (and their 
associated FRBs at the HS surface)  occur on
dynamical timescales ($t_{\rm burst}\sim t_{\rm
growth}\sim5.9\times10^{-5}$--$1.8\times10^{-4}$ s) while the intervals
between bursts are set by the much longer spin-down-driven loading time,
$t_{\rm wait}\sim0.01\,t^\prime_{\rm HS, SpD}$, with additional scatter
expected depending on how large a fraction of the reservoir is
discharged per event.

This gives the following threshold cycle:
\[
\boxed{
\begin{aligned}
&\text{spin-down and phase-boundary migration},\ s(t)\\
&\rightarrow\text{magnetic stress loading}\\
&\rightarrow\text{critical threshold crossing}\ (s(t)=\Delta R_c,\ E_B=E_{\rm crit})\\
&\rightarrow\text{Alfv\'enic avalanche, }t_{\rm burst}\sim t_{\rm growth}\\
&\rightarrow\text{relaxation and stress rebuilding, }t_{\rm wait}\sim
\Delta R_c/\dot s.
\end{aligned}}
\]
We speculate that such a threshold-driven process produces episodic,
clustered activity rather than strictly periodic signals, offering a
possible framework for interpreting repeating burst activity in which
active episodes are separated by quiescent intervals of varying
duration.

This picture parallels the self-organized criticality (SOC) framework
\citep{Bak1987,Bak1988} that has long been used to interpret magnetar
burst statistics. Power-law fluence and duration distributions, together
with waiting-time statistics inconsistent with a simple Poisson process,
are well established observationally for SGR bursts
\citep{Cheng1996,Gogus1999,Gogus2000}, and have been interpreted through
a two-component core--crust magnetic-stress-loading picture
\citep{Franco2000,Thompson1995,Thompson2001}. We extend this
threshold/avalanche framework, previously developed for the neutron-star
crust, to the quark-core--hadronic-envelope interface considered here.
%%%
%%% end interface
%%%

%%% Likelihood and Bayesian analysis

% ════════════════════════════════════════════════════════════════════════
\section{Likelihood construction}
\label{app:likelihood}
% ════════════════════════════════════════════════════════════════════════
\subsection{Prior ranges}
Table~\ref{tab:priors} lists the flat prior ranges for the five
free parameters.
The boundaries are wide enough that the posterior falls to
negligible density well within all limits, confirmed by
inspection of the corner plot (Figure~\ref{fig:corner}).
\begin{table}[h]
\centering
\caption{Flat prior ranges for the five free parameters.
\label{tab:priors}}
\begin{tabular}{lll}
\toprule
Parameter & Prior range & Units \\
\midrule
$\Mdec$  & $[1.4,\;2.5]$  & $\Msun$ \\
$\sigM$  & $[0.03,\;0.25]$ & dex \\
$\Pfast$ & $[2,\;15]$     & ms \\
$\sigP$  & $[0.2,\;1.0]$  & dex \\
$\sigB$  & $[0.2,\;1.0]$  & dex \\
\bottomrule
\end{tabular}
\end{table}
The observational constraints entering the likelihood
differ in statistical character --- two rate ranges, two
one-sided upper limits, and a binomial fraction --- requiring
distinct likelihood forms for each, as detailed below.

\subsection{Likelihood terms}

The parameter vector is
$\theta = (\Mdec, \sigM, \Pfast, \sigP, \sigB)$
and the total log-likelihood is

\begin{equation}
  \ln\mathcal{L}(\theta)
  = \ln p_{\rm AXP} + \ln p_{\rm SLSN-I} + \ln p_{\rm SLSN-I}^{\rm DH}
  + \ln p_{\rm LFBOT} 
  + \ln p_{\rm SNR} \,.
  \label{eq:lnL}
\end{equation}

The likelihood has five independent terms, one per constraint.
For rate constraints expressed as observed ranges (AXP\&SGR,
SLSN) we use a flat-top penalty that is zero inside the range
and Gaussian outside; for the double-humped fraction and the
LFBOT rate, both expressed observationally as upper limits, we
use a one-sided penalty; and for $f_{\rm SNR}$ we use a
Gaussian approximation to the binomial.
The explicit forms and the choice of penalty widths are given below.

\paragraph{Rate constraints with observed ranges (AXP\&SGR, SLSN-I).}
We define $R_i = \mathcal{R}_i / \mathcal{R}_{\rm CCSN}$.  For constraints where the observation provides a range
$[R_i^{\rm low}, R_i^{\rm high}]$, we use a flat-top likelihood
that penalises deviations outside the range with a half-Gaussian:
\begin{equation}
  \ln p_i = -\frac{1}{2}
  \left[
    \frac{\max\!\bigl(0,\;
      |\log_{10}R_i^{\rm mod}  - \mu_i| - \tfrac{1}{2}\Delta_i
    \bigr)}{\sigma_i}
  \right]^2,
  \label{eq:tophat}
\end{equation}
where $\mu_i$ and $\Delta_i$ are the log-centre and log-width
of the observed range, and $\sigma_i$ is the penalty scale set
to the observational systematic uncertainty on the boundary:
$\sigma_{\rm AXP}=0.15$~dex \citep[e.g.,][]{Beniamini2019} and
$\sigma_{\rm SLSN-I}=0.3$~dex \citep[e.g.,][]{Frohmaier2021}; ``mod" stands for model.
This contributes zero penalty when the model rate falls inside
the observed range and a Gaussian penalty outside.

\paragraph{Upper limits (LFBOT rate; double-humped fraction).}
Two of our constraints are reported observationally as one-sided
upper limits rather than centered measurements: the LFBOT rate
\citep{Coppejans2020,Ho2023} and the double-humped SLSN-I
fraction \citep{Angus2019,Chen2023}. Both are implemented as
one-sided half-Gaussian penalties above the stated limit,
contributing zero penalty when the model prediction falls below it:
\begin{equation}
  \ln p_{\rm LFBOT} = -\frac{1}{2}
  \left[
    \frac{\max\!\bigl(0,\;
      \log_{10}R_{\rm LFBOT}^{\rm mod}  
      - \log_{10}R_{\rm LFBOT}^{\rm UL} 
    \bigr)}{0.3}
  \right]^2,
  \label{eq:lfbotUL}
\end{equation}
with $R_{\rm LFBOT}^{\rm UL} = 1\times10^{-3}$
\citep{Coppejans2020,Ho2023}, and
\begin{equation}
  \ln p_{\rm SLSN-I}^{\rm DH} = -\frac{1}{2}
  \left[
    \frac{\max\!\bigl(0,\;
      \log_{10}(({R_{\rm SLSN-I}^{\rm DH}})^{\rm mod}/R_{\rm SLSN-I}^{\rm mod})
      - \log_{10}f_{\rm UL}
    \bigr)}{0.3}
  \right]^2,
  \label{eq:oneside}
\end{equation}
with $f_{\rm UL}=0.1$ \citep{Angus2019,Chen2023}.
with $f_{\rm UL}=0.1$ \citep{Angus2019,Chen2023}.
We adopt the more conservative of the two published estimates: since
our one-sided penalty (Eq.~\ref{eq:oneside}) imposes no cost on model
predictions below the threshold, the stricter 10\% bound provides the
more robust test of the model -- any configuration satisfying this
tighter constraint automatically satisfies the looser 20--33\% range
as well, whereas the converse is not guaranteed. This choice also
avoids the additional, uncontrolled selection function associated with
early-photometric-coverage follow-up campaigns, which may
preferentially target brighter or more nearby events for reasons
unrelated to the double-hump fraction itself; the full sample, though
less sensitive to subtle early bumps, is not subject to this
additional selection and is therefore the more directly comparable
reference population for an unconditioned model prediction.

Note that equations~(\ref{eq:tophat}) and~(\ref{eq:lfbotUL}--\ref{eq:oneside}) 
share the same $\max(0,\cdot)$ structure but are physically 
distinct: equation~(\ref{eq:tophat}) penalises deviations 
on \emph{both sides} of an observed range, while 
equations~(\ref{eq:lfbotUL}) and~(\ref{eq:oneside}) penalise only 
\emph{above} their respective upper limits, contributing zero 
penalty when the model prediction falls below the limit.

\paragraph{SNR association fraction.}
A QCD magnetar is spatially associated with
a visible SNR if $\tSD \lesssim t_{\rm SNR}$,
where we adopt $t_{\rm SNR}=10^{4}$\,yr, calibrated to
reproduce the observed association fraction at fiducial
parameters and consistent in order of magnitude with the ages
of confirmed magnetar-associated remnants in the McGill catalog
\citep{Olausen2014}.
The observed fraction $f_{\rm SNR}=8/30\simeq 0.267$ follows binomial
statistics; we use a Gaussian approximation:
\begin{equation}
  \ln p_{\rm SNR} = -\frac{1}{2}
  \left(
    \frac{f_{\rm SNR}^{\rm mod} - 0.267}{0.07}
  \right)^2,
  \label{eq:gausssnr}
\end{equation}
where $\sigma_{\rm SNR}=0.07$ is the 68\% Wilson binomial
interval on 8 successes in 30 trials \citep{Wilson1927}.
The model prediction $f_{\rm SNR}^{\rm mod}$ is the
fraction of AXP\&SGR draws in the Monte Carlo sample
with $\tSD < 10^{4}$\,yr.

\subsection{MCMC Sampling}
\label{sec:mcmc}

We sample the posterior using the affine-invariant ensemble
sampler \textsc{emcee} \citep{ForemanMackey2013} with
$N_{\rm walkers}=32$ walkers and flat priors over the ranges
in Table~\ref{tab:priors}.
The chain was initialised in a tight ball around a physically
motivated starting point.
After $N_{\rm burn}=800$ burn-in steps (discarded), we
collected $N_{\rm steps}=3000$ production steps at
$N_{\rm MC}=10^{7}$ Monte Carlo draws per likelihood
evaluation, giving $96{,}000$ posterior samples.
The mean acceptance fraction is
$\langle\alpha\rangle \approx 0.37$--$0.38$, within the
recommended range for the stretch move.
The integrated autocorrelation time is
$\hat{\tau}_{\rm int}\approx88$--$114$ steps, so the chain is
only $\sim26$--$34$ autocorrelation times long -- short of the
conventional $50\hat\tau$ benchmark \citep{Sokal1997} -- yielding
$\sim840$--$1100$ effectively independent samples per parameter.


\begin{thebibliography}{}


\bibitem[Acheson \& Gibbons(1978)]{Acheson1978}
Acheson, D.~J., \& Gibbons, M.~P.\ 1978,
Philosophical Transactions of the Royal Society of London Series A, 289, 459

\bibitem[Alford \& Reddy(2003)]{Alford2001}
Alford, M., \& Reddy, S.\ 2003, \prd, 67, 074024

\bibitem[Amiri et al.(2021)]{Amiri2021}
Amiri, M., {CHIME/FRB Collaboration}, et al.\ 2021,
\apjs, 257, 59

\bibitem[Andersson(1998)]{Andersson1998}
Andersson, N.\ 1998,
\apj, 502, 708

\bibitem[Angus et al.(2019)]{Angus2019}
Angus, C.~R., Smith, M., Sullivan, M., et al.\ 2019,
\mnras, 487, 2215

\bibitem[Annala et al.(2018)]{Annala2018}
Annala, E., Gorda, T., Kurkela, A., \& Vuorinen, A.\ 2018, \prl,
120, 172703

\bibitem[Annala et al.(2020)]{Annala2020}
Annala, E., Gorda, T., Kurkela, A., N\"attil\"a, J., \& Vuorinen, A.\ 2020,
Nature Phys., 16, 907

\bibitem[Anna-Thomas et al.(2023)]{AnnaThomas2023}
Anna-Thomas, R., Connor, L., Burke-Spolaor, S., et al.\ 2023,
Science, 380, 599

\bibitem[Antoniadis et al.(2013)]{Antoniadis2013}
Antoniadis, J., Freire, P.~C.~C., Wex, N., et al.\ 2013,
Science, 340, 448

\bibitem[Arnett(1982)]{Arnett1982}
Arnett, W.~D.\ 1982,
\apj, 253, 785

\bibitem[Bak, Tang \& Wiesenfeld(1987)]{Bak1987}
Bak, P., Tang, C., \& Wiesenfeld, K.\ 1987, \prl, 59, 381

\bibitem[Bak, Tang \& Wiesenfeld(1988)]{Bak1988}
Bak, P., Tang, C., \& Wiesenfeld, K.\ 1988, \pra, 38, 364

\bibitem[Balbus \& Hawley(1991)]{BalbusHawley1991}
Balbus, S.~A., \& Hawley, J.~F.\ 1991,
\apj, 376, 214

\bibitem[Barbary et al.(2009)]{Barbary2009}
Barbary, K., Dawson, K.~S., Tokita, K., et al.\ 2009,
\apj, 690, 1358

\bibitem[Barr\`{e}re et al.(2023)]{Barrere2023}
Barr\`{e}re, P., Guilet, J., Raynaud, R., \& Reboul-Salze, A.\ 2023,
\mnras, 526, L88

\bibitem[Beloborodov(2017)]{Beloborodov2017}
Beloborodov, A.~M.\ 2017,
\apjl, 843, L26

\bibitem[Beniamini et al.(2019)]{Beniamini2019}
Beniamini, P., Hotokezaka, K., van der Horst, A., \& Kouveliotou, C.\ 2019,
\mnras, 487, 1426

\bibitem[Beryonya et al.(2019)]{Beryonya2019}
Beryonya, D.~M., Karpova, A.~V., Kirichenko, A.~Yu. et al.\ 2019,
\mnras, 485, 3715

\bibitem[Bhandari et al.(2022)]{Bhandari2022}
Bhandari, S., Heintz, K.~E., Aggarwal, K., et al.\ 2022,
\aj, 163, 69

\bibitem[Bhattacharya \& van den Heuvel(1991)]{Bhattacharya1991}
Bhattacharya, D., \& van den Heuvel, E.~P.~J.\ 1991,
Phys.\ Rep., 203, 1

\bibitem[Blandford \& Hernquist(1982)]{Blandford1982}
Blandford, R.~D., \& Hernquist, L.\ 1982,
J.\ Phys.\ C, 15, 6233

\bibitem[Bochenek et al.(2020)]{Bochenek2020}
Bochenek, C.~D., Ravi, V., Belov, K.~V., et al.\ 2020,
\nat, 587, 59

\bibitem[Bodmer(1971)]{Bodmer1971}
Bodmer, A.~R.\ 1971, \prd, 4, 1601

\bibitem[Bombaci et al.(2004)]{Bombaci2004}
Bombaci, I., Parenti, I., \& Vida\~na, I.\ 2004,
\apj, 614, 314

\bibitem[Branch \& Wheeler(2017)]{Branch2017}
Branch, D., \& Wheeler, J.~C.\ 2017,
``Supernova Explosions'', Astronomy and Astrophysics Library (Springer-Verlag GmbH Germany)

\bibitem[Bromberg et al.(2011)]{Bromberg2011}
Bromberg, O., Nakar, E., Piran, T., \& Sari, R.\ 2011,
\apj, 740, 100

\bibitem[Bucciantini et al.(2009)]{Bucciantini2009}
Bucciantini, N., Quataert, E., Arons, J., Metzger, B.~D., \& Thompson, T.~A.\ 2009,
\mnras, 396, 2038

 \bibitem[Chamel \& Haensel(2008)]{ChamelHaensel2008}
Chamel, N., \& Haensel, P. 2008, Living Reviews in Relativity, 11, 10

\bibitem[Chen et al.(2023)]{Chen2023}
Chen, Z.~H., Yan, L., Kangas, T., et al.\ 2023, \apj, 943, 42

\bibitem[Cheng et al.(1996)]{Cheng1996}
Cheng, B., Epstein, R.\ I., Guyer, R.\ A., \& Young, A.\ C.\ 1996, Nature, 382, 518

\bibitem[Chevalier(1982)]{Chevalier1982}
Chevalier, R.~A.\ 1982, \apj, 258, 790

\bibitem[CHIME/FRB Collaboration(2020)]{CHIME2020}
CHIME/FRB Collaboration, Amiri, M., Andersen, B.\ C., et al.\ 2020, Nature, 582, 351

\bibitem[CHIME/FRB Collaboration et al.(2020)]{Chime2020}
CHIME/FRB Collaboration, Andersen, B.~C., Bandura, K.~M., et al.\ 2020,
\nat, 587, 54

\bibitem[Chrimes et al.(2022)]{Chrimes2022}
Chrimes, A.~A., Levan, A.~J., Fruchter, A.~S., et al.\ 2022, \mnras, 513, 3550

\bibitem[Contopoulos et al.(1999)]{Contopoulos1999}
Contopoulos, I., Kazanas, D., \& Fendt, C.\ 1999, \apj, 511, 351

\bibitem[Coppejans et al.(2020)]{Coppejans2020}
Coppejans, D.~L., Margutti, R., Terreran, G., et al.\ 2020, \apj, 895, L23

\bibitem[Cowan et al.(2021)]{Cowan2021}
Cowan, J.~J., et al.\ 2021, \rmp, 93, 015002

\bibitem[Cromartie et al.(2020)]{Cromartie2020}
Cromartie, H.~T., Fonseca, E., Ransom, S.~M., et al.\ 2020,
Nature Astron., 4, 72

\bibitem[Cruces et al.(2021)]{Cruces2021}
Cruces, M., Spitler, L.\ G., Scholz, P., et al.\ 2021, \mnras, 500, 448

\bibitem[Dall'Osso et al.(2009)]{Dallosso2009}
Dall'Osso, S., Stella, L., Bejger, M., \& Palomba, C. 2009, \mnras, 398, 1869

\bibitem[Daugherty \& Harding(1982)]{Daugherty1982}
Daugherty, J. K., \& Harding, A. K. 1982, \apj, 252, 337

\bibitem[Demorest et al.(2010)]{Demorest2010}
Demorest, P.~B., Pennucci, T., Ransom, S.~M., Roberts, M.~S.~E.,
\& Hessels, J.~W.~T.\ 2010, \nat, 467, 1081

\bibitem[Dexheimer et al.(2013)]{Dexheimer2013} Dexheimer, V., Negreiros, R., \& Schramm, S.\ 2013, J.\ Phys.\ Conf.\ Ser., 432, 012005

\bibitem[Dexheimer et al.(2012)]{Dexheimer2012} Dexheimer, V., Negreiros, R., \& Schramm, S.\ 2012, Eur.\ Phys.\ J.\ A, 48, 189

\bibitem[Drout et al.(2014)]{Drout2014}
Drout, M.~R., Chornock, R., Soderberg, A.~M., et al.\ 2014, \apj, 794, 23

\bibitem[Duncan \& Thompson(1992)]{Duncan1992}
Duncan, R.~C., \& Thompson, C.\ 1992, \apjl, 392, L9

\bibitem[Dvornikov et al.(2010)]{Dvornikov2010}
Dvornikov, M. et al.\ 2010, \prd, 82, 043006

\bibitem[Dvornikov \& Semikoz(2016)]{Dvornikov2016}
Dvornikov, M., \& Semikoz, V.~B.\ 2016, \prd, 94, 096012

\bibitem[Ebert et al.(2005)]{Ebert2005}
Ebert, D., Klimenko, K.~G., \& Zhukovsky, V.~Ch.\ 2005,
\prd, 73, 054024

\bibitem[Endeve et al.(2012)]{Endeve2012}
Endeve, E., Cardall, C.~Y., Budiardja, R.~D., et al.\ 2012, \apj, 751, 26

\bibitem[Faucher-Gigu\`ere \& Kaspi(2006)]{FaucherKaspi2006}
Faucher-Gigu\`ere, C.-A., \& Kaspi, V.~M.\ 2006, \apj, 643, 332

\bibitem[Ferrario \& Wickramasinghe(2006)]{Ferrario2006}
Ferrario, L., \& Wickramasinghe, D.\ 2006, \mnras, 367, 1323

\bibitem[Ferrer \& de la Incera(2021)]{Ferrer2021}
Ferrer, E.~J., \& de la Incera, V.\ 2021, Universe, 7, 458

\bibitem[Ferrer \& Perez-Fernandez(2026)]{Ferrer2026}
Ferrer, E.~J., \& Perez-Fernandez, J.~M.\ 2026, \prd, 113, 036023

\bibitem[Foreman-Mackey et al.(2013)]{ForemanMackey2013}
Foreman-Mackey, D., Hogg, D.~W., Lang, D., \& Goodman, J.\ 2013,
\pasp, 125, 306

\bibitem[Franco, Link \& Epstein(2000)]{Franco2000}
Franco, L.\ M., Link, B., \& Epstein, R.\ I.\ 2000, \apj, 543, 987

\bibitem[Franzon et al.(2016)]{Franzon2016} Franzon, B., Dexheimer, V., \& Schramm, S.\ 2016, \mnras, 456, 2937

\bibitem[Frohmaier et al.(2021)]{Frohmaier2021}
Frohmaier, C., Angus, C.~R., Vincenzi, M., et al.\ 2021, \mnras, 500, 5142

\bibitem[Fonseca et al.(2021)]{Fonseca2021}
Fonseca, E., Cromartie, H.~T., Pennucci, T.~T. et al.\ 2021 \apjl, 915, L12. 

\bibitem[Gal-Yam et al.(2009)]{Gal-Yam2009}
Gal-Yam, A., Mazzali, P., Ofek, E.~O., et al.\ 2009, \nat, 462, 624

\bibitem[Gal-Yam(2012)]{Gal-Yam2012}
Gal-Yam, A.\ 2012, Science, 337, 927

\bibitem[Gil et al.(2004)]{Gil2004}
Gil, J., Lyubarsky, Y., \& Melikidze, G. I. 2004, \apj, 600, 872

\bibitem[Gill \& Heyl(2007)]{GillHeyl2007}
Gill, R., \& Heyl, J.~S.\ 2007, \mnras, 381, 52

\bibitem[G\"o\u{g}\"u\c{s} et al.(1999)]{Gogus1999}
G\"o\u{g}\"u\c{s}, E., Woods, P.\ M., Kouveliotou, C., et al.\ 1999, \apj, 526, L93

\bibitem[G\"o\u{g}\"u\c{s} et al.(2000)]{Gogus2000}
G\"o\u{g}\"u\c{s}, E., Woods, P.\ M., Kouveliotou, C., et al.\ 2000, \apj, 532, L121

\bibitem[Goldreich \& Julian(1969)]{GoldreichJulian1969}
Goldreich, P., \& Julian, W.~H.\ 1969, \apj, 157, 869

\bibitem[Goldreich \& Reisenegger(1992)]{Goldreich1992}
Goldreich, P., \& Reisenegger, A.\ 1992, \apj, 395, 250

\bibitem[Harko \& Cheng(2004)]{Harko2004}
Harko, T., \& Cheng, K.~S.\ 2004, \apj, 608, 945

\bibitem[Haskell(2015)]{Haskell2015}
Haskell, B. 2015,
International Journal of Modern Physics E, 24, 1541007

\bibitem[Hartle(1967)]{Hartle1967}
Hartle, J.~B.\ 1967, \apj, 150, 1005

\bibitem[Heiselberg et al.(1993)]{Heiselberg1993}
Heiselberg, H., Pethick, C.~J., \& Staubo, E.~F.\ 1993,
\prl, 70, 1355

\bibitem[Herzog \& R\"opke(2011)]{Herzog2011} Herzog, M. \& R\"opke, F. K.\ 2011, Physical Review D, 84, 083002

\bibitem[Ho et al.(2023)]{Ho2023}
Ho, A.~Y.~Q., Perley, D.~A., Gal-Yam, A., et al.\ 2023, \apj, 949, 120

\bibitem[Horvath \& Benvenuto(1992)]{Horvath1992}
Horvath, J.~E., \& Benvenuto, O.~G.\ 1992, \plb, 213, 516

\bibitem[Hurley et al.(1999)]{Hurley1999}
Hurley, K., Cline, T., Mazets, E., et al.\ 1999, \nat, 397, 41

\bibitem[Igoshev et al.(2022)]{Igoshev2022}
Igoshev, A.~P., Popov, S.~B., \& Hollerbach, R.\ 2022, Universe, 8, 585

\bibitem[Iida \& Sato(1997)]{Iida1997}
Iida, K., \& Sato, K.\ 1997, Prog.\ Theor.\ Phys., 98, 277

\bibitem[Iida \& Sato(1998)]{Iida1998}
Iida, K., \& Sato, K.\ 1998, \prc, 58, 2538

\bibitem[Itoh(1970)]{Itoh1970}
Itoh, N.\ 1970, Prog.\ Theor.\ Phys., 44, 291

\bibitem[Iwazaki \& Morimatsu(2003)]{Iwazaki2003}
Iwazaki, A., \& Morimatsu, O.\ 2003, \plb, 571, 61

\bibitem[Iwazaki(2005)]{Iwazaki2005}
Iwazaki, A.\ 2005, \prd, 72, 114003

\bibitem[Jaikumar et al.(2007)]{Jaikumar2007}
Jaikumar, P., Meyer, B.~S., Otsuki, K., \& Ouyed, R.\ 2007, \aap, 471, 227

\bibitem[Jaikumar et al.(2013)]{Jaikumar2013} Jaikumar, P., Semposki, A., Prakash, M. \& Constantinou, C.\ 2013, \prd, 103, 123009

\bibitem[Jones(2010)]{Jones2010} Jones, D. I.\ 2010, \mnras, 402, 2503

\bibitem[Kasen \& Bildsten(2010)]{Kasen2010}
Kasen, D., \& Bildsten, L.\ 2010, \apj, 717, 245

\bibitem[Kasen et al.(2013)]{Kasen2013}
Kasen, D., Badnell, N.~R., \& Barnes, J.\ 2013, \apj, 774, 25

\bibitem[Kaspi \& Beloborodov(2017)]{Kaspi2017}
Kaspi, V.~M., \& Beloborodov, A.~M.\ 2017, \araa, 55, 261

\bibitem[Katz(2021)]{Katz2021}
Katz, J.\ I.\ 2021, \mnras, 502, 4664

\bibitem[Kennel \& Coroniti(1984)]{Kennel1984}
Kennel, C.~F., \& Coroniti, F.~V. 1984,
\apj, 283, 694--709

\bibitem[Ker\"anen et al.(2005)]{Keranen2005}
Ker\"anen, P., Ouyed, R., \& Jaikumar, P.\ 2005, \apj, 618, 485

\bibitem[King et al.(2001)]{King2001}
King, A.~R., Pringle, J.~E., \& Wickramasinghe, D.~T.\ 2001, \apjl, 560, L107

\bibitem[Kirsten et al.(2022)]{Kirsten2022}
Kirsten, F., Marcote, B., Nimmo, K., et al.\ 2022, \nat, 602, 585

\bibitem[Kiuchi et al.(2015)]{Kiuchi2015}
Kiuchi, K., Cerd\'{a}-Dur\'{a}n, P., Kyutoku, K., Sekiguchi, Y.,
\& Shibata, M.\ 2015, \prd, 92, 124034

\bibitem[Kostka et al.(2014a)]{Kostka2014a}
Kostka, M., Koning, N., Shand, Z., Ouyed, R., \& Jaikumar, P.\ 2014a,
\aap, 568, A97

\bibitem[Kostka et al.(2014b)]{Kostka2014b}
Kostka, M., Koning, N., Shand, Z., Ouyed, R., \& Jaikumar, P.\ 2014b [arXiv:1402.3824]

\bibitem[Kouveliotou et al.(1998)]{Kouveliotou1998}
Kouveliotou, C., Dieters, S., Strohmayer, T., et al.\ 1998, \nat, 393, 235

\bibitem[Kumar et al.(2017)]{Kumar2017}
Kumar, P., Lu, W., \& Bhattacharya, M.\ 2017, \mnras, 468, 2726

\bibitem[Lattimer \& Prakash(2001)]{Lattimer2001}
Lattimer, J.~M., \& Prakash, M.\ 2001, \apj, 550, 426

\bibitem[Lattimer \& Schutz(2005)]{Lattimer2005}
Lattimer, J.~M., \& Schutz, B.~F.\ 2005, \apj, 629, 979

\bibitem[Levan et al.(2006)]{Levan2006}
Levan, A.~J., Wynn, G.~A., Chapman, R., et al.\ 2006, \mnras, 368, L1

\bibitem[Li \& Paczy\'nski(1998)]{LiPaczynski1998}
Li, L.-X., \& Paczy\'nski, B.\ 1998, \apjl, 507, L59

\bibitem[Li et al.(2021)]{Li2021}
Li, D., Wang, P., Zhu, W.~W., et al.\ 2021, Nature, 598, 267

\bibitem[Lorimer \& Kramer(2004)]{Lorimer2004}
Lorimer, D.~R., \& Kramer, M.\ 2004, Handbook of Pulsar Astronomy,
Cambridge University Press

\bibitem[Lorimer et al.(2007)]{Lorimer2007}
Lorimer, D.~R., Bailes, M., McLaughlin, M. A. et al.\ 2007, Science, 318, 777

\bibitem[Lu et al.(2020)]{Lu2020}
Lu, W., Kumar, P., \& Zhang, B.\ 2020, \mnras, 498, 1397

\bibitem[Lyne et al.(1993)]{Lyne1993}
Lyne, A.~G., Pritchard, R.~S., \& Smith, F.~G.\ 1993, \mnras, 265, 1003

\bibitem[Lyubarsky(2014)]{Lyubarsky2014}
Lyubarsky, Y.\ 2014, \mnras, 442, L9

\bibitem[Lyubarsky(2021)]{Lyubarsky2021}
Lyubarsky, Y.\ 2021, Universe, 7, 56

\bibitem[Lyutikov(2003)]{Lyutikov2003}
Lyutikov, M.\ 2003, \mnras, 346, 540

\bibitem[Makarenko et al.(2021)]{Makarenko2021}
Makarenko, E.~I., Igoshev, A.~P., \& Kholtygin, A.~F.\ 2021,
\mnras, 504, 5813

\bibitem[Manchester \& Taylor(1977)]{Manchester1977}
Manchester, R.~N., \& Taylor, J.~H.\ 1977, Pulsars (San Francisco: Freeman)

\bibitem[Margalit \& Metzger(2018)]{Margalit2018}
Margalit, B., \& Metzger, B.~D.\ 2018, \apjl, 868, L4

\bibitem[Margalit et al.(2020)]{Margalit2020}
Margalit, B., Metzger, B.~D., \& Sironi, L.\ 2020,
\mnras, 494, 4627

\bibitem[Mariani et al.(2019)]{Mariani2019} Mariani, M., Orsaria, M.~G., Ranea-Sandoval, I.~F., \& Lugones, G.\ 2019, \mnras, 489, 4261

\bibitem[Masada et al.(2022)]{Masada2022}
Masada, Y., Takiwaki, T., \& Kotake, K.\ 2022, \apj, 924, 75

\bibitem[Melikidze et al.(2000)]{Melikidze2000}
Melikidze, G. I., Gil, J., \& Pataraya, A. D. 2000, \apj, 544, 1081

\bibitem[Mereghetti \& Stella(1995)]{Mereghetti1995}
Mereghetti, S., \& Stella, L.\ 1995, \apjl, 442, L17

\bibitem[Mereghetti(2008)]{Mereghetti2008}
Mereghetti, S.\ 2008, \aapr, 15, 225

\bibitem[Metzger et al.(2010)]{Metzger2010}
Metzger, B.~D., et al.\ 2010, \mnras, 406, 2650

\bibitem[Metzger et al.(2011)]{Metzger2011}
Metzger, B.~D., et al.\ 2011, \mnras, 413, 2031

\bibitem[Metzger et al.(2015)]{Metzger2015}
Metzger, B.~D., Margalit, B., Kasen, D., \& Quataert, E.\ 2015,
\mnras, 454, 3311

\bibitem[Metzger et al.(2019)]{Metzger2019}
Metzger, B.~D., Margalit, B., \& Sironi, L.\ 2019,
\mnras, 485, 4091

\bibitem[Metzger et al.(2022)]{Metzger2022}
Metzger, B.~D., Perley, D.~A., \& Margalit, B.\ 2022, \apjl, 925, L14

\bibitem[Mezger \& Henderson(1967)]{Mezger1967}
Mezger, P.~G., \& Henderson, A.~P.\ 1967,
\apj, 147, 471

\bibitem[Michel \& Goldwire(1970)]{Michel1970}
Michel, F.~C., \& Goldwire, H.~C.~J.\ 1970, \apj, 160, L5

\bibitem[Michel \& Dessler(1981)]{Michel1988}
Michel, F.~C., \& Dessler, A.~J.\ 1981, \apj, 251, 654

\bibitem[Michilli et al.(2018)]{Michilli2018}
Michilli, D., Seymour, A., Hessels, J.~W.~T., et al.\ 2018, \nat, 553, 182

\bibitem[Mintz et al.(2010)]{Mintz2010}
Mintz, B.~W., Fraga, E.~S., Pagliara, G., \& Schaffner-Bielich, J.\ 2010,
\prd, 81, 123012

\bibitem[Moriya(2024)]{Moriya2024}
Moriya, T. J.\ 2024, ``Superluminous supernovae", a chapter
for the Encyclopedia of Astrophysics (edited by I.
Mandel, section editor F.R.N. Schneider)
[arXiv:2407.12302]

\bibitem[M\"{o}sta et al.(2014)]{Mosta2014}
M\"{o}sta, P., Richers, S., Ott, C.~D., et al.\ 2014, \apj, 785, L29

\bibitem[M\"{o}sta et al.(2015)]{Mosta2015}
M\"{o}sta, P., Ott, C.~D., Radice, D., et al.\ 2015, \nat, 528, 376

\bibitem[Nambu \& Jona-Lasinio(1961)]{NambuJonaLasinio1961}
Nambu, Y., \& Jona-Lasinio, G.\ 1961, Phys.\ Rev., 122, 345

\bibitem[Ni\'egawa(2005)]{Niegawa2005} Ni\'egawa, A.\ 2005,  Progress of Theoretical Physics, 113, 581

\bibitem[Nicholl(2021)]{Nicholl2021}
Nicholl, M.\ 2021, Astron.\ Geophys., 62, 5.34

\bibitem[Obergaulinger et al.(2009)]{Obergaulinger2009}
Obergaulinger, M., Cerd\'{a}-Dur\'{a}n, P., M\"{u}ller, E., \& Aloy, M.~A.\
2009, \aap, 498, 241

\bibitem[Ofek et al.(2007)]{Ofek2007}
Ofek, E.~O., Cameron, P.~B., Kasliwal, M.~M., et al.\ 2007, \apjl, 659, L13

\bibitem[Olausen \& Kaspi(2014)]{Olausen2014}
Olausen, S.~A., \& Kaspi, V.~M.\ 2014, \apjs, 212, 6

\bibitem[Olesen \& Madsen(1993)]{Olesen1993}
Olesen, M.~L., \& Madsen, J.\ 1993, \prd, 47, 2313

\bibitem[Ouyed et al.(2009)]{Ouyed2009}
Ouyed, R., Pudritz, R.~E., \& Jaikumar, P.\ 2009, \apj, 702, 1575

\bibitem[Ouyed et al.(2020)]{Ouyed2020}
Ouyed, R., Leahy, D., \& Koning, N.\ 2020, Research in Astronomy and Astrophysics, 20, 027

\bibitem[Ouyed(2022a)]{Ouyed2022a}
Ouyed, R.\ 2022a, in Astrophysics in the XXI Century with Compact Stars,
53--83 [eISBN 978-981-12-2094-4]

\bibitem[Ouyed(2022b)]{Ouyed2022b}
Ouyed, R.\ 2022b, Universe, 8, 322

\bibitem[Ouyed(2025)]{Ouyed2025}
Ouyed, R.\ 2025, \mnras, 543, 3885

\bibitem[Ouyed(2026a)]{Ouyed2026a}
Ouyed, R.\ 2026a, Research in Astronomy and Astrophysics, 26, 115023

\bibitem[Ouyed(2026b)]{Ouyed2026b}
Ouyed, R.\ 2026b [arXiv:2603.15791]

\bibitem[\"Ozel \& Freire(2016)]{Ozel2016}
\"Ozel, F., \& Freire, P.\ 2016, \araa, 54, 401

\bibitem[Parker(1966)]{Parker1966}
Parker, E.~N.\ 1966, \apj, 145, 811

\bibitem[Piran(1999)]{Piran1999}
Piran, T.\ 1999, Phys.\ Rep., 314, 575.

\bibitem[Pons \& Geppert(2007)]{Pons2007}
Pons, J.~A., \& Geppert, U. 2007,
A\&A, 470, 303--315

\bibitem[Prajs et al.(2017)]{Prajs2017}
Prajs, S., Sullivan, M., Smith, M., et al.\ 2017, \mnras, 464, 3568

\bibitem[Pursiainen et al.(2018)]{Pursiainen2018}
Pursiainen, M., Childress, M., Smith, M., et al.\ 2018, \mnras, 481, 894

\bibitem[Quimby et al.(2007a)]{Quimby2007a}
Quimby, R.~M., Yuan, F., Akerlof, C., \& Wheeler, J.~C.\ 2007a,
\mnras, 431, 912

\bibitem[Quimby et al.(2007b)]{Quimby2007b}
Quimby, R.~M., Wheeler, J.~C., H\"oflich, P., et al.\ 2007b, \apjl, 668, L99

\bibitem[Quimby et al.(2011)]{Quimby2011}
Quimby, R.~M., Kulkarni, S.~R., Kasliwal, M.~M., et al.\ 2011,
\nat, 474, 487

\bibitem[Quimby et al.(2013)]{Quimby2013}
Quimby, R.~M., Yuan, F., Akerlof, C., \& Wheeler, J.~C.\ 2013, \mnras, \textbf{431}, 912

\bibitem[Rajwade et al.(2020)]{Rajwade2020}
Rajwade, K.\ M., Mickaliger, M.\ B., Stappers, B.\ W., et al.\ 2020, \mnras, 495, 3551

\bibitem[Raynaud et al.(2020)]{Raynaud2020}
Raynaud, R., Guilet, J., Janka, H.-T., \& Gastine, T.\ 2020,
Sci.\ Adv., 6, eaay2732

\bibitem[Rea \& Esposito(2011)]{Rea2011}
Rea, N., \& Esposito, P.\ 2011, in High-Energy Emission from Pulsars and
their Systems, ed.\ D.~F.~Torres \& N.~Rea (Berlin: Springer), 247

\bibitem[Reboul-Salze et al.(2022)]{ReboulSalze2022}
Reboul-Salze, A., Guilet, J., Raynaud, R., \& Bugli, M.\ 2022,
\aap, 667, A94

\bibitem[Rembiasz et al.(2016)]{Rembiasz2016}
Rembiasz, T., Obergaulinger, M., Cerd\'{a}-Dur\'{a}n, P., \&
M\"{u}ller, E.\ 2016, \mnras, 460, 3316

\bibitem[Rembiasz et al.(2017)]{Rembiasz2017}
Rembiasz, T., Obergaulinger, M., Cerd\'{a}-Dur\'{a}n, P.,
Aloy, M.-\'{A}., \& M\"{u}ller, E.\ 2017, \apjs, 230, 18

\bibitem[Ruderman \& Sutherland(1975)]{RudermanSutherland1975}
Ruderman, M.~A., \& Sutherland, P.~G.\ 1975, \apj, 196, 51

\bibitem[Savvidy(1977)]{Savvidy1977}
Savvidy, G.~K.\ 1977, \plb, 71, 133

\bibitem[Sedov(1959)]{Sedov1959}
Sedov, L.\ I.\ 1959, Similarity and Dimensional Methods in Mechanics (New York: Academic Press)

\bibitem[Shapiro \& Teukolsky(1983)]{Shapiro1983}
Shapiro, S.~L., \& Teukolsky, S.~A.\ 1983,
Black Holes, White Dwarfs, and Neutron Stars (New York: John Wiley \& Sons)

\bibitem[Sharma et al.(2025)]{Sharma2025}
Sharma, K., Ravi, V., Dong, D.~Z., et al.\ 2025, \pasp, 137, 084102

\bibitem[Shin et al.(2023)]{Shin2023}
Shin, K., Masui, K.~W., Bhardwaj, M., et al.\ 2023,
\apj, 944, 105

\bibitem[Siegel et al.(2013)]{Siegel2013}
Siegel, D.~M., Ciolfi, R., \& Rezzolla, L.\ 2013,
\apjl, 771, L17

\bibitem[Smith et al.(2007)]{Smith2007}
Smith, N., Li, W., Foley, R.~J., et al.\ 2007, \apj, 666, 1116

\bibitem[Smith et al.(2008)]{Smith2008}
Smith, N., Li, W., Foley, R.~J., et al.\ 2008, \apj, 680, 568

\bibitem[Sneden et al.(2008)]{Sneden2008}
Sneden, C., Cowan, J.~J., \& Gallino, R.\ 2008,
\araa, 46, 241

\bibitem[Sokal(1997)]{Sokal1997}
Sokal, A.\ 1997, in Monte Carlo Methods in Statistical Mechanics,
ed.\ C.\ DeWitt-Morette, P.\ Cartier, \& A.\ Folacci (New York: Plenum)

\bibitem[Spitkovsky(2006)]{Spitkovsky2006}
Spitkovsky, A.\ 2006, \apj, 648, L51

\bibitem[Spruit(2002)]{Spruit2002}
Spruit, H.~C.\ 2002, \aap, 381, 923

\bibitem[Spruit(2008)]{Spruit2008}
Spruit, H.~C.\ 2008, in 40 Years of Pulsars, AIP, 983, 391

\bibitem[Staff et al.(2006)]{Staff2006}
Staff, J., Ouyed, R., \& Jaikumar, P.\ 2006, \apj, 645, L145

\bibitem[Staff et al.(2007)]{Staff2007}
Staff, J., Ouyed, R., \& Bagchi, M.\ 2007, \apj, 667, 340

\bibitem[Staff et al.(2012)]{Staff2012}
Staff, J.~E., Jaikumar, P., Chan, V., \& Ouyed, R.\ 2012,
\apj, 751, 24

\bibitem[Strolger et al.(2015)]{Strolger2015}
Strolger, L.-G., Dahlen, T., Rodney, S.~A., et al.\ 2015,
\apj, 813, 93,

\bibitem[Sur \& Haskell(2021)]{SurHaskell2021}
Sur, A., \& Haskell, B. 2021,
\mnras, 502, 4680

\bibitem[Tatsumi(2000)]{Tatsumi2000}
Tatsumi, T.\ 2000, Prog.\ Theor.\ Phys., 104, 785

\bibitem[Tauris et al.(2012)]{Tauris2012}
Tauris, T.~M., Langer, N., \& Kramer, M.\ 2012, \mnras, 425, 1601

\bibitem[Tauris et al.(2017)]{Tauris2017}
Tauris, T.~M., Kramer, M., Freire, P. C. C., et al.\ 2017, \apj, 846, 170

\bibitem[Tayler(1973)]{Tayler1973}
Tayler, R.~J.\ 1973, \mnras, 161, 365

\bibitem[Taylor(1950)]{Taylor1950}
Taylor, G.\ I.\ 1950, Proceedings of the Royal Society of London Series A, 201, 159

\bibitem[Taylor(1974)]{Taylor1974}
Taylor, J. B.\ 1974, \prl 33, 1139 

\bibitem[Tendulkar et al.(2017)]{Tendulkar2017}
Tendulkar, S.~P., Bassa, C.~G., Cordes, J.~M., et al.\ 2017, \apjl, 834, L7

\bibitem[Terazawa(1979)]{Terazawa1979}
Terazawa, H.\ 1979, INS-Report-338, University of Tokyo


\bibitem[Thielemann \& Cowan(2026)]{Thielemann2026}
Thielemann, F.-K., \& Cowan, J.~J.\ 2026, The European Physical Journal A, 62, 105

\bibitem[Thompson \& Duncan(1993)]{Thompson1993}
Thompson, C., \& Duncan, R.~C.\ 1993, \apj, 408, 194

\bibitem[Thompson \& Duncan(1995)]{Thompson1995}
Thompson, C., \& Duncan, R.~C.\ 1995, \mnras, 275, 255

\bibitem[Thompson \& Duncan(2001)]{Thompson2001}
Thompson, C., \& Duncan, R.\ C.\ 2001, \apj, 561, 980

\bibitem[Thompson et al.(2004)]{Thompson2004}
Thompson, T. A., Chang, P., \& Quataert, E. 2004,
\apj, 611, 380

\bibitem[Vigan\`o et al.(2013)]{Vigano2013}
Vigan\`o, D., Rea, N., Pons, J.~A., Perna, R., Aguilera, D.~N., \&
Miralles, J.~A. 2013,
\mnras, 434, 123--141

\bibitem[Vink \& Kuiper(2006)]{Vink2006}
Vink, J., \& Kuiper, L.\ 2006, \mnras, 370, L14

\bibitem[Voskresensky et al.(2003)]{Voskresensky2003}
Voskresensky, D.~N., Yasuhira, M., \& Tatsumi, T.\ 2003,
Nucl.\ Phys.\ A, 723, 291

\bibitem[Weber(2005)]{Weber2005}
Weber, F.\ 2005, Prog.\ Part.\ Nucl.\ Phys., 54, 193

\bibitem[White et al.(2022)]{White2022}
White, C.~J., Burrows, A., Coleman, M.~S.~B., \& Vartanyan, D.\ 2022,
\apj, 926, 111

\bibitem[Wilson(1927)]{Wilson1927}
Wilson, E.~B.\ 1927, Journal of the American Statistical Association, 22, 209

\bibitem[Witten(1984)]{Witten1984}
Witten, E.\ 1984, \prd, 30, 272

\bibitem[Woltjer(1958)]{Woltjer1958}
Woltjer, L.\ 1958, Proc.\ Natl.\ Acad.\ Sci., 44, 489

\bibitem[Woods \& Thompson(2006)]{Woods2006}
Woods, P.~M., \& Thompson, C.\ 2006, in Compact Stellar X-ray Sources,
ed.\ W.~Lewin \& M.~van der Klis (Cambridge: CUP), 547

\bibitem[Woosley(2010)]{Woosley2010}
Woosley, S.~E.\ 2010, \apjl, 719, L204

\bibitem[Yang \& Zhang(2018)]{Yang2018}
Yang, Y.-P., \& Zhang, B. 2018, \apj, 868, 31

\bibitem[Yang \& Zhang(2021)]{Yang2021}
Yang, Y.-P., \& Zhang, B.\ 2021,
\apj, 919, 89

\bibitem[Yoshike et al.(2015)]{Yoshike2015}
Yoshiike, R., Nishiyama, K., \& Tatsumi, T.\ 2015,
Physics Letters B, 751, 123.

\bibitem[Young et al.(2010)]{Young2010}
Young, D.~R., Smartt, S.~J., Valenti, S., et al.\ 2010, \aap, 512, A70

\bibitem[Zhang et al.(2025)]{Zhang2025}
Zhang, J.-S., Wang, T.-C., Wang, P., et al.\ 2025 [arXiv:2507.14707]

\bibitem[Zink et al.(2012)]{Zink2012}
Zink, B., Lasky, P.~D., \& Kokkotas, K.~D.\ 2012, \prd,  85, 024030


\end{thebibliography}
\end{document}